\documentclass[lettersize,journal]{IEEEtran}
\usepackage{supertabular, ragged2e, tabularx}
\usepackage{amsfonts}
\usepackage{amsmath,amsfonts}
\usepackage[utf8]{inputenc}
\usepackage[dvipsnames]{xcolor}
\usepackage[colorlinks,bookmarksopen,bookmarksnumbered,citecolor=Maroon,urlcolor=NavyBlue,breaklinks]{hyperref}
\usepackage{graphicx}
\usepackage[backend=biber,style=ieee,sorting=none,citestyle=numeric-comp]{biblatex}
\usepackage{textgreek}
\usepackage{longtable} 
\usepackage{geometry} 
\usepackage{amsmath}
\usepackage{caption} 
\usepackage{datetime}
\usepackage{authblk}
\usepackage{multirow}
\usepackage{array}
\usepackage{booktabs}
\usepackage{makecell}
\usepackage{float}
\usepackage{relsize}
\usepackage[capitalize]{cleveref}
\usepackage{tikz}
\usetikzlibrary{arrows.meta, shapes.geometric, positioning}
\usepgflibrary{shadings}
\usetikzlibrary{positioning}
\usetikzlibrary{shapes}
\usepackage{ragged2e} 
\usepackage{siunitx}
\usepackage{tcolorbox}
\usepackage{soul}
\usepackage{fancyhdr}
\fancypagestyle{copyright}{%
    \fancyhf{}
    \cfoot{\footnotesize%
        \textcopyright~2025 IEEE. Personal use of this material is permitted.  Permission from IEEE must be obtained for all other uses, in any current or future media, including reprinting\slash republishing this material for advertising or promotional purposes, creating new collective works, for resale or redistribution to servers or lists, or reuse of any copyrighted component of this work in other works. This material is referenced by DOI: {\href{https://doi.org/10.1109/MGRS.2025.3546527}{10.1109/MGRS.2025.3546527}}
    }
}

\tcbset{
    highlightstyle/.style={colback=yellow!20, colframe=yellow!50!black, sharp corners, boxrule=0pt, left=5pt, right=5pt, top=5pt, bottom=5pt}
}

\definecolor{ForestGreen}{RGB}{34,139,34} 

\begin{document}

\title{Lossy Neural Compression for Geospatial Analytics: A Review}
\author[1]{Carlos Gomes}
\author[1]{Isabelle Wittmann}
\author[6]{Damien Robert}
\author[1]{Johannes Jakubik}
\author[5]{Tim Reichelt}
\author[4,9]{Stefano Maurogiovanni}
\author[2]{Rikard Vinge}
\author[7]{Jonas Hurst}
\author[4]{Erik Scheurer}
\author[4]{Rocco Sedona}
\author[1]{Thomas Brunschwiler}
\author[4]{Stefan Kesselheim}
\author[3]{Matej Bati\v c}
\author[5]{Philip Stier}
\author[6]{Jan Dirk Wegner}
\author[4,9]{Gabriele Cavallaro}
\author[7]{Edzer Pebesma}
\author[2]{Michael Marszalek}
\author[8]{Miguel A. Belenguer-Plomer}
\author[4,9]{Kennedy Adriko}
\author[1]{Paolo Fraccaro}
\author[1]{Romeo Kienzler}
\author[4]{Rania Briq}
\author[4]{Sabrina Benassou}
\author[8]{Michele Lazzarini}
\author[2]{Conrad M Albrecht}

\affil[1]{IBM Research - Europe, Switzerland}
\affil[2]{German Aerospace Center, Germany}
\affil[3]{Sinergise Solutions, Slovenia}
\affil[4]{Forschungszentrum Jülich, Germany}
\affil[5]{University of Oxford, United Kingdom}
\affil[6]{University of Zurich, Switzerland}
\affil[7]{University of Münster, Germany}
\affil[8]{European Union Satellite Centre, Spain}
\affil[9]{University of Iceland, Iceland}


\markboth{IEEE GEOSCIENCE AND REMOTE SENSING MAGAZINE}%
{Gomes \MakeLowercase{\textit{et al.}}: Lossy Neural Compression for Geospatial
Analytics: A Review}
\maketitle
\begin{abstract}

Over the past decades, there has been an explosion in the amount of available Earth Observation (EO) data. 
The unprecedented coverage of the Earth's surface and atmosphere by satellite imagery has resulted in large volumes of data that must be transmitted to ground stations, stored in data centers, and distributed to end users. Modern Earth System Models (ESMs) face similar challenges, operating at high spatial and temporal resolutions, producing petabytes of data per simulated day.

Data compression has gained relevance over the past decade, with neural compression (NC) emerging from deep learning and information theory, making EO data and ESM outputs ideal candidates due to their abundance of unlabeled data.

In this review, we outline recent developments in NC applied to
geospatial data. We introduce the fundamental concepts of NC including seminal works in its traditional applications to image and video compression domains with focus on lossy compression. We discuss the unique characteristics of EO and ESM data, contrasting them with ``natural images'', and explain the additional challenges and opportunities they present. Additionally, we review current applications of NC across various EO modalities and explore the limited efforts in ESM compression to date.
The advent of self-supervised learning (SSL) and foundation models (FM) has advanced methods to efficiently distill representations from vast unlabeled data. We connect these developments to NC for EO, highlighting the similarities between the two fields and elaborate on the potential of transferring compressed feature representations for machine--to--machine communication.
Based on insights drawn from this review, we devise future directions relevant to applications in EO and ESM.  

\end{abstract}
\thispagestyle{copyright}

\newpage
{\small\tableofcontents}
\newpage


\section{Introduction}
\label{sec:intro}
\subsection{Motivation \& Approach}

\IEEEPARstart{E}{arth} Observation (EO) is the process of capturing data about the Earth's surface and atmosphere, carried out through instruments on board of satellites, airborne vehicles, ships, or through ground stations. 
Owing to their constant activity and wide coverage, the bulk of these data is produced by satellites, with the Copernicus system alone delivering a reported 16 terabytes of data per day~\cite{copernicus2022}.
As large as this amount of data already is, it is only set to increase, with over 100 new EO satellites launched in 2021, over 150 in 2022, and almost 250 in 2023~\cite{clauson_etal_2024}.
Earth System Models (ESMs) simulate the evolution of components of the Earth system to predict future climate and air pollution. 
These systems also produce large volumes of data, and, driven by the need for higher resolutions to predict increasingly complex phenomena, these volumes are certain to increase with next-generation ESMs. In fact, data output and storage have already become a major bottleneck for high-resolution climate modeling.

Among others, this situation sparked projects to utilize AI methodologies to come at rescue, cf.\ ESA's MajorTOM dataset with embeddings~\cite{czerkawski2024global}, ESA's project  \textit{CORSA}\footnote{\url{https://remotesensing.vito.be/about}}, the EU Horizon project \textit{Embed2Scale}\footnote{\url{https://embed2scale.eu/vision-strategy}}, as well as the \textit{Earth Index}\footnote{\url{https://www.earthgenome.org/earth-index}} solution developed by the start-up Earth Genome. In general, and beyond EO and ESM, a vibrant and innovative start-up community for NC emerged in recent years, with notable examples such as Deep Render\footnote{\url{https://deeprender.ai}}.
\begin{figure}[t]
    \hspace{-3.5ex}\includegraphics[width=1.2\linewidth]{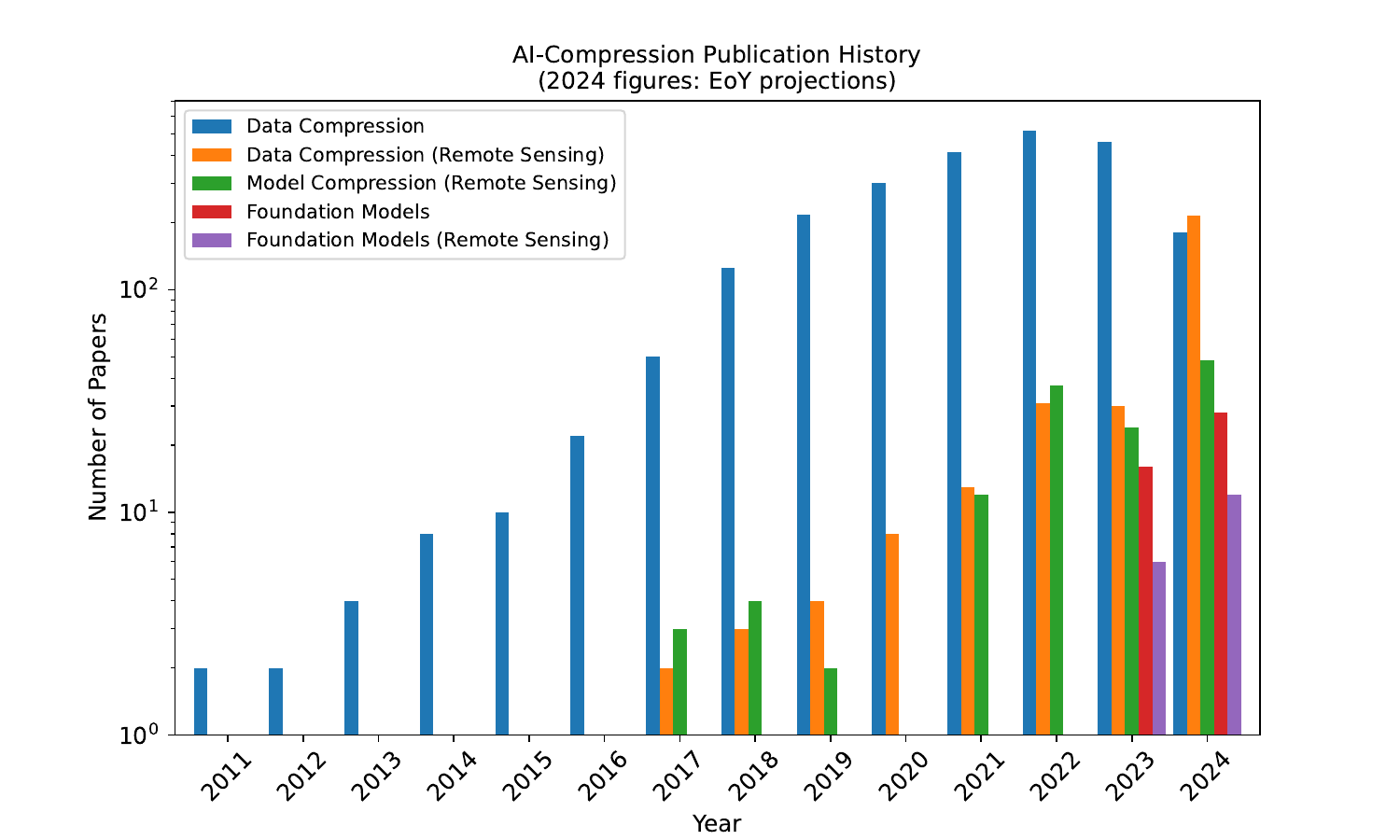}
    \caption{Literature on neural compression (NC) summarized for the past 15 years. We plot separate bars for remote sensing (RS) and recent developments in foundation model (FM) methodology. Data Source: Queries to the \textit{Web of Science}~\cite{WoCGraphic2000-2024AICompressorHistory}.}
    \label{fig:AICompressorPublicationHistory}
\end{figure}

The importance of accessing EO and ESM data for analysis cannot be overstated.
ESMs are vital for predicting the course of climate change and its potential impacts across the Earth.
For the next generation of high-resolution climate models, it is currently no longer possible to store the full range of simulated parameters. Hence, only a small subset is stored for future analysis, making computationally expensive re-runs necessary.  

To best utilize EO data, it is critical to store it for long-term usage, enabling comparative studies, as well as distribute it to end users effectively.
There are two main bottlenecks in doing this:
\begin{enumerate}
    \item Bandwidth between satellite and base stations required for transferring the observations for storage and analysis on the ground is limited. This is a well-known problem in the community, referred to as the data downlink bottleneck~\cite{downlink_latency}.
    While close-to-lossless transmission is desired for comprehensive EO data archives, dedicated (nano-)satellites with focus on specialized applications open the opportunity to implement lossy compression for near real-time geospatial analytics.  
    \item Storage of such a volume of data on a physical medium is expensive, and its transfer through a network (typically from data center to research institutions distributed globally) causes egress costs and delays research.
\end{enumerate}
\begin{figure}[t]
    \centering
    \includegraphics[width=\linewidth]{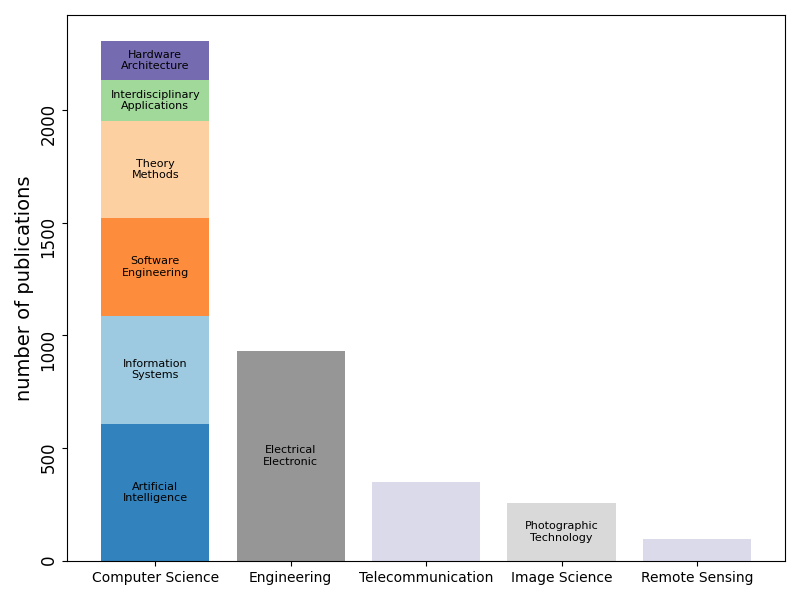}
    \caption{Domain-specific shares in publications in NC methodologies from years 2000 through 2023 for the ten biggest (sub)categories as per \textit{Web of Science}~\cite{WoCGraphic2024AICompressor}. The stacked subcategories for \textit{Computer Science} have been ordered from largest to smallest bottom-up.}
    \label{fig:AICompressorDomainShare}
\end{figure}

\textit{Traditional data compression} methods, mostly based on the JPEG2000 standard~\cite{yeh_new_2005}, have long been used to reduce the required storage. However, the growth in data volume demands new approaches to more efficiently store only those aspects of the data that are required for their reconstruction or usage for geospatial analytics.

\textit{Neural compression} (NC) emerges as a natural candidate to improve compression algorithms.
It has been shown in the literature to outperform traditional, hand-designed compression algorithms 
on curated datasets across several fields such as image compression~\cite{autoregressive_hyperprior}, video compression~\cite{mentzer2022vct}, and audio compression~\cite{zeghidour2021soundstream}.
NC uses deep neural networks to perform data compression.
It seeks to identify and efficiently store the critical aspects of the data, discarding irrelevant or repetitive information, by learning directly from the data. Here, we associate the notion of \textit{irrelevant information} with lossy compression. In contrast to traditional approaches, loss of information is not set by an external parameter in NC. Relevant information is inherently learned from the data distribution itself. 
NC algorithms are mostly based on autoencoders~\cite{Hinton1994} and do not require labels to be trained.
However, they do rely on large datasets which representatively sample the underlying distribution of the data.
This creates an inviting environment for adapting and applying NC techniques to EO and ESM.
\textit{Neural Compression for Geospatial Analytics} embraces computational algorithms employing artificial neural networks to reduce the storage required to digitize geospatial data while comprehending its (relevant) information content.
The present review focuses on this approach, aiming to foster research on NC for EO and ESM data. 
Specifically, we focus on lossy NC, where higher rates of compression can be obtained as some loss of information is allowed to occur.

\cref{fig:AICompressorPublicationHistory} reveals the historical growth in popularity of NC in research publications. 
The plot also demonstrates that applying NC to RS developed since 2017---with a lag of about 6 years to general NC.
However, \cref{fig:AICompressorDomainShare} illustrates that it is still a relatively unexplored topic in remote sensing (RS), making up only $3\%$ of publications in NC methodologies.

\textit{Foundation Models} (FMs) became an emergent and related topic to NC, recently (cf.\ \cref{fig:AICompressorPublicationHistory}): large pretrained neural networks that seek to learn embedding spaces can be leveraged for multiple downstream tasks in a domain.
In contrast to NC, FMs have been quickly adopted in RS domains~\cite{lu2024ai}, and a community forms around FMs for ESM~\cite{mukkavilli2023ai}.
Despite not being directly optimized to compress data, FMs share similarities with NC, with both being trained on very large unlabeled datasets to extract fundamental features from data.

The emergence of FMs represents a fundamental paradigm shift in deep learning. This shift has primarily resulted from three factors: (1)~the availability of vast amounts of \textit{unlabeled (geospatial) data}, (2)~the emergence of self-supervised learning (SSL)\footnote{a concept that allows deep learning models to learn from unlabeled data} in RS~\cite{wang2022self}, and (3)~a significant increase of computational power enabling large models to be trained at scale~\cite{bommasani2021opportunities}. The absence of human-annotated labels in such large-scale training processes results in \textit{task-agnostic} deep learning models that are generally referred to as FMs. Within a couple years, a zoo of models for various satellite data modalities and weather models has been developed, \cite{jakubik2023foundation,lessig2023atmorep,sun2022ringmo,liu2024remoteclip,wang2022advancing,guo2024skysense,hong2023spectralgpt,bodnar2024aurora,braham2024spectralearth}.
\begin{table*}[t!]
    \centering
    \begin{tabular}{lll}
    \toprule \textbf{Aspect}     & \textit{Neural Compression}                                  & \textit{Traditional Compression}\\
    \midrule \textbf{Approach}   & neural networks \& machine learning~\cite{yang2023introduction}    & predefined algorithms\\
    \textbf{Adaptability}        & learn from data, flexible~\cite{Zhang2024CVPRa}              & fixed algorithms tailored to data format\\
    \textbf{Performance}         & potentially higher compression efficiency~\cite{Kim2023CVPR} & consistent performance\\ \textbf{Comput.\ Complexity} & high~\cite{Duan2024CVPR}                                     & low\\
    \textbf{Lossy vs.\ Lossless} & handles both~\cite{Zhang2024CVPRb}                           & design specialized to either\\ \textbf{Costs} (today)       & high (active research, model training)~\cite{Khoshkhahtinat2024CVPR}& low (established field, algorithms optimized)\\
    \textbf{Use Cases} (today)   & image\slash video compression, specialized tasks~\cite{Li2024ECCV}& general-purpose\\
    \bottomrule
    \end{tabular}
    \caption{Comparing key aspects of compression for traditional and neural approaches. For NC we picked a 2024 sample publication. A more comprehensive list of relevant papers for lossy NC provides \cref{tab:neural_overview,fig:AICompressorApplicationSpace}}
    \label{fig:NeuralVSTradCompress}
\end{table*}

Empirical evidence demonstrates several improved capabilities of FMs for RS~\cite{wang2023ssl4eo} and atmospheric modeling compared to supervised deep learning models. For example, recent work has shown a significant acceleration in solving tasks in RS based on pretrained large-scale SSL models (e.g.,~\cite{sun2022ringmo}). In addition, fine-tuning pretrained FMs can significantly reduce the required amount of task-specific, often human-annotated, data, thereby improving data efficiency compared to traditional supervised deep learning~\cite{jakubik2023foundation}.
Furthermore, recent research demonstrated that FMs for RS benefit from their pre-training when generalizing to other, unseen geographical regions. For example, models have performed better on segmenting flood extents in regions that have not been part of the pre-training data compared to other supervised deep learning approaches~\cite{li2023assessment}. 

Despite various benefits resulting from FMs for RS of land and atmosphere, several challenges remain---especially regarding their significant data consumption, creating significant data transmission and storage bottlenecks. While FMs can be seen as performing a certain compression of the raw data in the embedding space, those embeddings are still relatively large. This makes the NC of embeddings particularly interesting~\cite{furutanpey2024fool,gomes2024compressed}, especially in an upcoming era of large growth in data generation. We discuss FMs in their combination with NC to generate compressed features in \cref{sec:lossy_compression} and \cref{sec:compression_RS}.

\begin{figure*}[ht!]
\centering
\usetikzlibrary{arrows.meta, shapes.geometric, positioning}
\begin{tikzpicture}[node distance=0.5cm and 1cm, every node/.style={font=\small, align=center}]

    \node (compression) [rectangle, draw, rounded corners, text width=2.5cm, align=center, fill=blue!10] {Compression};
    \node (lossless) [above right=of compression, rectangle, draw, rounded corners, text width=2.5cm, align=center, fill=gray!10] {Lossless};
    \node (lossy) [below right=of compression, rectangle, draw, rounded corners, text width=2.5cm, align=center, fill=blue!10] {Lossy};
    \node (hand) [above right=of lossy, rectangle, draw, rounded corners, text width=3.5cm, align=center, fill=gray!10] {Hand-Crafted};
    \node (learned) [below right=of lossy, rectangle, draw, rounded corners, text width=2.5cm, align=center, fill=blue!10] {Learned};

    \node (transforms) [right=2cm of learned, yshift=1.25cm, rectangle, draw, rounded corners, text width=2.5cm, align=center, fill=red!10] {Transforms};
    \node (quant) [below=0.5cm of transforms, rectangle, draw, rounded corners, text width=2.5cm, align=center, fill=green!10] {Quantization Strategies};
    \node (entropy) [below=0.5cm of quant, rectangle, draw, rounded corners, text width=2.5cm, align=center, fill=yellow!10] {Entropy Models};
    \node (objective) [below=0.5cm of entropy, rectangle, draw, rounded corners, text width=2.5cm, align=center, fill=brown!10] {Optimization Objectives};

    \draw[->] (compression) -- (lossless);
    \draw[->] (compression) -- (lossy);
    \draw[->] (lossy) -- (learned);
    \draw[->] (lossy) -- (hand);

    \draw[-] (learned.east) to[out=0, in=180] (transforms.west);
    \draw[-] (learned.east) to[out=0, in=180] (quant.west);
    \draw[-] (learned.east) to[out=0, in=180] (entropy.west);
    \draw[-] (learned.east) to[out=0, in=180] (objective.west);

\end{tikzpicture}

\caption{Taxonomy of compression methods. This review focuses on the family of methods defined by following the blue nodes. We group all variations within this family into transforms, quantization strategies, entropy models, and optimization objectives.}
\label{fig:compression_taxonomy}
\end{figure*}
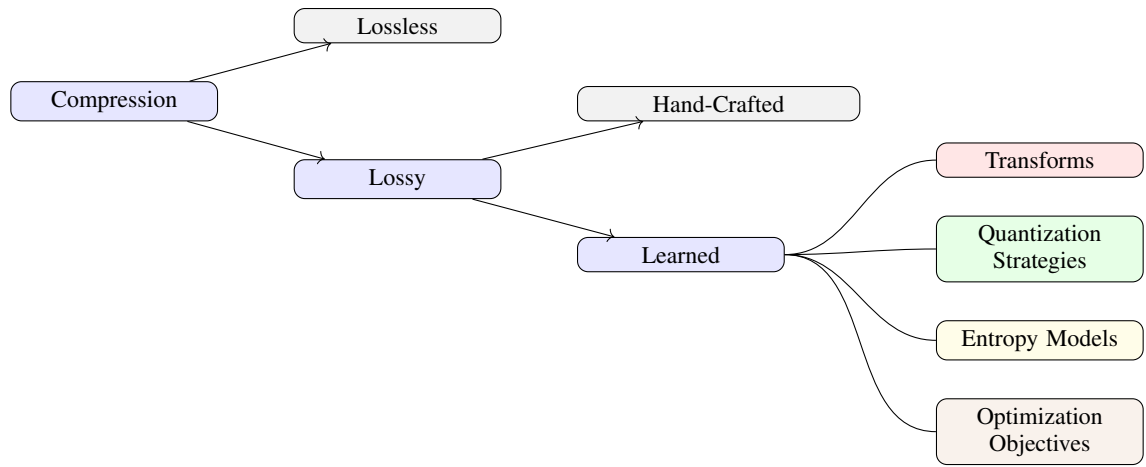

\subsection{Compression}
\label{sec:neural_traditional_compression}
Before diving into a more formal approach in \cref{sec:lossy_compression:transform_coding}, we share an intuitive picture of compression. When it comes to NC, it is worth noting that our review concerns elements of data compression. There exists an entire field of \textit{model compression}~\cite{deng2020model} that revolves around pruning the size of artificial neural networks. We will not touch this domain here.

\subsubsection{Neural vs.\ Traditional Compression}

Compression algorithms aim to encode a signal in as few symbols (and ultimately bits) as possible.
These algorithms are core enablers of modern computing infrastructures, allowing different types of data to be stored and transmitted without prohibitively large costs.

Traditionally, compression algorithms (\textit{codecs}) consist of a pipeline of components that have been hand-engineered by experts to compress signals of a specific type.
We denote them as engineered \textit{by hand} in the sense that they are not the direct result of data-driven algorithms, but rather human-crafted.
These rely on signal processing and information theory, with each codec requiring a large number of human hours of work, often organized through consortia.
Currently, virtually all codecs seeing widespread use belong to this category, such as MP3~\cite{mp3}, H.264~\cite{H.264}, HEVC~\cite{hevc} or JPEG~\cite{jpeg}, to name only a few.

Learning-based methods, including artificial neural networks, have been explored for compression since at least the 1980s~\cite{early_nn_compression}.
However, with the recent emergence of deep learning, promising results~\cite{theis2022lossy, balleEndtoendOptimizedImage2016} led to a growth of research in the field of \textit{neural data compression}.
The main premise of NC is to replace traditionally hand-designed components of codecs with data-driven modules, usually neural networks, typically learned over a large representative dataset.
Ultimately, not just individual components are replaced, but rather the whole pipeline, leading to a codec that is learned fully end-to-end. \Cref{fig:NeuralVSTradCompress} provides a high-level overview of pros and cons for both neural and traditional compression methods.

Two main benefits emerge from learned approaches.
The first is the reduction in expensive expert hours required for elaborating algorithms, relying on data-driven processes to determine the transformations applied to the data.
By modifying the loss function used to train the neural network, the codec can explicitly prioritize different aspects of the data, depending on its type and use case, as opposed to manually tuning the parameters of different components in a traditional codec.

The second is the potential for improved compression, in particular, due to the joint optimization of all learned components.
Especially when the domain of data is known \textit{a priori} (e.g. optical imagery from satellites) and fixed, a neural codec can specialize to that domain simply through the design of its training dataset, granting it an advantage compared to traditional codecs designed to offer stable performance across many domains.

Despite the complexity of modern hand-designed codecs, compression algorithms can essentially achieve their goal in two main ways, both of which deep learning proposes to improve.

\textbf{Encoding the signal using fewer symbols:}
An accurate model of the distribution of the data (in particular one that takes into account the context surrounding a given symbol) is a crucial building block to be able to cleverly encode data.
Leveraging neural networks allows us to learn complex models of the underlying data distribution, leading to more effective encoding schemes.

\textbf{Allowing for some loss of information:}
In lossy compression, some parts of a signal may be deemed as unimportant or too costly to encode and thus may be dropped, leading to a potentially large reduction in the length of the encoded signal.
Understanding which parts of the signal to drop to minimize the impact on its reconstruction is critical in designing such algorithms.
By leveraging deep neural networks to learn the structure of the data, better trade-offs between message length and reconstruction quality can be discovered.

\subsubsection{Neural Compression for Imagery}
\begin{figure*}[t!]
    \centering
    \includegraphics[width=\textwidth]{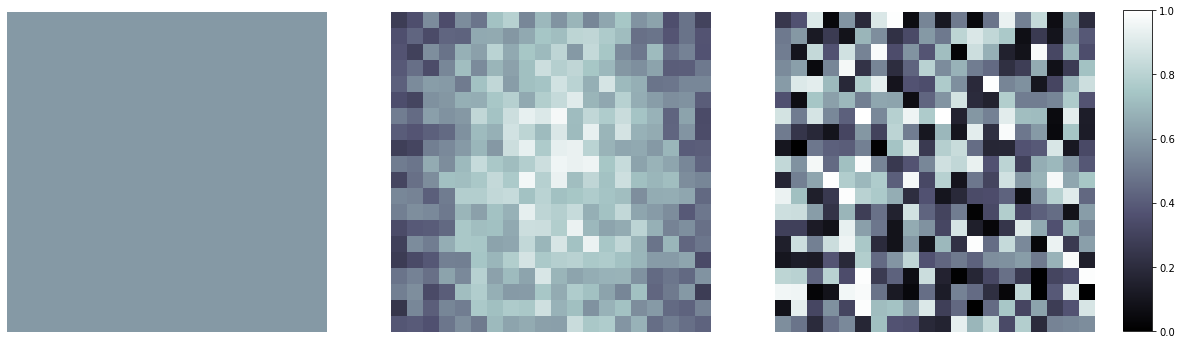}   
    \caption{Image compression from the perspective of pixel correlations, \textbf{Left} to \textbf{Right}: constant image, correlated noise, uncorrelated noise. A detailed description is provided in the main text.}
    \label{fig:ImageCompressionFromInformationTheory}
\end{figure*}

As introduced above, we distinguish traditional and NC methods. While the former easily allows to design algorithms with perfect reconstruction (\textit{Lossless}), the latter utilizes deep neural networks with millions to billions of learnt parameters that are, as of today, black box models hard to explain (\textit{Learned}). Correspondingly, NC is primarily lossy; although there also exist initial approaches of designing lossless codecs with neural approaches~\cite{Zhang2024CVPRb}. \Cref{fig:compression_taxonomy} summarizes our taxonomy categorizing NC (\textit{Learned}) further as detailed in \cref{sec:compression_taxonomy}.

Data compression exploits knowledge about the distribution $p(\mathbf x)$ of data $\mathbf x$ when encoding a sequence of such data. \Cref{fig:ImageCompressionFromInformationTheory} illustrates the concept of compressibility of image data for an intuition on lossy compression. Let us consider the sequence of $N$ data points being generated from a gray-scale image ($\mathbf x_i=\text{const}\in[0,1]$) of dimensions $\sqrt N \times\sqrt N$ read out row-by-row (or column-by-column). In the extreme case of all pixels having the same constant value (\cref{fig:ImageCompressionFromInformationTheory}, left), a neural network $f(\mathbf x)=z$ can learn to compress the input down to a single scalar $z$ in feature space without loss. All image pixels are highly correlated by assuming exactly the same grayscale value. The entropy of the image is zero. On the other end of the spectrum, we may encounter random images (\cref{fig:ImageCompressionFromInformationTheory}, right) where every single pixel is completely uncorrelated to all the other ones. Any compression to $N'<N$ symbols will imply information loss. Therefore,  dimensional reduction is impossible.

Notice that the ability to losslessly compress is related to the smoothness (or sharpness) of an image. Certainly, if neighboring pixels are totally uncorrelated, on average, their random grayscale value generates sharp contrasts. The introduction of correlation smoothens and causes the emergence of recognizable patterns, as exemplified by \cref{fig:ImageCompressionFromInformationTheory}, center: A compressor may use the pattern that pixels towards the center of the image tend to become more bright compared to the boundary regions. In contrast to hand-crafted base functions---e.g., planar waves (Fourier transform) or wavelets (JPEG2000 encoder)~\cite{mallat1999wavelet}---NC bears the potential to automatically learn complex correlation patterns in images (and their associated base functions) from the data itself.

\subsection{Outline}

The goal of this review is to introduce the current status of \textit{Neural Compression (NC) for Earth Observation (EO) and Earth System Modeling (ESM) data} with a focus on lossy image and video compression. We target an academic audience while staying as self-consistent as possible. More specifically,
\begin{itemize}
    \item \Cref{sec:lossy_compression} will provide relevant background and an overview of state-of-the-art NC for lossy image and video compression,
    \item \Cref{sec:compression_RS} continues by exploring NC in EO whereas \cref{sec:compression_Climate} is dedicated to the compression of the outputs from ESMs, with both sections laying out corresponding challenges in applying NC. 
   \item \Cref{sec:implementation_application} provides an introduction on how NC for geospatial analytics may integrate with corresponding data platforms for implementation.
\end{itemize}
Using concrete examples, we discuss how this integration democratizes geospatial applications in domains such as global vegetation monitoring, maritime awareness, climate modeling, and agriculture management.
We close our review in \cref{sec:perspectives}, highlighting relevant future directions for the field of NC for geospatial analytics.

\section{(Lossy) Neural Compression}
\label{sec:lossy_compression}
We begin with an introduction to lossy compression, presenting NC as an extension of transform coding.
We then provide an overview of the NC literature, propose a taxonomy for navigating the field, and detail works most relevant to our geospatial focus.
For a more theoretical and in-depth introduction to the field, we refer readers to \textcite{yang2023introduction}. 

\subsection{Background}
For the sake of simplicity, let us assume a finite set of data we wish to transmit, with each independent element indexed by $i=1\dots M$. In a sequence of data of length $N$, we count the frequency $n_i$ of data point $\mathbf x_i$ such that $N=\sum_in_i$ or equivalently $\sum_ip_i=1$. The corresponding (Shannon) entropy, which quantifies the amount of information (or the uncertainty) associated with the data,
\begin{equation}
    H(\mathbf p=(p_1,\dots,p_M))=-\sum_ip_i\log p_i
\end{equation}
is maximal for the uniform distribution when $p_i=1/M=p,~i=1\dots M$, i.e.\ all data points $\mathbf x_i$ are equally likely. A simple encoding scheme for $M$ distinct data points can be represented as a \textit{key-value} (lookup) table $T$ where the \textit{values} refer to the $\mathbf x_i$, and the \textit{keys} represent a binary sequence of bits. Using a simple encoding scheme, we can number all M data points using $b=\log M(=-\log p)$ bits for each key, such that $2^b=M$. However, we can notice that in the extreme case where $p_I=1$ for a fixed $I$ and $p_{i\neq I}=0$ the entropy $H$ drops to zero.
Given that we certainly know we are just transmitting a single data point $N$ times, we do not need to transmit any bit, i.e. we can use a different encoding scheme where we use 0 bits to encode data point $I$ such that $b=0=\log1(=-\log p_I)$. If we more rigorously generalize this thought to include cases outside these two extremes, Shannon's source coding theorem~\cite{shannon_source_coding} tells us that in the optimal encoding scheme, the length of the bit sequence indexing each data point $\mathbf x_i$ should be inversely related to its frequency $n_i$. Precisely, \
\begin{equation}
    b_i = \log \frac{1}{n_i} = -\log p_i\quad.
\end{equation}
In this more generic scenario, we observe that the entropy $H$ expresses the average bit length required to encode $\mathbf x_i$ based on its probability $p_i$:
\begin{equation}
    H(\mathbf p)=\sum_ip_ib_i=\mathbb E[b]\quad.
\end{equation}
Using the table $T$ with keys whose lengths follow Shannon's source coding theorem (rounding up to the nearest integer) enables the encoding and decoding of the sequence of $N$ data points without any loss---it is lossless compression.
    
NC replaces the lookup table $T$ with a separate encoder artificial neural network $f$ and a decoder artificial neural network $g$. The finite size $M$ is enforced by a discretization step (\textit{quantization}) in embedding/ feature space $\mathbf z=f(\mathbf x)$. For the purpose of illustration $g\circ f$ may represent an autoencoder (AE), or its probabilistic sibling, a \textit{Variational Autoencoder} (VAE)~\cite{bank2023autoencoders}. In case such a VAE is trained alongside with $K$ learnable feature vectors $\{\mathbf z^{(k)}\}_{k=1\dots K}$ in embedding space such that every encoding $\mathbf z=f(\mathbf x)$ is \textit{snapped} to one of these vectors (quantization step), the resulting neural compressor is termed \textit{Vector Quantized}--VAE (VQ--VAE)~\cite{van2017neural}.

\subsubsection{Lossy Compression}

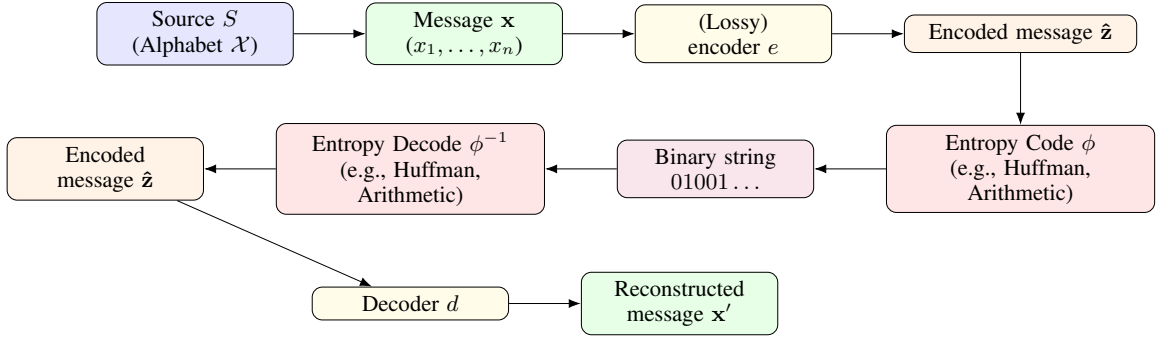
\begin{figure*}[ht!]
\centering
\scalebox{.95}{
\begin{tikzpicture}[node distance=1cm and 1cm, every node/.style={font=\small, align=center}, >=Latex]

    \node (source) [rectangle, draw, rounded corners, text width=2.5cm, align=center, fill=blue!10] {Source $S$\\(Alphabet $\mathcal{X}$)};
    \node (message) [right=of source, rectangle, draw, rounded corners, text width=2.5cm, align=center, fill=green!10] {Message $\mathbf{x}$\\($x_1, \ldots, x_n$)};
    \node (encoder) [right=of message, rectangle, draw, rounded corners, text width=2.5cm, align=center, fill=yellow!10] {(Lossy)\\encoder $e$};
    \node (encoded) [right=of encoder, rectangle, draw, rounded corners, text width=3cm, align=center, fill=orange!10] {Encoded message $\mathbf{\hat{z}}$};
    \node (entropy) [below=of encoded, rectangle, draw, rounded corners, text width=3.5cm, align=center, fill=red!10] {Entropy Code $\phi$\\(e.g., Huffman, Arithmetic)};
    \node (binary) [left=of entropy, rectangle, draw, rounded corners, text width=2.5cm, align=center, fill=purple!10] {Binary string\\$01001\ldots$};
    \node (entropy2) [left=of binary, rectangle, draw, rounded corners, text width=3.5cm, align=center, fill=red!10] {Entropy Decode $\phi^{-1}$\\(e.g., Huffman, Arithmetic)};
    \node (encoded2) [left=of entropy2, rectangle, draw, rounded corners, text width=2.5cm, align=center, fill=orange!10] {Encoded message $\mathbf{\hat{z}}$};
    \node (decoder) [below=of entropy2, rectangle, draw, rounded corners, text width=2.5cm, align=center, fill=yellow!10] {Decoder $d$};
    \node (decoded) [right=of decoder, rectangle, draw, rounded corners, text width=2.5cm, align=center, fill=green!10] {Reconstructed message $\mathbf{x'}$};

    \draw[->] (source) -- (message);
    \draw[->] (message) -- (encoder);
    \draw[->] (encoder) -- (encoded);
    \draw[->] (encoded) -- (entropy);
    \draw[->] (entropy) -- (binary);
    \draw[->] (binary) -- (entropy2);
    \draw[->] (entropy2) -- (encoded2);
    \draw[->] (encoded2) -- (decoder);
    \draw[->] (decoder) -- (decoded);

\end{tikzpicture}
}
\caption{Depiction of typical compression pipeline.}
\label{fig:compression_pipeline}
\end{figure*}
Lossy compression techniques reduce data size by allowing some loss of information. This section outlines the typical steps involved in a compression pipeline, as illustrated in~\cref{fig:compression_pipeline}.
In lossy compression, perfect signal reconstruction is not required. Instead, an approximate reconstruction is acceptable, enabling higher compression ratios.
Let $S$ be a random variable, known as the source, producing symbols that form strings $\mathbf{x} := (x_1, x_2, \ldots, x_n)$ from an alphabet $\mathcal{X}$. We refer to $\mathbf{x}$ as the signal or message to be compressed.
To compress $\mathbf{x}$, we require an encoder $e$ that maps it to a string of symbols $\mathbf{\hat z} := (z_1, z_2, \ldots, z_m)$ from a different alphabet $\mathcal{Z}$.
A decoder $d$ then seeks to reconstruct $\mathbf{x}$ as $d(\mathbf{\hat z}) = \mathbf{x^\prime}$. 
To efficiently transmit $\mathbf{z}$, the encoder and decoder agree on an encoding scheme $\phi$, known as an \textit{entropy code}, which losslessly encodes $\mathbf{\hat z}$ into a binary string.
Examples of entropy coding schemes include \textit{Huffman Coding}~\cite{huffman} or \textit{Arithmetic Coding}~\cite{arithmetic_coding}.
Intuitively, $\phi$ assigns shorter binary codes to frequently occurring symbols, to reduce the overall length of the encoded message without losing information.

Together, the encoder $e$, decoder $d$, and entropy code $\phi$ constitute a \textit{codec}. 
The objective of $e$ is to minimize the length of the encoded string $\phi(\textbf{z})$, known as the \textit{code-length}, while minimizing the information loss in the reconstructed signal $\mathbf{x^\prime}$. This introduces a trade-off between compression rate and the distortion of the reconstruction.

We can express this trade-off as a Lagrangian optimization problem.
\begin{align}
    \min \lambda D + R~.
    \label{eq:lagrangian_lossy}
\end{align}
Here, $R$ represents the rate, defined as the expected number of bits required to transmit a data point, and $D$ denotes the distortion, the expected error between a data point and its reconstruction. By varying $\lambda$, different codecs achieve various trade-offs between $R$ and $D$, which can be visualized using a Rate--Distortion plot.

\begin{figure*}[t]
    \centering
    \includegraphics[width=\linewidth]{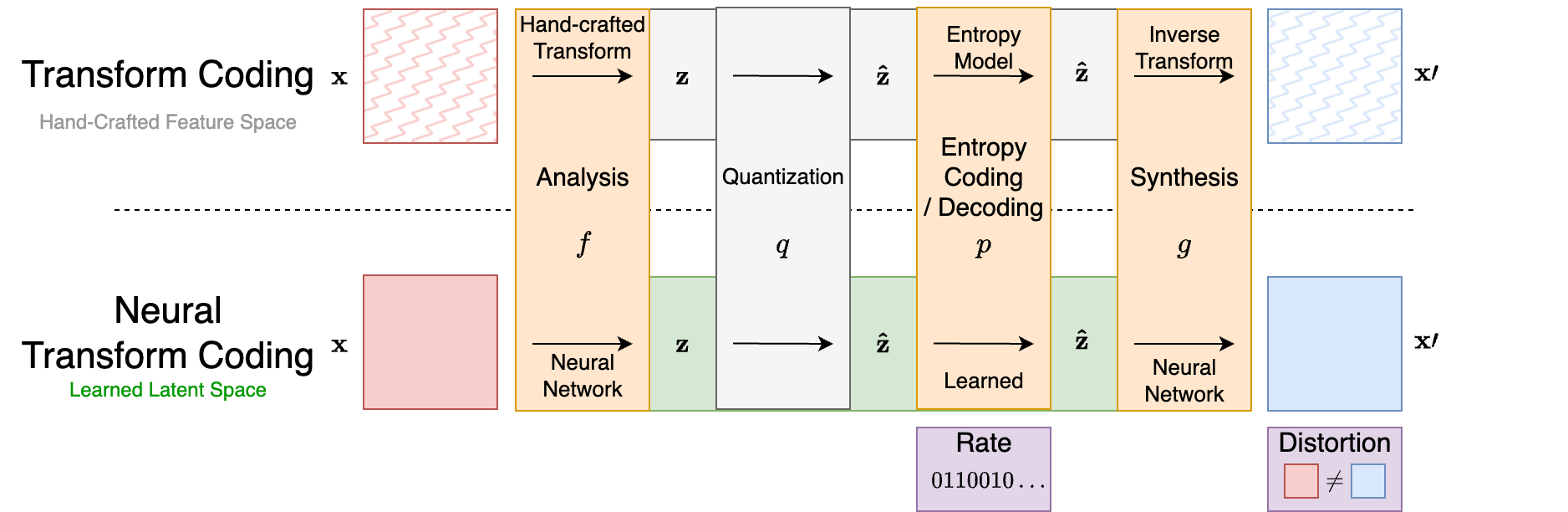}
    \caption{Illustration of the transform coding framework and its adaptation to neural transform coding. Components within orange boxes are replaced with learned counterparts. Terms within purple boxes are used in the loss.}
    \label{fig:neural_transform_coding}
\end{figure*}

The rate term $R$ is grounded in Shannon's source coding theorem~\cite{shannon_source_coding}, which provides a lower bound on the number of bits required to encode $\mathbf{\hat z}$ losslessly: 
\begin{align}
    -\log_2 p(\mathbf{\hat z})~,
    \label{eq:shannon_source_coding}
\end{align}
Here, $p$ is the probability mass function of the distribution of $\mathbf{\hat z}$. The entropy, defined as the expectation of this quantity:
\begin{align}
    \mathbb{E}[-\log_2 p(\mathbf{\hat z})]~,
    \label{eq:entropy}
\end{align}
measures the average minimum number of bits required per symbol for optimal encoding.  It characterizes how difficult it is to compress samples drawn from $p$.

While the entropy value represents the theoretical lower bound on the rate, it is not the actual number of bits achieved by a specific algorithm (known as the \textit{operational rate}). However, encoding schemes such as arithmetic coding~\cite{arithmetic_coding} can achieve operational rates very close to this theoretical limit~\cite{blelloch2001introduction}.
This makes entropy a good approximation which is useful in optimization frameworks to compute gradients for loss functions. It is important to note, that the accuracy of this approximation depends on the quality of the modeled distribution $p$.

The distortion term $D$ relies on an underlying error function $\rho(\mathbf{x}, \mathbf{x^\prime})$, which quantifies the difference between the input and reconstruction. Common choices for image and video compression are the Mean Squared Error (MSE) and the Structural Similarity Index Measure (SSIM)~\cite{ssim}.

\subsubsection{Transform Coding}
\label{sec:lossy_compression:transform_coding}

Most NC methods can be viewed as learned, non-linear variations of the transform coding paradigm~\cite{neural_transform_coding}.
Transform coding, as illustrated in Figure~\ref{fig:neural_transform_coding}, is fundamental to codecs like JPEG or HEVC. The encoder first applies a transform to the raw data, followed by quantization. This transform is designed to map the data to a space where it can be compressed more efficient.

\textbf{Transform.}
In traditional compression, the transform $f$ is typically a handcrafted, linear, and invertible mapping known as the \textit{analysis transform}.  Commonly used in image compression are the Fourier Transform~\cite{bracewell1986fourier} and the Discrete Wavelet Transform~\cite{wavelets}. These transforms aim to decorrelate the input data to facilitate the quantization and entropy modeling steps that follow.

\begin{figure*}[t]
  \includegraphics[width=.49\linewidth]{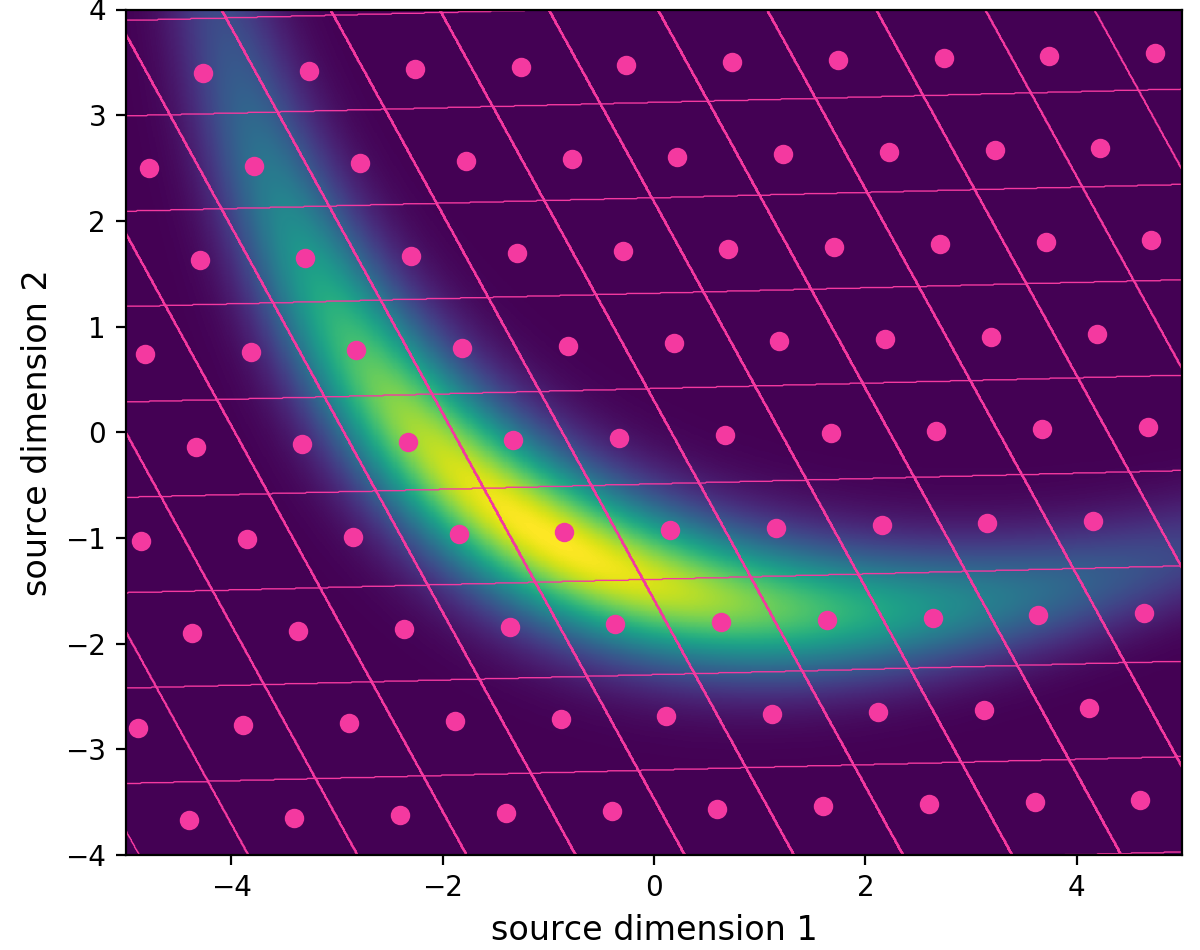}\hfill%
  \includegraphics[width=.49\linewidth]{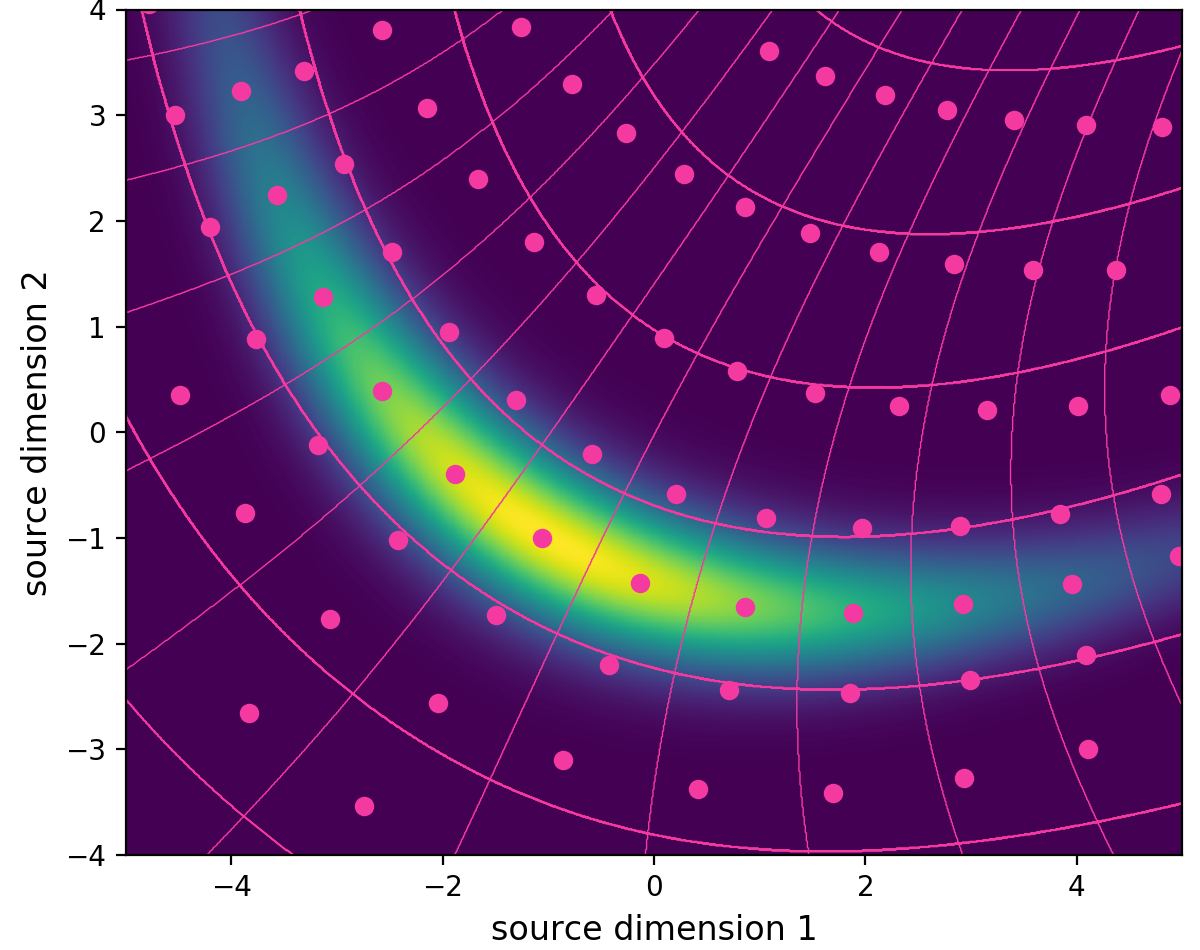}
  \caption{Linear transform code (LTC, left), and nonlinear transform code (NTC, right) of a banana-shaped source distribution, both obtained by minimizing the rate--distortion Lagrangian. Lines represent quantization bin boundaries, while dots indicate code vectors. NTC adapts more closely to the source distribution, resulting in better compression performance compared to LTC. Figure and caption taken from~\cite{balle2020nonlinear}.}
  \label{fig:nonlinear_transform}
\end{figure*}

The input data $\mathbf{x}$ is often expressed as a vector whose coordinates are correlated. 
For example, adjacent pixels in natural images tend to have similar values.
These correlations introduce redundancies in the input signal: knowing part of the signal allows one to predict other parts.
Hence, discarding such correlations is desirable in a compression framework.
To address this, a \textit{transform} $f$ is applied to map the data into new representation $\mathbf{z} = f(\mathbf{x})$, where coordinates are less correlated and ideally independent.
In NC, $f$ is an artificial neural network, trained to map $\textbf{x}$ to an \textit{embedding} $\mathbf{z}$ within a continuous latent space. Although ambiguous in the field of compression, this neural network $f$ is often referred to as the \textit{encoder network}.
The ability to learn complex, non-linear transforms directly from dataset statistics puts NC approaches at a great advantage compared to traditional handcrafted methods.
Figure~\ref{fig:nonlinear_transform} illustrates this advantage with a toy example.
    
\textbf{Quantization.} 
The output of the transform is embedded in a continuous latent space and must be quantized to allow compression with a finite number of bits. By quantization, we broadly mean any mapping from a continuous space to a discrete and countable set.
Beyond this necessity, quantization introduces information loss into the compression process, and is thus also the desirable mechanism by which rate is traded for distortion. 
The chosen quantization method $q$ is applied to $\mathbf{z}$, resulting in $\mathbf{\hat z} = q(\mathbf{z})$.
A neural network-based transform $f$ can learn to warp the embedding space to effectively manipulate which information is lost through quantization.

\textbf{Entropy Coding.}
After quantization, the discrete representation $\mathbf{\hat z}$ can be losslessly compressed using entropy coding.
Assuming an encoding scheme that can approach the theoretical lower bound given by Shannon's source coding theorem, such as arithmetic coding~\cite{arithmetic_coding}, the efficiency of this compression is determined by the \textit{entropy model} $p^\prime$.
Since the true data distribution $p$ is unknown, it is modeled using $p^\prime$. 
An accurate approximation $p^\prime$ is crucial for assigning shorter codes to more frequently occurring symbols, thereby minimizing the average length of the compressed representations. The closer $p^\prime$ matches the true distribution $p$, the nearer the final encoding length will be to the lower bound established by Shannon's theorem (Expression~\ref{eq:shannon_source_coding}).
In NC, the entropy model also takes on an additional role during training, by providing differentiable estimates of the bit cost for encoding a batch of data.
This estimation is incorporated into the model's loss function, enabling end-to-end rate--distortion optimization.
On the receiver side, the entropy decoding process reconstructs $\mathbf{\hat z}$ from the compressed binary string.

\textbf{Inverse Transform.} 
To reconstruct the data, the inverse transform $g$, often referred to as the \textit{synthesis transform}, is applied to the quantized representation $\mathbf{\hat z}$. 
In NC, an analytical inverse of the encoder $f$ is typically unavailable.
Instead, a second neural network, the \textit{decoder}, is trained to reconstruct the original data from $\mathbf{\hat z}$.
This results in $\mathbf{x^\prime} = g(\mathbf{\hat z})$, an approximation of the original input $\mathbf{x}$, subject to some loss.

NC aims to optimize the Lagrangian from Expression~\ref{eq:lagrangian_lossy} end-to-end using deep learning techniques.
By defining $f$ and $g$ as neural networks, and flexibly modeling $p$, NC can learn non-linear transform functions and complex entropy models. This approach leads to superior rate--distortion performance compared to traditional compressors, as demonstrated for tasks such as image~\cite{autoregressive_hyperprior}, video~\cite{mentzer2022vct}, audio~\cite{zeghidour2021soundstream}, and 3D scene compression~\cite{bird20213d}.

Two important aspects for optimizing the Lagrangian, which we postpone to Section~\ref{sec:entropy_models}, are the specific quantization methods used and how a continuous model $p$ can be used to fit the resulting discrete distribution.

We can now express the complete loss function for a single input $\mathbf{x}$ as:

\begin{align*}
    L(\mathbf{x}) = 
    \lambda \cdot \overbrace{
        \rho \bigg(
            \mathbf{x}, ~g \big(
                q(f(\mathbf{x}))
            \big)	
        \bigg)
    }^{D} 
    \\
    \underbrace{
        ~-~\log_2 p^\prime \bigg(
            q \big(
                f(\mathbf{x})
            \big)
        \bigg)
    }_{R}~.
    \label{eq:full_rd_loss}
\end{align*}

For the distortion term $D$, only the encoder $f$ and decoder $g$ participate. 
The gradients of this loss will push the encoder to produce quantization-robust representations that the decoder can accurately reconstruct.

The rate term $R$, involving the entropy model $p^\prime$ and the encoder $f$, pushes the encoder to create compressible representations by minimizing the entropy of $\mathbf{z}$.
The entropy model $p^\prime$ serves as an approximation of the true distribution $p$, aiming to assign high probability to $\mathbf{\hat z}$ to minimize the loss, under the constraint that it must be a valid probability density function.
As $p^\prime$ only contributes to the loss through the entropy term, the gradients of its parameters with respect to the distortion term will be $0$.
Thus, using the same loss, $f$ and $g$ can be trained while jointly fitting $p^\prime$ to $\mathbf{\hat z}$.

\subsection{Compression Taxonomy}
\label{sec:compression_taxonomy}
To categorize compression approaches and research areas, we propose a framework for classifying compression techniques, as illustrated in \cref{fig:compression_taxonomy}.
Within this framework, we distinguish lossy compression techniques, which are the focus of this work, from lossless approaches. 
Next, we further classify methods into explicitly engineered approaches and those that are learned from data. Most widely used compression algorithms today, such as JPEG~\cite{jpeg}, MP3~\cite{mp3}, and HEVC~\cite{hevc}, fall into the explicitly engineered category.
However, we focus on learned compression methods, which have consistently outperformed handcrafted approaches in rate--distortion metrics, demonstrating their potential for advancing the field.

Within this framework, we identify four primary axes of variation in learned compression methods:

\begin{itemize}
    \item \textbf{Transforms:} The design and architecture of the encoder and decoder networks. 
    \item \textbf{Quantization strategies:} How the continuous latent representations are discretized, and how quantization can be made compatible with end-to-end learning.
    \item \textbf{Entropy models:} The assumptions and implementation used to model the probability distribution of the latent representation.
    \item \textbf{Optimization objectives:} The optimization framework, particularly the choice of distortion measure used in the rate--distortion loss.
    %
\end{itemize}

\begin{figure}[t]
\begin{center}
\includegraphics[width=\linewidth]{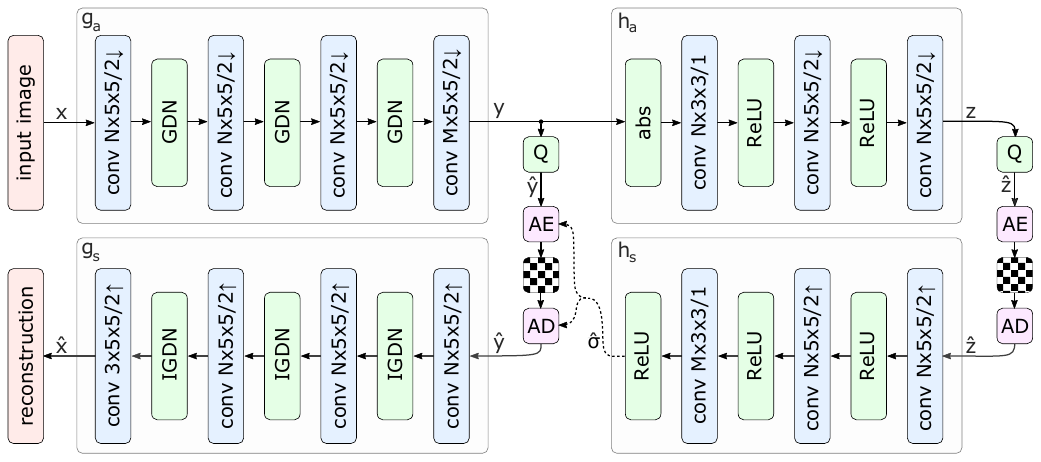}
  \caption{Network architecture of the hyperprior model~\cite{balleVariationalImageCompression2018}. The left side shows an image autoencoder architecture, the right side corresponds to the autoencoder implementing the hyperprior. The factorized-prior model uses the identical architecture for the analysis and synthesis transforms $g_a$ and $g_s$. Q represents quantization, and AE, AD represent arithmetic encoder and arithmetic decoder, respectively. Convolution parameters are denoted as: number of filters $\times$ kernel support height $\times$ kernel support width $/$ down- or upsampling stride, where $\uparrow$ indicates upsampling and $\downarrow$ downsampling. $N$ and $M$ were chosen dependent on $\lambda$, with $N=128$ and $M=192$ for the 5 lower values, and $N=192$ and $M=320$ for the 3 higher values. Figure and caption taken from \textcite{balleVariationalImageCompression2018}.}
\label{fig:variational_architecture}
\end{center}
\end{figure}

\subsection{Methods in Neural Compression}
\label{sec:neural_compression_literature} 
The present section explores the functionalities, benefits, and limitations of different methodological approaches in the literature for each of the four axes of NC proposed in \ref{sec:compression_taxonomy}.
A summary can be found in \cref{tab:neural_overview}.
\subsubsection{Transforms}
In learned image and video compression, the synthesis and analysis transforms are typically implemented as two halves of a deep convolutional auto-encoder~\cite{Masci2011} as popularized by \textcite{balleEndtoendOptimizedImage2016} and \textcite{theis2022lossy}.

The encoder gradually downsamples the spatial dimensions with a repeating pattern of convolutional layers and non-linear activations, while increasing the number of channels (embedding dimension). 
As shown in \cref{fig:variational_architecture}, the decoder mirrors this architecture to recover the original input.

Architectural innovations in deep learning and computer vision have introduced improvements to these transforms, such as the integration of attention mechanisms~\cite{bahdanau2014neural} and residual connections~\cite{he2016deep} into the network.

Building on the framework of neural transform coding, researchers have explored alternative architectures beyond conventional convolutional networks.
Indeed, unstructured data compression has leveraged fully connected feed-forward neural networks~\cite{balle2020nonlinear} and earlier works employed recurrent neural networks~(RNNs)~\cite{toderici2015variable} as encoder and decoder architectures.
More recent works have also explored the use of transformers~\cite{zhu2022transformerbased, li2024frequency} and denoising diffusion models~\cite{yang2024lossy}.

A distinct paradigm within NC that has also emerged is that of \textit{Implicit Neural Representations} (INRs).
Popularized in large part through their use in NeRF~\cite{mildenhall2021nerf} for 3D scene representation, INRs have shown promise as an alternative way of representing and storing 3D geometry~\cite{park2019deepsdf, mescheder2019occupancy, chen2019learning}, audio~\cite{sitzmann2020implicit}, images~\cite{sitzmann2020implicit}, video~\cite{du2021neural}, amongst others.
INRs aim to represent any signal as an implicitly defined function. 
For instance, we may represent an image as a function $f(x, y): \mathbb{R}^2 \to \mathbb{R}^3$ mapping from pixel coordinates $x$ and $y$ to an RGB value.
In practice, this function is learned by overfitting a neural network on a single input such that it can be recovered through inference on the network, essentially storing the input in the weights of the network.
This representation allows for the leveraging of model compression literature to achieve general signal compression. 
It has successfully been employed for compressing 3D scenes~\cite{bird20213d}, images~\cite{dupont2021coin, strumpler2022implicit} and videos~\cite{chen2021nerv, Gomes_2023_CVPR}, although their use together with an entropy penalty is still not ubiquitous, with many methods relying instead on more typical model compression techniques.

We note that INRs can also be interpreted as a pair of transforms.
The encoder is replaced by the training process, mapping the input into the space of the neural network parameters, and the decoder is replaced by the forward pass of the network itself.
The remaining 3 axes are then still fully applicable to INRs.
INRs show competitive {R-D} performance as well as versatility in the types of signals they can encode.
In particular, since they encode a single sample, they are not affected by the out-of-distribution problem that other NC methods may face and thus do not require a large dataset to be collected.
Additionally, since decoding is simply a forward pass on the network, it typically can be fully parallelized, granting it great performance advantages in fields such as video~\cite{chen2021nerv, Gomes_2023_CVPR}, image~\cite{guo2023compression, he2024recombiner}, and NeRF~\cite{pham2024neural} compression.
However, the lengthy training process required for compressing each sample makes INRs impractical for deployment in many real-world applications.

\subsubsection{Quantization Strategies}
\label{sec:quantization_strategies}
\begin{figure}
    \centering
    \includegraphics[width=\linewidth]{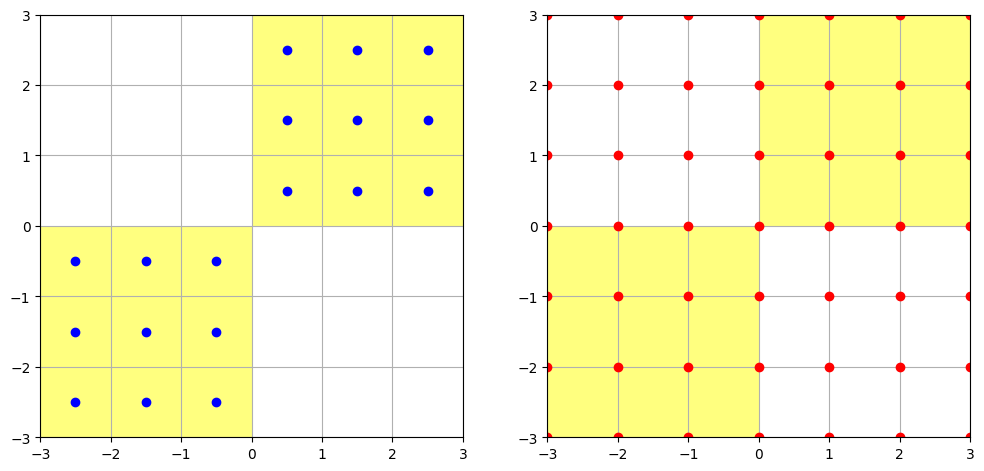}
    \caption{Illustration of vector vs scalar quantization for a given data distribution, shown in yellow. \textbf{Left:} Vector quantization can efficiently cover the space by freely building a codebook of vectors, shown in blue. \textbf{Right:} Scalar quantization quantizes each dimension individually. This potentially leads to an inefficient coverage of the space, with many quantization points covering areas that are outside the distribution of the data. In this case, uniform quantization to the integers is shown.}
    \label{fig:quantization}
\end{figure}

It can be shown that the optimal rate for a given distortion can be achieved through vector quantization~\cite{elements_information_theory}.
In vector quantization, all dimensions of the space are jointly discretized, usually by mapping the given $\mathbf{z}$ to its nearest neighbor in a codebook.
However, as the dimension of $\mathbf{z}$ grows, vector quantization becomes infeasible, with the curse of dimensionality requiring exponentially more entries in the codebook to optimally quantize the space, along with more data and compute to optimize them.

As an alternative, the most popular form of quantization in the non-linear transform coding paradigm is scalar uniform quantization, as introduced by \textcite{balle2016end} and \textcite{theis2022lossy}. Here, each dimension of the transformed data is quantized independently, typically by rounding each value to the nearest integer.
This scheme can be seen as a constrained form of vector quantization where the grid is fixed and equal to the set of integers~\cite{yang2023introduction}.
Despite its simplicity, scalar quantization is effective due to the flexibility of the non-linear transform, which can in essence warp this grid as desired. Figure~\ref{fig:quantization} illustrates these two approaches.
The main obstacle introduced by quantization is its non-differentiability, as backpropagating gradients is necessary for an end-to-end optimization of the network. 
%
%
Quantization has a gradient of $0$ almost everywhere, preventing any components before it from receiving gradients.
Two main techniques to address non-differentiability are:

\textbf{Straight Through Estimator (STE)}: The STE~\cite{bengio2013estimating} treats non-differentiable components as identity functions during backpropagation, fixing their gradient to $1$ and thus allowing gradients to pass through unchanged.
\textcite{theis2022lossy} applied this approach to the quantization function of NC models. For the forward pass, the quantization process remains unchanged.
\textbf{Uniform Noise}: \textcite{balle2016end} propose the replacement of quantization with additive uniform noise during training.
In the case of quantization to the nearest integer, this noise has a range of $[-0.5, 0.5]$ and thus the same width as the quantization bins.

Empirically, the combination of both of these methods seems to be optimal for training~\cite{9506497}, with STE used for calculating the distortion term and additive uniform noise used for the entropy term.

Beyond scalar quantization, other forms of quantization have been explored, despite being less popular. 
Early works employed binarization, reducing every element of $\mathbf{z}$ to two possible values~\cite{toderici2015variable, todericiFullResolutionImage2017a, li2018learning}.
Vector quantization (VQ) has also been successfully employed in neural transform coding~\cite{agustsson2017soft, mentzer2018conditional} with adaptations to mitigate its computational complexity problems.
Promisingly, recent work in generative modeling combines these approaches, leveraging a \textit{Vector Quantized Generative Adversarial Network} (VQ-GAN)~\cite{esser_taming_2021} for vector quantization and binarization to make this quantization more computationally feasible~\cite{yu2023language}. Despite the work focusing on video generation, it demonstrates competitive performance in video compression.
VQ-GANs extend the \textit{Vector Quantized Variational Autoencoder} (VQ-VAE) framework by employing an adversarial training strategy~\cite{creswell2018generative} to discriminate between real input images and the reconstructed outputs of the VQ-VAE decoder. Moreover, VQ-GANs enable the synthesis of high-resolution images (i.e., in the megapixel range) by modeling the learned (quantized) embeddings and their codebook through a transformer-based model. This interplay between GAN-enhanced autoencoder-based compression and transformer-based synthesis outperforms equivalent state-of-the-art approaches using plain autoencoders, thus opening the way for more context-rich compression strategies. \textcite{yang2024lossy} takes a different generative approach with diffusion models and outperforms a GAN benchmark on four metrics.

For INRs, quantization techniques often derive from general neural network compression, such as weight quantization or pruning~\cite{dupont2021coin, chen2021nerv}.
These methods can be applied after optimization or, often achieving better results, throughout training.
Quantization-aware training~\cite{jacob2018quantization} 
can be used to obtain INRs that are more robust to the error introduced by quantization, and finetuning after pruning can reduce its effect on distortion~\cite{chen2021nerv}.

\begin{table*} 
\captionsetup{justification=centering} 
\captionof{table}{Collection of neural compression papers in this section, aligned by contributions along the axes described in Section~\ref{sec:compression_taxonomy}.}
\centering
\begin{tabularx}{\textwidth}{>{\bfseries}l X X} 
    \toprule
    \textbf{Axis} & \textbf{Approach} & \textbf{Papers} \\
    \midrule
    \multirow{4}{*}{Transforms} 
        & CNN                                 & \RaggedRight \cite{balleEndtoendOptimizedImage2016, balleVariationalImageCompression2018, theis2022lossy, autoregressive_hyperprior} \\
        & RNN                                 & \RaggedRight \cite{toderici2015variable} \\
        & Transformer                         & \RaggedRight \cite{zhu_research_2022, li2024frequency} \\
        & Diffusion                           & \RaggedRight \cite{yang2024lossy} \\
        & INR                                 & \RaggedRight \cite{bird20213d, dupont2021coin, chen2021nerv, strumpler2022implicit, Gomes_2023_CVPR, guo2023compression, he2024recombiner, pham2024neural} \\
    \midrule
    
    \multirow{4}{*}{Quantization Strategies} 
        & Scalar Uniform Quantization         & \RaggedRight \cite{balle2016end, theis2022lossy} \\
        & Binarization                        & \RaggedRight \cite{toderici2015variable, todericiFullResolutionImage2017a, li2018learning} \\
        & Vector Quantization                 & \RaggedRight \cite{agustsson2017soft, mentzer2018conditional} \\
        & Weight Quantization                 & \RaggedRight \cite{bird20213d, Gomes_2023_CVPR} \\
    \midrule

    \multirow{3}{*}{Entropy Models} 
        & Fully Factorized                    & \RaggedRight \cite{balle2016end, balleVariationalImageCompression2018} \\
        & Hyperprior                          & \RaggedRight \cite{balleVariationalImageCompression2018,qian2022entroformer} \\
        & Autoregressive and Transformer-based & \RaggedRight \cite{balleVariationalImageCompression2018, minnen2020channel, mentzer2022vct, defossez2024high, deletang2023language, yu2023language, bird20213d, Gomes_2023_CVPR, jiang2023mlic} \\
    \midrule
    
    \multirow{3}{*}{Optimization Objectives} 
        & Rate-Distortion-X                   & \RaggedRight \cite{zhang2021universal, agustsson2023multi} \\
        & Downstream Embedding                & \RaggedRight \cite{singh2020end, dubois2021lossy} \\
        & Split Computing                     & \RaggedRight \cite{matsubara2022supervised, furtuanpey2024frankensplit} \\
    \bottomrule
    \label{tab:neural_overview}
\end{tabularx}
\end{table*}

\subsubsection{Entropy Models}
\label{sec:entropy_models}
The objective of the entropy model is to provide accurate approximations of $p({\mathbf{\hat z}})$ for two purposes:

\begin{enumerate}
    \item[\it a)] Estimating the rate during training to be used in the loss. 
    
    \item[\it b)] Entropy coding and decoding in operational use after the network has been trained.
\end{enumerate}
Both of these uses impose a demand for reasonable computational efficiency. Additionally, differentiability is required to enable end-to-end training.
A common approach to achieve this is to employ uniform quantization
and define $p(\mathbf{\hat{z}})$ in terms of an underlying continuous density $p^\prime(\mathbf{\hat{z}})$~\cite{yang2023introduction}:

$$
        p({\mathbf{\hat z}}) := \int_{[-0.5, 0.5)^n} p^\prime({\mathbf{\hat z}} + \mathbf{v}) d \mathbf{v},  \quad \forall {\mathbf{\hat z}} \in \mathbb{Z}^n.
$$
The assumptions imposed on $p({\mathbf{\hat z}})$ \textit{a priori}, to make the integral tractable and ensure computational efficiency and differentiability, determine the possible architectures for the entropy model. 

\textbf{Fully factorized model.} One of the stronger simplifying assumptions is that each element of $p({\mathbf{\hat z}})$ is independent, allowing for a fully factorized model. 
Each marginal distribution can be modeled with varying degrees of complexity, from a simple parametric distribution such as a Gaussian or Laplacian to neural networks that estimate the cumulative distribution function (CDF)~\cite{balleVariationalImageCompression2018}.

\textbf{Hyperprior model.} The assumption of full independence is likely too strong. 
An alternative approach to modeling dependencies between variables is to assume conditional independence given some other latent variable~\cite{Bishop1998}.
\textcite{balleVariationalImageCompression2018} extend their fully-factorized model to a latent variable model by introducing a \textit{hyperprior}:
\begin{align}
    \mathbf{z} \sim p(\mathbf{z} | \mathbf{\hat h}) \\
    \mathbf{h} \sim p(\mathbf{h})
\end{align}

Here, the hyperprior $p(\mathbf{h})$ is modeled as fully factorized, while $p(\mathbf{z} | \mathbf{\hat h})$ is conditioned on the quantized hyperprior. Typically, $p(\mathbf{z} | \mathbf{\hat h})$ is represented as a $0$-mean Gaussian, with the standard deviation for each dimension of $\mathbf{z}$ derived from $\mathbf{\hat h}$.
Although the hyperprior $\mathbf{h}$ introduces additional side information that must be transmitted, its size is negligible compared to ${\mathbf{\hat z}}$, and the added flexibility tends to significantly improve rate--distortion performance.
Intuitively, the hyperprior allows the entropy model to adjust to the specific ${\mathbf{\hat z}}$ being transmitted.

\textbf{Autoregressive and transformer-based models.} More sophisticated entropy models capture complex dependencies between elements of ${\mathbf{\hat z}}$, often at the cost of added computational complexity.
Autoregressive models~\cite{minnen2020channel} predict each element of $\mathbf{\hat{z}}$ based on the previously encoded elements, while 
transformer-based models leverage self-attention mechanisms to model complex relationships in the latent space of ${\mathbf{\hat z}}$, particularly in video compression.
\textcite{mentzer2022vct} greatly simplify video compression pipelines by relying on the modeling power of a transformer model, while~\textcite{defossez2024high} use a small transformer to reason on quantized units for efficient audio compression.
Recent advancements in large language models~(LLMs) have inspired their exploration for lossless~\cite{deletang2023language} as well as lossy~\cite{yu2023language} text, image, and video compression, demonstrating the potential of leveraging their predictive power and large context modeling for next-generation compression algorithms.

\textbf{Implicit Neural Representations (INRs).} 
In the case of INRs, the weights of the network themselves are treated as $\mathbf{z}$.
After quantization, the distribution of these weights is modeled by the entropy model.
This method has shown competitive results in compressing 3D scene~\cite{bird20213d} and videos~\cite{Gomes_2023_CVPR}.

\subsubsection{Optimization Objectives}
\label{sec:optimization_objectives}
Traditionally, NC methods aim to minimize distortion as perceived by humans, using loss functions like MSE and SSIM as proxies for human perception.
Developing loss functions that more accurately reflect human perception remains an active area of research, both within NC and for generative visual models in general.

Recent studies have explored new trade-offs by reinterpreting the concept of distortion.
Notably, the introduction of rate--distortion--perception~\cite{zhang2021universal} and rate--distortion--realism~\cite{agustsson2023multi} frameworks allows for more nuanced optimization strategies.

\begin{figure}[t]
    \centering
    \includegraphics[width=.95\linewidth]{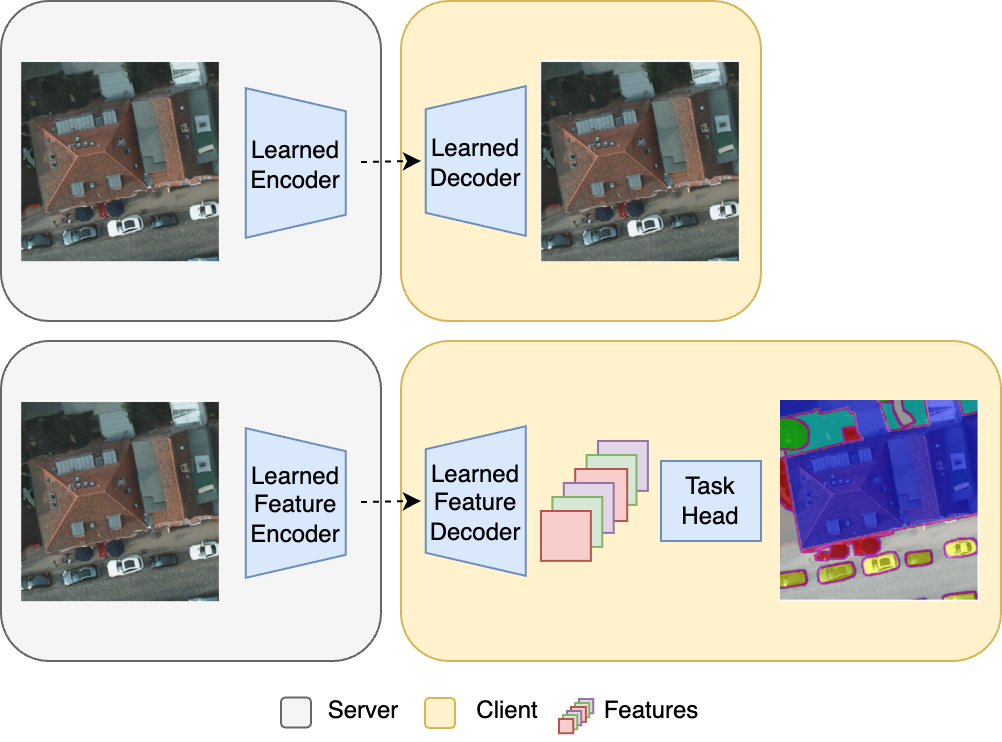}
    \caption{Illustration of feature compression. \textbf{Top:} the usual scenario where a client wishes to reconstruct the input compressed by the server. \textbf{Bottom:} the server sends a compressed feature representation of the input, which the client can directly feed to a task-specific head (e.g. semantic segmentation in this case). Images from the Potsdam dataset~\cite{potsdam}.}
    \label{fig:feature_compression}
\end{figure}

The continuing growth and deployment of deep learning algorithms in real-world applications introduces a new use case, which can be viewed as a further reinterpretation of distortion. 
This begs the question whether compression designed for human perception is the best choice when the end consumers of the data are algorithms (e.g. neural networks) instead of humans.
From this perspective, recent works propose reframing distortion from an algorithmic point of view.
As illustrated in \cref{fig:feature_compression}, the goal is not necessarily to recover the original data such that it is minimally affected for a human observer, but rather to produce compressed feature representations that enable algorithms (e.g. classification, image segmentation, object detection) to perform well when using them as input~\cite{singh2020end}.
Under this setting, the distortion metric is not a function of some reconstruction of the original data but rather based on the performance of such feature representations when fed to models for different downstream tasks.

A more general setting that makes the task somewhat more complex is that one may not know \textit{a priori} the type of downstream tasks the embeddings will be used for, or labeled data for those tasks may not be available during pre-training.
Instead, the process of learning general-purpose, compressible features must rely on proxy losses which may borrow from SSL~\cite{gomes2024compressed} to identify which aspects of the data may safely be lost during compression without affecting downstream performance~\cite{dubois2021lossy}.
A similar idea has been applied to transfer data in the reverse direction, from an edge device collecting data to a powerful server where it can be analyzed, in a paradigm known as \textit{Split Computing}.
\textcite{matsubara2022supervised} combine \textit{Knowledge Distillation} with NC to train a small encoder that can run on edge devices and produce compressible features that are fed to a larger network on a server with more compute resources. They demonstrate improved R-D performance compared to neural image compression methods that focus on reconstruction.
\textcite{furtuanpey2024frankensplit} further develop the framework, conducting a thorough analysis of bottleneck placement within the network. 
They further introduce a saliency-guided loss and design blueprints for leveraging feature compression with different backbone architectures, showing improved R-D performance.

Finally, while FMs for vision have been shown to generate embeddings that can generalize to several downstream tasks~\cite{radford2021learning,caron2021emerging,wang2023ssl4eo}, the dimension of their output feature space may result in embeddings that are larger than the original data, making them impractical for storage or transmission.
How to best generate such general-purpose compressed embeddings is an open question. 
However, a system capable of doing so could have a large-scale impact, democratizing both data and powerful models by enabling the widespread distribution of powerful, ready-to-use features.
We deem this line of research to be particularly important as FMs become increasingly prevalent in EO~\cite{jakubik2023foundation, wang2023ssl4eo, satmae2022, smith2023earthpt}.


\section{Neural Compression for Remote Sensing}
\label{sec:compression_RS}
Advances in RS technologies have led to an increasing number of EO satellite acquisitions with enhanced spatial resolution, broader spectral bands, and higher temporal frequency~\cite{clauson_etal_2024}.
These data volumes present challenges for data transmission, storage, and processing~\cite{guo_big_2017, wilkinson_environmental_2024}.
This section explores the application of NC to raster EO data from air- and space-borne instruments, drawing parallels to NC techniques used for natural images (\cref{sec:neural_compression_literature}).
We discuss the specific challenges of compressing RS data (\ref{sec:remote_sensing_challenges}), review existing compression research (\ref{sec:remote_sensing_compression_research}), and explore future developments (\ref{sec:remote_sensing_compression_future}).

\subsection{Challenges in Compressing Remote Sensing Data}
\label{sec:remote_sensing_challenges}

RS data presents unique challenges and opportunities for compression, stemming from its distinctive data characteristics, acquisition constraints, and usage.
These factors influence the design and applicability of existing image compression techniques~\cite{kuester_approach_2020}.

\begin{figure*}[t]
    \centering
\includegraphics[width=.99\linewidth]{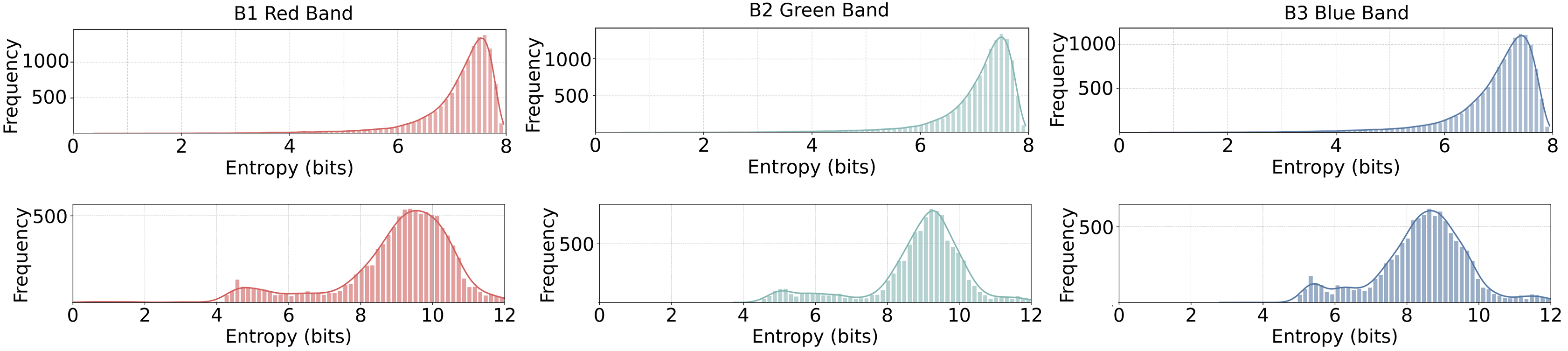}
    \caption{
    Comparison of per-image entropy distributions for RGB bands across two datasets. \newline
    \textbf{Top}: ImageNet dataset -- Random sample of 10,000 images~\cite{deng2009imagenet}. \newline
    \textbf{Bottom}: BigEarthNet dataset -- Random sample of 10,000 Sentinel-2 images~\cite{sumbul2019bigearthnet}.}
    \label{fig:within_img_entropy}
\end{figure*}

\subsubsection{Data Characteristics}
~\\
\textbf{Spectral resolution}. 
RS images often comprise multiple spectral bands beyond the visible range, such as near-infrared (NIR) and short-wave infrared (SWIR), enabling detailed surface and atmospheric analysis~\cite{vali_deep_2020}. 
While multispectral data typically captures a limited number of broad bands, hyperspectral data covers hundreds of narrow wavebands. 
This richer spectral dimension leads to larger data volumes and increased correlations between adjacent spectral bands~\cite{bajwa_hyperspectral_2004,qian_hyperspectral_2021}. 
Compression techniques need to effectively exploit spectral correlations while avoiding excessive computational cost~\cite{mijaresiverduScalableReducedComplexityCompression2023}. 
In contrast to optical sensor data, Synthetic Aperture Radar (SAR) captures the amplitude, phase, and polarization of radar signals reflected from the Earth's surface, making it suitable for applications like surface and moisture analysis~\cite{zribi_analysis_2019}.

\textbf{Spatial resolution}. 
The spatial resolution in RS varies depending on the physics of data acquisition. 
For example, the ESA Sentinel-2 satellite provides images with a spatial resolution of 10 to 60 meters per pixel~\cite{berger2012esa}. 
Compared to other imaging domains, RS generally operates at lower spatial resolutions over large geographical areas. 
Consequently, individual pixels can contain highly relevant information for downstream tasks, and images exhibit complex textures with rich information~\cite{xiang_discrete_2023, zhu_research_2022}.  

\textbf{Temporal resolution}.
Most satellites capture data for specific geographic regions at regular intervals, generating time series that can support the monitoring of dynamic processes and environmental changes~\cite{miller_deep_2024}. 
These successive images often exhibit temporal correlations, reflecting gradual landscape transformations and seasonal or weather-dependent variations. 
Unlike static image compression, where there are no temporal relationships, or video compression, where frames are closely spaced in time, RS imagery involves wider temporal gaps, often spanning several days between acquisitions~\cite{liu_survey_2015}.

\textbf{Radiometric resolution}.
Radiometric resolution indicates a sensor's ability to measure the intensity of reflected radiation within a specified wavelength range. 
For instance, Sentinel-series satellites employ 12-bit resolution~\cite{berger2012esa}, providing higher precision than the 8-bit standard common in natural images. 
Greater precision enables more detailed measurements but increases the complexity for compression algorithms as it expands the input alphabet. 

The varying resolutions in RS data make it difficult to define a 'typical' RS image. 
Designing compression algorithms that address the diverse spectral, spatial, temporal, and radiometric characteristics is inherently complex. 
As a result, compression methods in RS are often dataset-specific with limited generalizability across different types of RS data~\cite{lieberman_neural_2023}.

\subsubsection{Data Acquisition and Application}
~\\
\textbf{Data Downlink Bottleneck.} 
A core practical challenge is the limited bandwidth to transmit satellite imagery to Earth~\cite{furutanpey2024, du2024earth+, downlink_latency}.
To facilitate transmission, images are often compressed onboard using compression algorithms such as the Consultative Committee for Space Data Systems (CCSDS) standards~\cite{yeh_new_2005}.

\textbf{Onboard Processing Limitations.}
In addition to bandwidth constraints, satellites have limited computational and storage resources, restricting the complexity of compression approaches that can be deployed onboard. 
Such constraints can limit the usability of NC techniques for onboard satellite applications.

\textbf{Preservation of Critical Information.}
RS imagery is used in numerous scientific and operational applications, including environmental monitoring of above-ground biomass~\cite{li_forest_2020}, agricultural mapping of oil palm density~\cite{rodriguez_mapping_2021}, and flood detection for disaster management~\cite{oms_2024}.
Today, many EO tasks leverage machine learning models and rely on the analysis of specific aspects in the data.
Compression techniques should therefore maintain relevant features for downstream task, rather than focusing only on perceptual reconstruction~\cite{furutanpey2024fool}.

\textbf{Comparison with Natural Images and Entropy Analysis.}
Unlike natural photos, which prioritize visual appeal and are often post-processed for aesthetic purposes~\cite{ramanath_color_2005}, RS data undergoes radiometric, atmospheric, and geometric adjustments to ensure scientific accuracy. Moreover, EO data exclusively captures Earth landscapes, opposed to natural image datasets made up of diverse scenes and objects. To highlight how these differences translate into distinct compression demands, we conducted an entropy analysis of two representative datasets: ImageNet~\cite{deng2009imagenet} (natural images) and BigEarthNet~\cite{sumbul2019bigearthnet} RGB bands (Sentinel-2 Level 2A satellite images).

Per-image entropy quantifies the average amount of information contained and gives us an indication of the pixel variability within a single image. Our analysis, shown in \cref{fig:within_img_entropy}, involved calculating a pixel value histogram for each image and band, and computing its entropy. We randomly sampled 10,000 images from each dataset for comparison. The results indicate that ImageNet images have an on average higher per-image entropy. These values can be due to the general greater variety in colors and patterns, as well as to post-processing that increases contrast. Sentinel-2 L2A images show lower entropy, despite covering diverse landscapes. Pixel values within individual bands tend to be more concentrated, leading to lower per-image entropy. The very low entropy of some images can be explained by certain scenery classes. For example, sea images often have similar pixel values across the whole image. This is a property that may be exploited by domain-specific compression algorithms.

These findings underscore the need for compression techniques tailored to the statistical properties of EO data. Compression leverages bias in a dataset, allowing short bit rates to be used for redundant elements in the input data. The design of a compression algorithm is therefore always subject to a fundamental trade-off between broad applicability and data specificity. Conventional compression methods optimized for natural images might not fully exploit the redundancies and correlations prevalent in RS data. Even within RS, the variety of spectral, spatial, temporal, and radiometric resolutions of the instruments complicate the development of a single effective algorithm. Research therefore tends to focus on a specific type of RS data, as we describe in \cref{sec:remote_sensing_compression_research}.

\subsection{Classification of Compression Methodologies}
\label{sec:remote_sensing_compression_research}

This section reviews compression approaches for RS data, categorizing them into traditional hand-crafted methods and NC techniques.

\subsubsection{Traditional Approaches}
~\\
\textbf{Transform Coding} methods, such as JPEG and JPEG2000, first convert data into the frequency domain before applying entropy coding. Common transformations include the discrete cosine transform~(DCT)~\cite{ahmed1974} and the discrete wavelet transform (DWT)~\cite{wavelets}. 
Their efficiency make transform-based methods widely used in RS applications~\cite{lee_hyperspectral_2002,yu_image_2009, hacihaliloglu_dct_2004}. For example, \textcite{hou_improving_2000} adapt JPEG to detect and simplify cloud features, while \textcite{gonzalez-conejero_jpeg2000_2010} modify JPEG2000 to handle EO image areas without useful information. The CCSDS has developed international compression standards based on JPEG which are commonly used for onboard applications~\cite{yang_compression_2005, machairas_133_2020, garcia-vilchez_extending_2009}. Transform-based methods decorrelate input along the spatial dimension but require extensions to leverage spectral redundancy in multi- and hyperspectral RS data. 
\textcite{markman_hyperspectral_2001} extend the DCT to the spectral dimension, while \textcite{lim_compression_2001} and \textcite{luigi_dragotti_compression_2000} apply three-dimensional wavelet transforms using the SPIHT algorithm. \textcite{du_hyperspectral_2007} and \textcite{du_low-complexity_2008} combine JPEG2000 with PCA for spectral decorrelation.

\begin{figure}[t]
    \centering
    \includegraphics[width=0.95\linewidth]{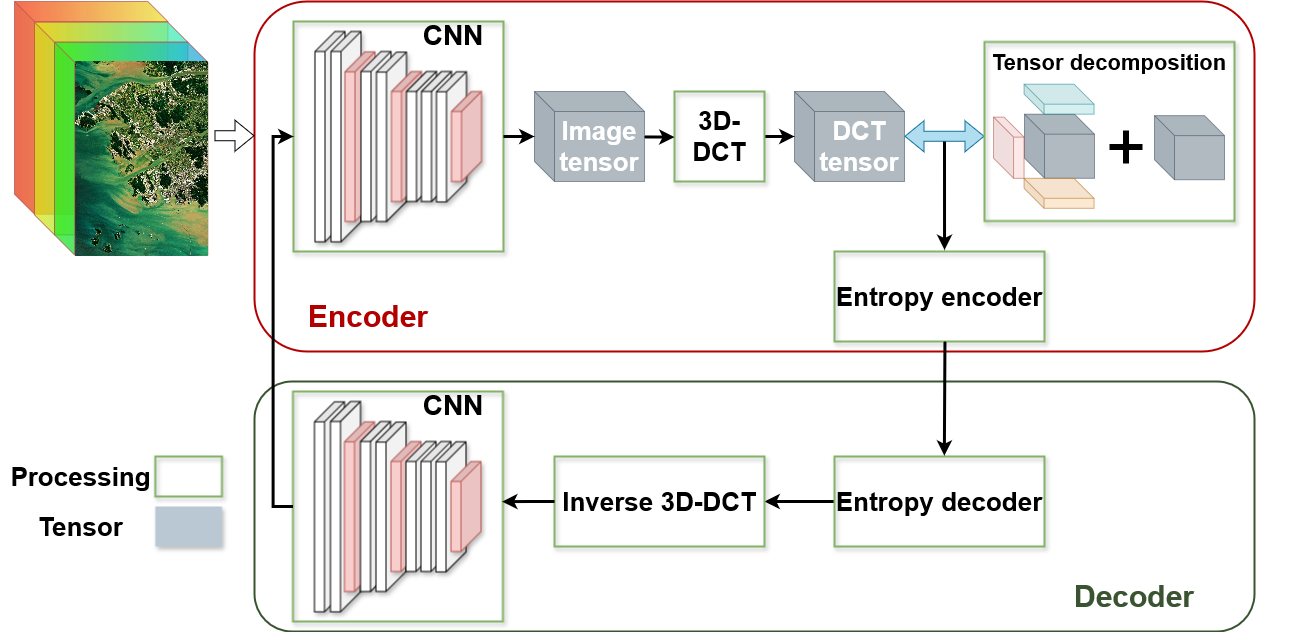}
    \caption{CNN-based transformation as part of a tensor decomposition framework introduced by \textcite{li_multispectral_2019}.}
    \label{fig:transform_with_cnn}
\end{figure}

\textbf{Tensor Decomposition} techniques address the multispectral nature of RS data by decomposing multidimensional matrices into low-rank components. Notable methods include the Tucker decomposition~\cite{tucker1966}, which approximates a tensor with factor matrices and weight coefficients as a reduced core tensor. These methods are particularly effective for high-dimensional data like hyperspectral images, achieving high compression rates while preserving multidimensional structures~\cite{zhang_compression_2015, chen_low-rank_2016, du_pltd_2017}.
Ongoing research aims to lower the computational complexity of tensor decomposition approaches for practical applications in RS~\cite{li_tensor_2010, akarami_hyperspectral_2010, karami_hyperspectral_2011, karami_compression_2012, sidiropoulos_multi-way_2012, li_compression_2014, li_compression_2017, zhang_compressing_2024}. In the context of SAR, raw radar echos are typically compressed onboard using Block Adaptive Quantization (BAQ)~\cite{Kwok1989_baq}. BAQ adapts quantization levels per block to meet bitrate requirements. Extensions of BAQ~\cite{Attema2010, Martone2019} dynamically adjust bitrates to improve compression rates. 

\subsubsection{Neural Approaches} 

We now introduce NC techniques for RS, summarized in Table~\ref{tab:RS_neural_overview_eo}.

\textbf{Neural transformation.} As an early contribution, \textcite{li_multispectral_2019} combine a CNN-based transformation with tensor decomposition (Figure~\ref{fig:transform_with_cnn}). The encoder CNN, optimized for MSE, is combined with a DCT to produce a compact representation, reducing the computational cost of tensor decomposition. 

\begin{figure}
    \centering
    \includegraphics[width=.95\linewidth]{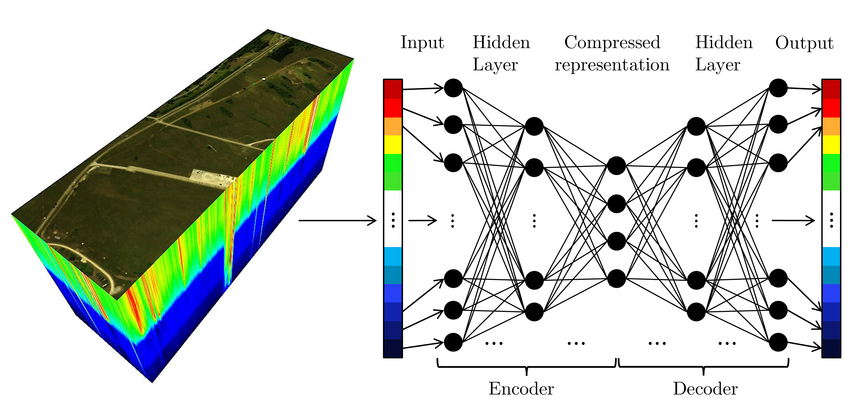}
    \caption{Autoencoder for hyperspectral data by \textcite{kuester_approach_2020}. Figure taken from the original paper.}
    \label{fig:hyperspectral_AE}
\end{figure}

\textbf{Autoencoders} compress data by encoding it into a lower-dimensional latent space and minimizing a reconstruction error, making them a nonlinear transform coding method. \textcite{yang_compression_2005} introduce an early autoencoder for satellite imagery, replacing sigmoid activations with a ridgelet transform~\cite{candes1999} to improve compression performance. \textcite{kuester_approach_2020} introduce an autoencoder for hyperspectral data by compressing only the spectral component (Figure~\ref{fig:hyperspectral_AE}).

\textbf{Rate--Distortion Autoencoders.}
Rather than relying solely on dimensionality reduction, rate--distortion autoencoders are optimized end-to-end for bitrate and reconstruction quality.
Following~\textcite{balleEndtoendOptimizedImage2016}, several studies have applied CNN-based rate--distortion models to optical satellite imagery~\cite{alves_de_oliveira_reduced-complexity_2021,kong_spectralspatial_2021,cao_spectralspatial_2022, chien_tensor-factorized_2018}, aerial imagery~\cite{zhu_research_2022} and SAR data~\cite{xu_synthetic_2022, Pilikos2022, Reza2023}. 

\textbf{Reduced--complexity Rate--Distortion Autoencoders.} CNNs capture spatial features well, and larger kernel sizes enables them to capture information over broader image ranges. However, increasing kernel size also increases model complexity, limiting their suitability for onboard compression applications. \textcite{alves_de_oliveira_reduced-complexity_2021} propose a reduced-complexity VAE that outperforms JPEG2000~\cite{yeh_new_2005}. They reduce network parameters and simplify the entropy model by using a parametric estimation of a Laplacian distribution. While this demonstrates that lower-complexity models can compete with the previously mentioned neural methods, their computational costs remain higher than those of traditional onboard techniques.
\textcite{mijaresiverduScalableReducedComplexityCompression2023} adapt a hyperprior VAE for hyperspectral data by clustering spectral bands and applying separate, smaller compression models to each cluster.

\begin{figure*}[ht]
    \centering
    \includegraphics[width=1.\linewidth]{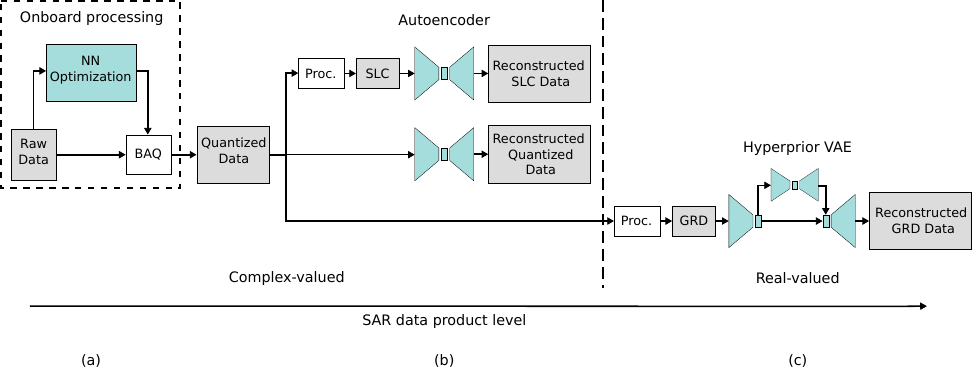}
    \caption{Overview of the SAR data processing stages where compression methods have been developed. In (a) the raw SAR data is compressed before wireless transmission of the data. Most end-users have access to (b) the compressed data or complex-valued data products calculated from the compressed data, e.g. Single Look Complex (SLC) products, or (c) real-valued data products projected onto an Earth ellipsoid model, e.g. Ground Range Detected (GRD) products.}
    \label{fig:SAR_nn_compression}
\end{figure*}

\textbf{Instrument diversity.} 
Compression approaches within the RS domain must address diverse data types, including modalities, like SAR, which are significantly different from ordinary images in computer vision.
For aerial imagery, \textcite{zhu_research_2022} incorporate radiation calibration into a rate-distortion compression model and exploit interspectral redundancies using $1 \times 1$ convolutions.
For SAR, compression methods have been developed for different stages of data processing (Figure~\ref{fig:SAR_nn_compression}).
For onboard compression,~\cite{Gollin2023} extend the BAQ algorithm by a CNN-based optimization of the bitrate used for quantization. 
\textcite{Martone22} use backscatter statistics to improve resource allocation and image quality.
Other research for SAR focuses on accurate phase and amplitude reconstruction. \textcite{Paras23} introduce a complex-valued hybrid approach that combines HEVC for encoding with a neural decoder to precisely reconstruct phase and amplitude characteristics. \textcite{Pilikos2022} propose two architectures for complex-valued SAR compression based on Vector Quantized VAEs (VQ-VAE), which use real-valued convolutional layers, while activation functions, batch normalization, and backpropagation are complex-valued. Conversely,~\cite{Reza2023} use complex-valued convolutional layers to decode the HV polarization.

\textbf{Architectural adaptations.} Many studies build on the hyperprior model~\cite{balleVariationalImageCompression2018} but modify the transform architecture.
\textcite{xiang_remote_2023} integrate attention and long-range convolution to capture spatial redundancy. \textcite{kong_end--end_2021} modify the encoder to extract spatial-spectral features at multiple scales and adaptively adjust the weights of the features from different branches of the encoder network. \textcite{fu_remote_2023} introduce a mixed hyperprior net with two prior models: a transformation-based prior to capture global redundancy and a CNN-based prior to capture local redundancy. \textcite{gao_mixed_2023} extend the hyperprior model with an enhanced residual attention module (ERAM) that applies spatial attention to create importance masks for adjusting bit distribution across latent channels. 
For SAR, \textcite{Fu2023_sar_compression} use multi-layered residual blocks and hyperpriors with local and global context information, while \textcite{di_learned_2022} utilize pyramidal feature extraction. Their approach involves a single Gaussian hyperprior framework and the pyramidal encoder to capture both coarse and fine structures. 

\begin{figure}[ht]
    \centering
    \includegraphics[width=0.95\linewidth]{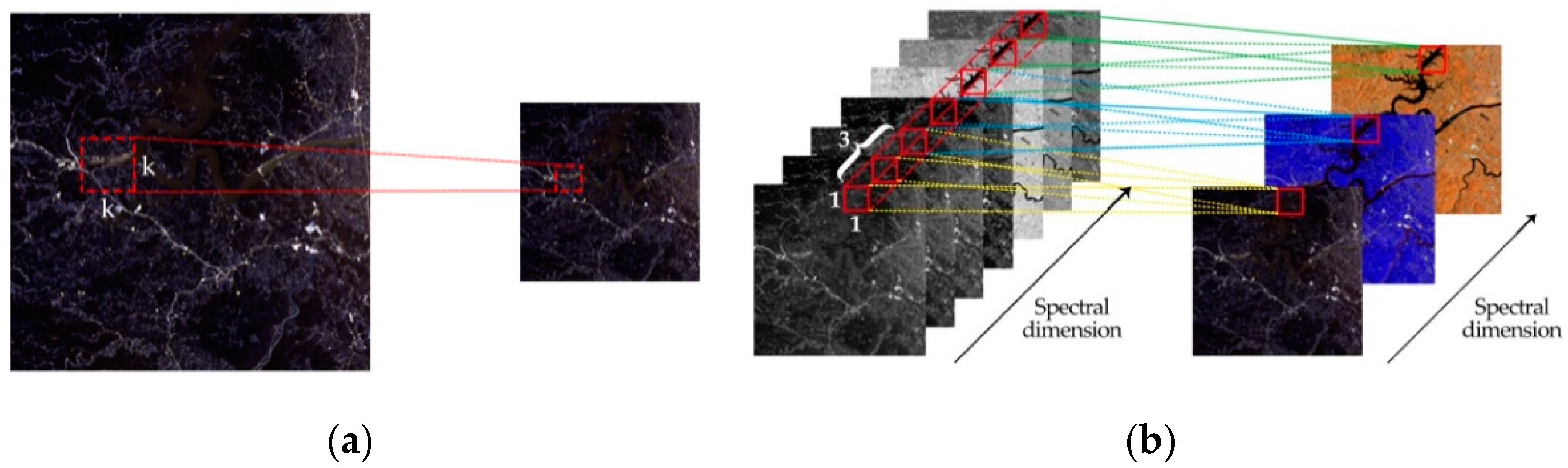}
    \caption{Spectral convolution taken from  \textcite{kong_spectralspatial_2021}. (a) 2D convolution; (b) 1D spectral convolution. Figure taken from the original paper.}
    \label{fig:spectral_conv}
\end{figure}

A common approach in RS compression is to decompose input images into spatial and spectral components. \textcite{kong_spectralspatial_2021} propose a feature extraction module that extracts spectral and spatial features separately, using spectral convolution (Figure~\ref{fig:spectral_conv}), and fuses them later for further processing. Similarly, \textcite{cao_spectralspatial_2022} extract spectral and spatial features separately but without fusing them at a later point. They incorporate Tucker decomposition through tensor layers~\cite{chien_tensor-factorized_2018} for better decomposition of the multi-way data representations. 
Although not always aimed at reducing complexity, spatial-spectral decomposition can enhance computational efficiency, particularly for datasets with many spectral bands.
Besides adaptations to the transform model, novel entropy models in the RS domain, such as Gaussian mixture models (GMM), refine the estimation of latent distributions~\cite{cheng2020, xiang_discrete_2023}. 

\begin{figure}
    \centering
    \includegraphics[width=0.95\linewidth]{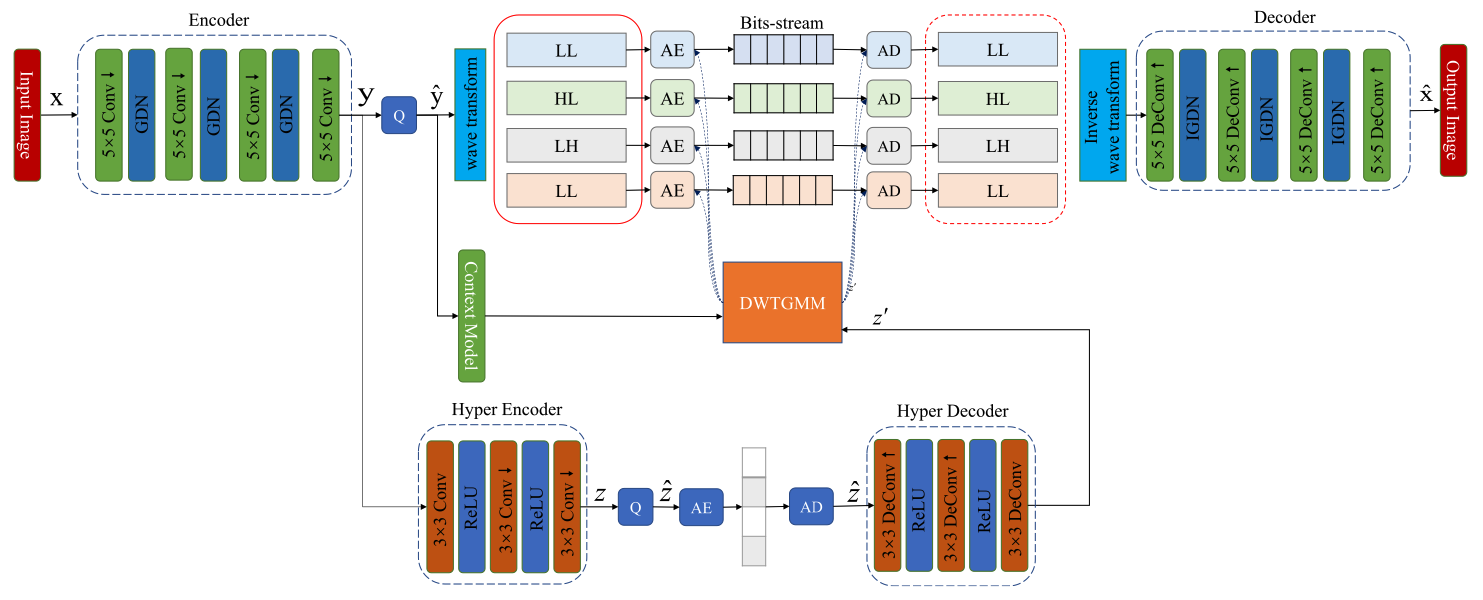}
    \caption{NC architecture with an incorporated non-learnable wavelet transform introduced by \textcite{xiang_discrete_2023} as a Discrete Wavelet Transform-Based Gaussian Mixture Model. Figure taken from the original paper.}
    \label{fig:dwtgmm_architecture}
\end{figure}

\textbf{Hybrid methods.} Several approaches combine non-learnable wavelet transforms with end-to-end compression methods. 
\textcite{anuradha_efficient_2024} combine DWT for spatial-spectral decorrelation with LSTM networks for hyperspectral data. \textcite{xiang_discrete_2023} apply a DWT to latent representations, separating high- and low-frequency features. Gaussian mixture models are then used to estimate entropy models for the high- and low-frequency components separately (Figure \ref{fig:dwtgmm_architecture}). \textcite{xiang_remote_2024} also addresses the challenge of reconstructing high-frequency information in RS images, which often leads to edge-blurred artifacts. They introduce a two-branch architecture that employs a DWT to separate the input data into high-frequency and low-frequency components, which are then processed separately in dedicated sub-networks.

\textbf{Explicit bitrate allocation.} Some approaches add a mechanism to explicitly control the code length allocated to different regions of the input. This is achieved through the introduction of importance maps and attention modules, weighing the compression of certain areas of an input image.
\textcite{ye_gfscompnet_2024} employ an image segmentation approach to create semantic maps before compression, thereby ensuring enhanced detail fidelity. The compression architecture incorporates an attention mechanism and a rate allocation technique that assigns higher compression rates to regions with smaller-sized details. \textcite{deng_synthetic_2024} incorporate a quality map from a pretrained ViT network to preserve high information content in the regions of interest while reducing redundancy in non-target areas.

\begin{figure}[t]
    \centering
    \includegraphics[width=0.95\linewidth]{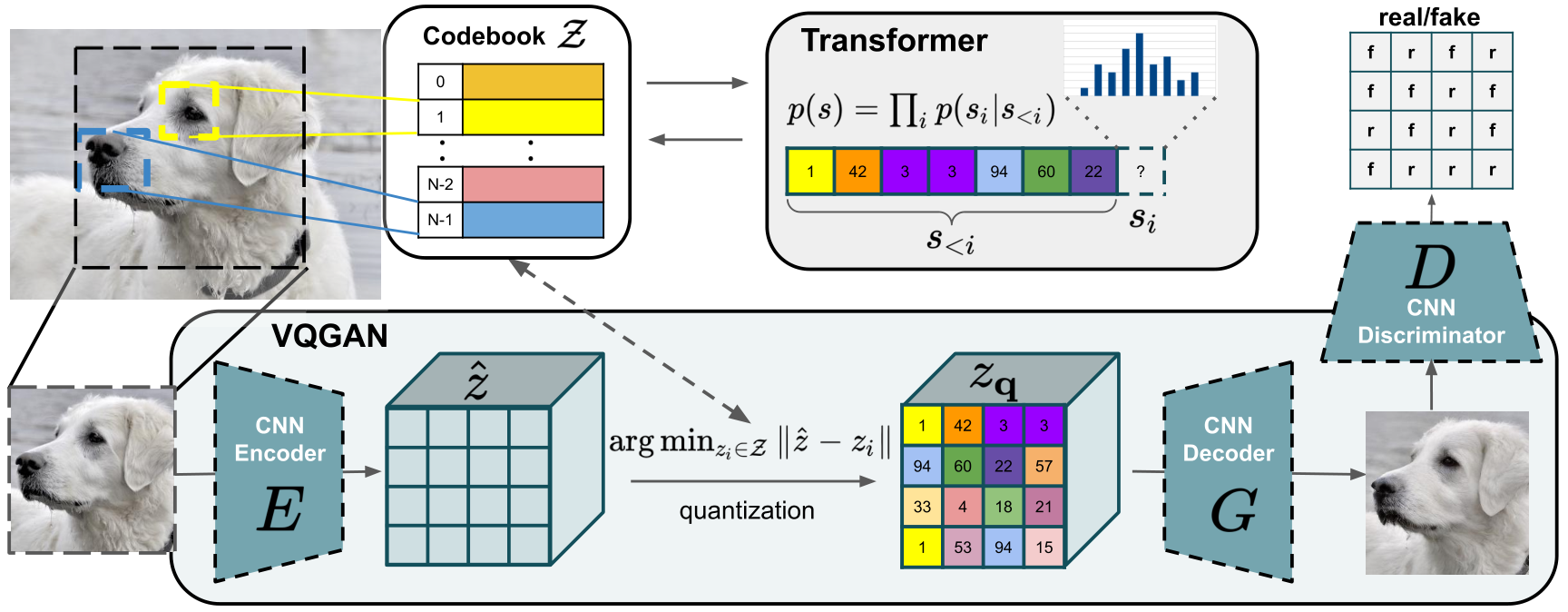}
    \caption{
        \textit{Vector-Quantized Generative Adversarial Network} (VQ-GAN) architecture introduced by \textcite{esser_taming_2021} as an extension of the \textit{Vector-Quantized Variational Autoencoder} (VQ-VAE) replacing the pixel-based image reconstruction loss by a discriminator network $D$. Figure taken from the original paper.
    }
    \label{fig:vqgan}
\end{figure}

\textbf{Generative Adversarial Networks.} 
GAN-based compression models have shown impressive performance at lower bitrates.
Leveraging the VQGAN~\cite{esser_taming_2021} architecture (Figure~\ref{fig:vqgan}), such models usually consist of autoencoders with GANs serving as decoder modules. The associated adversarial loss is tailored to favor either specialized~\cite{li_intelligent_2023} or generalist~\cite{zhao_sparse_2021} compressed representations. 
\textcite{zhao_sparse_2021} optimizes for the visual realism of reconstructed images by including a perceptual similarity term within the adversarial loss of a Conditional GAN~\cite{mirza_conditional_2014} decoder. To improve the quality of decompressed edges, textures, or contours \textcite{ZHAO2021} propose a Laplacian of Gaussian loss for a symmetric lattice GAN. 
On the other hand, \textcite{li_intelligent_2023} focus on generating generalist compressed representations. Using Least Squares GANs~\cite{mao_least_2017}, they reconstruct (dense) low-frequency components from (sparse) high-frequency components of the original images.

\textbf{Implicit Neural Representations} do not rely on autoencoder backbones and have demonstrated their potential in RS~\cite{strümpler2022implicitneural}, outperforming JPEG2000 on both multispectral~\cite{li_remote_2023} and hyperspectral~\cite{zhang_compressing_2024} datasets. 
INR-based methods regress the channel values for each pixel of a given image based on corresponding pixel coordinates, or transformations thereof. By optimizing the fidelity of these regressed values, a neural network encodes the implicit mapping between spectral values and pixel locations. The trained weights then undergo quantization and entropy coding, thus serving as a compressed representation of the images.
\textcite{li_remote_2023} successfully apply INRs to multi-spectral image compression. They train an MLP with equally sized residual layers to predict pixel values from longitude and latitude coordinates associated with pixel locations. Given the size of the input images and residual blocks, an upper bound to the width of the MLP hidden layers is derived that allows for effective compression. This method matches the quality of reconstruction of JPEG2000 while using half the bits per pixel.
A similar approach for hyperspectral imagery is implemented in the FHNeRF~\cite{zhang_compressing_2024} model, which regresses pixel values from transformed pixel coordinates using Neural Radiance Fields~\cite{mildenhall2021nerf}. The proposed model nearly doubles the reconstruction quality of traditional and autoencoder-based NC at comparable compression ratios.

The main advantage of INR-based compressors is that they are agnostic to certain image features, such as native resolution. In principle, the produced representations are invariant of the scale of the original image, and their size uniquely depends on the architecture of the model performing the regression task. In other words, rather than data compression, methods relying on INRs perform model compression. While these methods stand out as more generalist alternatives to autoencoder-based NC, the latter still achieve higher compression efficiencies for multi-spectral images and are not bounded by the size of the compressor’s backbone model.

\textbf{Feature Compression} focuses on compressing representations for downstream tasks rather than reconstructing the input and has seen some preliminary exploration in the RS domain.
\textcite{furutanpey2024fool} leverage this approach to mitigate the bandwidth bottleneck between satellites and base stations. 
They design an end-to-end pipeline for onboard feature compression capable of producing task-agnostic features and perceptually similar reconstructions of the input data. 
Their evaluation on benchmarks for object detection from aerial images demonstrates improved performance compared to neural image codecs and existing neural feature compressors~\cite{matsubara2022supervised, furtuanpey2024frankensplit}.
\textcite{gomes2024compressed} use the same idea tailored to the transmission of features from data centers to end users hosting models for training or inference.
They adopt a rate--distortion objective that combines masked auto-encoding as a form of SSL~\cite{he2022masked} with an entropy penalty to encourage compressible, general-purpose features. They further leverage an existing FM and show that fine-tuning a small portion of the pretrained weights is sufficient to create a general feature compressor for classification and segmentation.

\textbf{Dictionary Learning} involves learning a set of basis elements from the data and representing the data as sparse combinations of these elements, enabling efficient compression. In RS, \textcite{Wu_2015} propose a double sparsity model for hyperspectral images. Their method, involving entropy coding with Differential Pulse Code Modulation (DPCM) and arithmetic coding, demonstrates superior rate--distortion performance and improved spectral information preservation compared to 3D-SPIHT and JPEG2000. \textcite{Wang_2017} propose a dictionary learning approch that induces sparse coefficients through online learning. The sparse coefficients are quantized and entropy-coded to generate the final bit stream.
\textcite{Ertem_2020} also employ dictionary learning to improve hyperspectral image compression. Their method generates superpixel maps for adaptive spatial--spectral representation, computes an optimal dictionary, and determines sparse coefficients using Simultaneous Orthogonal Matching Pursuit (SOMP). Notable innovations include a modified dictionary learning step, an ordering scheme that eliminates the need to transmit the superpixel map as side information, as well as using DPCM to reduce sparse coefficient magnitudes. 
\begin{table*} 
\captionsetup{justification=centering} 
\captionof{table}{Contributions to the field of neural compression for remote sensing described in Section~\ref{sec:remote_sensing_compression_research} ordered along the axes of the taxonomy described in Section~\ref{sec:compression_taxonomy}.}
\centering
\begin{tabularx}{\textwidth}{>{\bfseries}l X X} 
    \toprule
    \textbf{Axis} & \textbf{Approach} & \textbf{Papers} \\
    \midrule
    \multirow{8}{*}{Transforms} 
        & Complexity Reduction             & \RaggedRight \cite{alves_de_oliveira_reduced-complexity_2021, mijaresiverduScalableReducedComplexityCompression2023} \\
        & Novel spatial extraction         & \RaggedRight \cite{kong_end--end_2021,di_learned_2022, zhu_research_2022, xiang_remote_2023, gao_mixed_2023} \\
        & Novel spectral extraction        & \RaggedRight \cite{kong_end--end_2021, kuester_approach_2020} \\
        & Separate spectral/spatial extraction & \RaggedRight \cite{kong_spectralspatial_2021,cao_spectralspatial_2022} \\
        & Incorporate Wavelet Transform    & \RaggedRight \cite{xiang_discrete_2023, xiang_remote_2024, anuradha_efficient_2024} \\
        & Bitrate Allocation               & \RaggedRight \cite{ye_gfscompnet_2024, deng_synthetic_2024} \\
        & Image-specific (INRs)            & \RaggedRight \cite{li_remote_2023,zhang_compressing_2024} \\
    \midrule

    \multirow{1}{*}{Entropy Models} 
        & Hyperprior with attention        & \RaggedRight \cite{gao_mixed_2023} \\
        & Multiple hyperpriors             & \RaggedRight \cite{fu_remote_2023} \\
        & Split latent space               & \RaggedRight \cite{cao_spectralspatial_2022, xiang_discrete_2023, xiang_remote_2024} \\
    \midrule

    \multirow{1}{*}{Optimization Objectives} 
        & Adversarial loss (GANs)          & \RaggedRight \cite{zhao_sparse_2021,li_intelligent_2023, ZHAO2021} \\
        & Downstream Embedding             & \RaggedRight \cite{furutanpey2024fool, gomes2024compressed} \\
    \bottomrule
    \label{tab:RS_neural_overview_eo}
\end{tabularx}
\end{table*}

\subsection{Summary}
\label{sec:remote_sensing_compression_future}

The variety in EO instruments produces diverse datasets, resulting in a wide application space for compression techniques. However, this data diversity also leads to a fragmented field. Studies often train and evaluate models on different datasets, which complicates method comparisons. 
While hand-crafted methods remain prevalent, recent research has shifted towards neural methods that directly optimize a rate--distortion objective. In particular, the hyperprior framework introduced by \textcite{balleVariationalImageCompression2018} has been widely adopted and adapted for RS data.
The flexibility of this approach in designing synthesis and analysis networks allows research to explore diverse architectures, with most studies emphasizing innovations along the ``Transform'' axis identified in Section~\ref{sec:compression_taxonomy}.

Future research directions include exploring data-specific characteristics, such as exploiting temporal correlations inherent in the relatively static nature of consecutive EOs. Recent studies adapt traditional video compression techniques to satellite image sequences. With reference-based coding, historical images of the same region are used to compress only the temporal changes instead of individual images~\cite{video_earth_observation, du2024earth+}.
The exploitation of spectral redundancy remains a research focus. For instance, INRs pose an interesting research direction for hyperspectral datasets, which often suffer from limited training samples.

As machine learning becomes increasingly important for processing RS data, it is essential to align compression techniques with the requirements of downstream tasks. Consequently, research is starting to shift from compression strategies optimized for human perception—such as minimizing MSE—toward methods that preserve data integrity for machine processing~\cite{furutanpey2024fool}. Neural feature compression shows promise in two distinct scenarios: the transmission of features from satellites to ground stations, overcoming the data downlink bottleneck, and the transmission of features from data centers to analysts for model training and inference.

Due to the limited resources onboard, only compression methods with low computational and storage complexity can be used. Nevertheless, the onboard application of machine learning represents an important future aspect, for example, to carry out geophysical tasks~\cite{Meoni2024}. With this in mind, the European Space Agency (ESA) launched the $\Phi$-Sat-2 mission to demonstrate onboard AI for various use cases. For instance, the first neural-based onboard compression using convolutional autoencoders was demonstrated, opening up new opportunities to save bandwidth and storage~\cite{Guerrisi2023} as well as to support the growing amount of satellite data. \textcite{lu_onboard_2024} developed an onboard AI model able to detect bushfire smoke much faster than traditional ground-based processing, which highlights the potential of onboard AI for real-time geophysical applications.


\section{Neural Compression for Climate Data}
\label{sec:compression_Climate}
Earth system models (ESMs) are one of the key tools for understanding the impact of anthropogenic climate change on the Earth.
ESMs model the dynamics of the earth's atmosphere on a discretized spherical grid; the horizontal grid spacing in current climate models is usually on the scale of around 100 km.
However, with such a coarse resolution many important processes, such as precipitation and deep convection, cannot be explicitly resolved, which motivated the development of the next generation of climate models with grid spacings on the scale of 1-5 km~\cite{stevens2013climate,palmer2014climate,schneider2017climate,stevens2019dyamond,bauer2021digital}.
The increased resolution of these models, in turn, leads to a significant increase in the size of datasets produced by ESMs~\cite{kay2015community,hersbach2020era5}.
For example, the recently launched Destination Earth initiative~\cite{bauer2021digital} generates around 1 petabyte of data \emph{per day}; making it infeasible to store all the generated data on disk for long time-scales.\footnote{\url{https://stories.ecmwf.int/the-digital-twin-engine/}}
With storage costs now making up a significant factor of computing center budgets~\cite{kunkel2014exascale,cappello2020fulfilling}, there is a pressing need for compression algorithms for climate data.

\subsection{Challenges}

\subsubsection{Data Characteristics}

Data generated by climate simulators has multiple key characteristics that set it apart from other data modalities and emphasize the need for bespoke compression tools and algorithms:

\textbf{Multidimensional data.} 
The output of models includes multiple variables, e.g. wind, atmospheric pressure, temperature, etc., which are localized in space and time and stored in multidimensional arrays~\cite{eaton2003netcdf,cherian2023cfxarray}. 
While natural images and videos commonly only have a small number of channels, i.e. 3 for an RGB image, climate models can have hundreds of different output variables.
A key feature of atmospheric data is that there is generally a high correlation in space and time, and between some of the different variables.
Additionally, different climate models do not necessarily use the same grid projection. 
So while the output data is generally saved as a multi-dimensional array, the corresponding coordinate reference system might vary between different climate models~\cite{roberts2018climate,danabasoglu2020community,jungclaus2022icon}.

\textbf{Importance of high-frequency signals.} 
The dynamics represented by a climate model are inherently chaotic; accurately modeling the evolution of the dynamics at smaller scales is important for accurately modeling the larger scale dynamics. The representation of variables at small scales also encodes the physics of key climate processes, such as clouds~\cite{stevens2019dyamond,schar2020kilometer}.
This makes compressing climate data fundamentally different from compression methods for other datasets such as natural images.
For natural images, blurring at smaller scales might be desirable because it does not create images that are visually distinguishable for humans~\cite{wang2004image}.
However, the importance of small-scale features for climate processes means that compression algorithms that overly smooth data over small scales may result in undesirable or unpredictable downstream effects~\cite{palmer2014real}.

\textbf{Extreme events.}
A key goal for climate modeling is to predict the probability of extreme events such as floods or storms, for which small-scale variability of precipitation is important~\cite{palmer2019scientific}.
It is therefore imperative that the statistics of these extreme events are preserved when compressing the data, which may require additional explicit constraints, due to their unlikely nature in the scope of entire datasets~\cite{mirowski2024neural}.

\textbf{Lack of quantitative metrics.}
As outlined in Section~\ref{sec:lossy_compression}, compression methods are usually evaluated based on how well they reconstruct the input data.
This requires a distortion metric to compare the original data, $\mathbf{x}$, and the reconstructed data, $\mathbf{x^\prime}$.
However, for climate models classical metrics such as the pixel-wise mean-squared error are often insufficient to capture the structural differences between inputs and reconstructions that scientists are interested in~\cite{rasp2024weatherbench,mandorli2024assessment,freischem2024multifractal}.
Ideally, any reconstruction should conserve the physical properties of the input in space and time, e.g. individual clouds should have the same mass in the input and reconstructed data and spatio-temporal structures should be preserved.
However, developing metrics that capture the distortions relevant for climate data is still an active area of research~\cite{baker2017toward,cappello2020fulfilling,klower2021compressing,baker2023structural}.

\subsubsection{Data Acquisition and Application}

Modern climate models are executed on the world's largest supercomputers. 
A single forecast run often requires carefully orchestrating and integrating multiple sub-components, such as ocean and atmospheric models~\cite{danabasoglu2020community,mauritsen2019developments,jungclaus2022icon}. 
Even on a supercomputer, completing a single run can take weeks to months. 
Consequently, data generated from these models is typically produced once and then stored for subsequent access by scientists. 
However, researchers often need only a subset of the data for their analysis---restricted to a specific time period, geographical region, or selection of variables---so they often want to avoid downloading the entire dataset.
Given that data is usually generated only once, it is then logical to invest significant computational resources towards compression if this reduces the bandwidth required for transmitting the data to scientists for further analysis.

\subsection{Classification of Compression Methodologies}
\label{sec:compression_methods_climate}

\subsubsection{Traditional Approaches}

Most existing codecs for multi-dimensional arrays are hand-engineered (in the sense of Section~\ref{sec:neural_traditional_compression}) and employ the transform-based compression approach described in Section~\ref{sec:lossy_compression}.
A key to achieving good compression ratios is to exploit correlations in the data; many codecs divide the input data into sub-blocks\footnote{For a $d$-dimensional array, each sub-block has size $n^d$ where $n$ is the sub-block size.} and try to identify correlations within a sub-block.
ZFP~\cite{lindstrom2014fixed} decorrelates individual sub-blocks using an orthogonal transform.
SZ3~\cite{liang2022sz3} uses a spline-based interpolator to identify correlations in a given sub-block.
TTHRESH~\cite{ballester2019tthresh} uses a generalization of the singular value decomposition (SVD) to tensors with more than three dimensions to transform the data.

Arguably, an even more simple approach to compression is to reduce the number of bits used for the floating point representation of individual output variables.
Most output of climate models is stored in 64-bit precision but to or many variables, the least significant bits in the mantissa of the IEEE floating point representation are effectively random noise, i.e. they are not useful in predicting the future state of the system~\cite{vavna2017single,tinto2019use}.
One can further formalize this by deriving a measure of the ``real information'' contained at a given bit location for a given variable~\cite{klower2021compressing}.
\textcite{klower2021compressing} find that in data from the Copernicus Atmospheric Monitoring Service~\cite{inness2019cams} most variables contain fewer than 7 bits of real information, i.e. the remaining 57 bits in the floating point representation are purely random noise which can be safely discarded.
This has been the motivation for a couple of compression schemes~\cite{zender2016bit,klower2021compressing} which discard a certain number of least significant bits in the mantissa of the floating point representation.
The advantage of this approach is that it is complementary to the compression schemes presented in the previous paragraph because it can simply be run as a pre-processing step before passing the dataset to a compressor. 

\subsubsection{Neural Approaches}

\begin{table*} 
\captionsetup{justification=centering} 
\captionof{table}{Contributions to the field of neural compression for climate data described in Section~\ref{sec:compression_methods_climate} ordered along the axes of the taxonomy described in Section~\ref{sec:compression_taxonomy}.}
\centering
\begin{tabularx}{\textwidth}{>{\bfseries}l X l} 
    \toprule
    \textbf{Axis} & \textbf{Approach} & \textbf{Papers} \\
    \midrule
    \multirow{2}{*}{Transforms} 
        & Pre-process input with HealPIX projection & \RaggedRight \cite{mirowski2024neural} \\
        & Pre-process input using Random Fourier Features  & \RaggedRight \cite{huang2023compressing} \\
    \midrule

    \multirow{2}{*}{Entropy Models} 
        & Hyperprior                  & \RaggedRight \cite{mirowski2024neural} \\
        & Hyperprior with Attention   & \RaggedRight \cite{han2024cra5} \\
    \midrule

    \multirow{1}{*}{Optimization Objectives} 
        & Rate-Distortion and/or Adversarial loss (GANs)          & \RaggedRight \cite{han2024cra5,mirowski2024neural} \\
    \bottomrule
    \label{tab:climate_overview}
\end{tabularx}
\end{table*}

Compared to other data modalities such as images, text, or video there has been relatively little work on developing NC methodologies for climate data.
Most work focuses on adapting existing architectures to the inherently spherical geometry of the data domain.

\textbf{Implicit neural representations} (INRs) have been shown to be able to compress ERA5 temperature data at higher compression ratios than JPEG2000~\cite{dupont2022coin++}.
\textcite{huang2023compressing} demonstrate that representing the data with random Fourier features~\cite{rahimi2007random,tancik2020fourfeat} leads to improved compression ratios.
However, these INR codecs can struggle to accurately capture extreme events in the input data.
In a case study examining the reconstructed ERA5 geopotential variable during Hurricane Matthew in October 2016, \textcite{huang2023compressing} found that the INR reconstructions failed to preserve the extreme values at the center of the hurricane.
Further work is needed to address these shortcomings in order to make INR a viable compression method for climate data.

\textbf{Autoencoders} \textcite{mirowski2024neural} evaluate a variety of neural architectures for compressing the temperature, pressure, wind velocity, geopotential, and humidity variables in the ERA5 data.
The architectures they consider are VQ-VAE, VQ-GAN, factorized prior and hyperprior models (see Section~\ref{sec:neural_compression_literature} for a description of these models).
As a pre-processing step, they use the HEALPix projection~\cite{gorski2005healpix} to re-grid the data which leads to pixels representing equal area and allows to efficiently compute spherical harmonics transforms.
Their experimental results indicate that the hyperprior model gives the best compression results and is able to preserve the power spectrum of the data more accurately than the alternative architectures; even though, the hyperprior model is trained only using a mean squared error reconstruction loss.
While the hyperprior model is found to be better at preserving extreme events compared to the INR approach of~\cite{huang2023compressing}, it does not provide any theoretical guarantees on bounding the reconstruction error compared to traditional approaches such as SZ3.
Concurrently, \textcite{han2024cra5} also use a hyperprior architecture to compress ERA5 data and develop a novel window-based attention mechanism for atmospheric data; they report compression ratios of around $\sim$ 300$\times$ while \textcite{mirowski2024neural} report ratios of $\sim$ 1000-3000$\times$.

\subsection{Summary}

NC methods have shown encouraging results in being able to compress weather and climate data.
As highlighted in Table~\ref{tab:climate_overview}, most existing work focuses on adapting architectures developed for the image compression domain and adapt them to the climate domain by adding pre-processing steps to account for the underlying spherical geometry of the input data.
While existing work reports impressive compression ratios on the order of $~1000\times$, more work is needed to establish trust in these novel neural codecs to ensure they do not erase extreme events such as hurricanes from the input data. 

Overall, the use of lossy compression for climate data is relatively under-explored both with traditional and neural approaches.
This can be attributed partly due to a lack of trust in lossy compression algorithms in the climate community which tends to have high standards for data integrity~\cite{baker2016evaluating,cappello2020fulfilling}.
There is a growing body of work that addresses this issue by developing quality metrics that lossy compression algorithms need to pass in order to be suitable for climate data ~\cite{baker2016evaluating,baker2017toward,poppick2018statistical,underwood2022understanding,sather2022can}.
Fundamentally, the goal of these evaluation metrics is to ensure that \emph{lossy compression does not change scientific conclusions} drawn from the data~\cite{cappello2020fulfilling}.
How to exactly assess this goal is still an area of open research. 
Suggested quality checks include: asking climate scientists to identify the (lossily) reconstructed members of an ensemble of ESM runs~\cite{baker2016evaluating}; adapting the SSIM metric for scientific datasets~\cite{baker2023structural}; using the change in real information content~\cite{klower2021compressing} of variables to detect reconstruction artifacts~\cite{sather2022can}; and defining thresholds on quantitative metrics~\cite{underwood2022understanding}.

However, so far, these quality metrics designed by climate scientists have seen little adoption in the NC literature.
Generally, neural approaches tend to mainly evaluate the trained codec based on mean-squared error and by assessing the power spectrum of the original and reconstructed data~\cite{huang2023compressing,mirowski2024neural}.
To ensure the adoption of neural codecs, it is therefore important that algorithm designers work hand-in-hand with climate scientists to build trust. 




\section{Neural Compression: Implementation \& Application}
\label{sec:implementation_application}
\subsection{Neural Compression for Geospatial Analytics Platforms}

The exponential growth of data from EO missions has led to significant challenges in transfer, storage, and processing, resulting in substantial resource requirements~\cite{wilkinson_environmental_2024}. Given the broad relevance of EO data across multiple fields~\cite{KANSAKAR201646}, innovative real-world compression methods can have a considerable impact.
As discussed in \cref{sec:compression_RS}, NC can significantly reduce the volume of EO data that needs to be transferred. Furthermore, the sheer scale of data collection means that most of this data will never be analyzed by humans. Instead, automated processing using computer vision algorithms—primarily neural networks—must handle the bulk of the data.
Given this context, despite most NC research focusing on human perception, as seen in \cref{sec:neural_compression_literature,sec:remote_sensing_compression_research}, we emphasize the emerging field of feature compression as a promising solution to these challenges.

A setting in which an image's feature vector is compressed instead of the image itself may allow for much greater compression ratios: The neural compressor may deem perceptual information unnecessary when maximizing performance on downstream tasks (e.g., scene classification, semantic segmentation, object detection) instead of reconstruction.
Using feature compression, \textcite{dubois2021lossy} demonstrate a 1000x rate gain compared to JPEG for ImageNet classification adapting a pretrained FM for compression. While segmentation or detection tasks remain relatively unexplored, even a fraction of this rate gain would be significant.
FMs~\cite{lu2024ai_EOFMreview}, pretrained by SSL, may be key in this context, as they provide generic features, which can be utilized for multiple downstream tasks. A platform like ESA's Copernicus Data Spaces Ecosystem (CDSE), which hosts the data from Copernicus Sentinel missions could compute and host compressed features generated by a FM trained with a compression objective. Customers would request those compressed embeddings, rather than raw data, for their various downstream applications and benefit from reduced i) egress cost, ii) data transfer latency, iii) storage requirements, and lower compute requirements for iv) model training, v) model inference, as well as from fast vi) search in data sets.

\textbf{Training:} In a feature compression system, training datasets for downstream tasks could be transferred as compressed embeddings. This approach reduces the volume of data that must be transferred and stored and allows the client to receive feature vectors directly representing the requested data. As a result, the client is relieved from the need to train a large backbone network as a feature extractor.

\textbf{Inference:} The client’s trained downstream model can then use these compressed embeddings for inference. This is particularly advantageous in scenarios requiring real-time user interactions, where latency is primarily dictated by the time needed to transfer input data. It also benefits large-scale inference tasks, such as global-scale analyses, where the reduced costs of data transfer and storage enable more frequent evaluations (e.g., monthly instead of yearly). In both cases, since embeddings are computed server-side, only a lightweight network is necessary on the client side. This empowers users without access to powerful computational resources to train models and enables edge devices to perform inference. 

\begin{figure}[ht]
    \centering
    \includegraphics[width=1.0\linewidth]{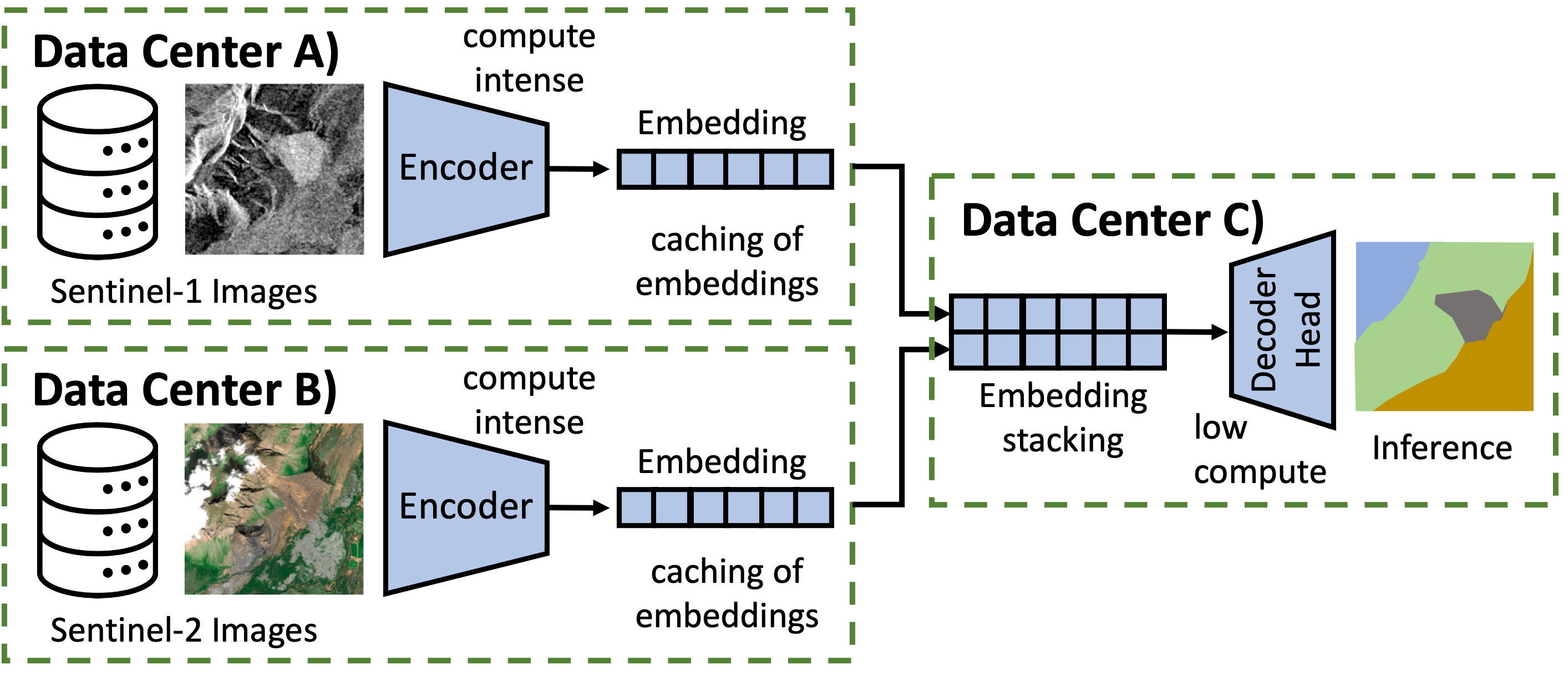}
    \caption{
        Concept of data federation through compressed embedding sharing between data centers. 
    }
    \label{fig:embeddingsharing}
\end{figure}

\textbf{Data Federation / Data Fusion:} Larger downstream models may also benefit from feature compression. The significantly reduced amount of data that must be transferred when working with NC allows for building downstream machine learning models based on fused data, hosted in different data storage facilities, e.g.: fusing optical imagery from Sentinel-2 hosted on the Amazon Web Services (AWS) in Europe (\texttt{AWS-eu-central-1}) with Landsat8 multi-spectral satellite imagery hosted in the United States (\texttt{AWS-us-west-2}), with large spatial and temporal extent \cref{fig:embeddingsharing}).

\textbf{Search:} Similarity search driven by embeddings has demonstrated significant potential in the EO domain, both in academic research~\cite{blumenstiel2024multi} and in practical applications. 
This method enables efficient querying of large datasets by identifying entries that are similar to an input query, bypassing the need for metadata-based searches. Typically, this is achieved using a vector database—a specialized system designed to retrieve similar vectors from vast datasets based on a similarity metric in vector space. Implementing a compressed, server-side store of these embeddings could facilitate the global scalability of this approach over extended periods.

Feature compression offers an effective balance between storage and processing demands: the computationally intensive task of generating embeddings can be handled by high-performance computing (HPC) systems, while lightweight decoding is reserved for end applications. This allows intermediate processing steps to operate on the reduced data size provided by the embeddings.
Moreover, using embeddings derived from FMs has shown great promise in few-shot learning~\cite{allen2023fewshot}, which can significantly lower the resources needed for data labeling while maintaining high task accuracy.

While deep NC of EO data addresses the challenges of storage, transfer, and processing, it introduces new challenges that must be managed when working with neurally compressed data:

\textbf{Tiling:} The optimal strategy for tiling the underlying data is unclear. Standard tile sizes in EO (e.g. 110x110 km in Sentinel-2) are generally too large to be used as inputs to neural network models. Additionally, embeddings are taken to represent the entire tile atomically and cannot (in general) be subdivided, making large tile sizes insufficiently granular for many EO tasks. Conversely, smaller tile sizes increase the number of embeddings that must be computed and stored.

\textbf{Caching / Pre-computing:} An additional advantage may arise from intelligently caching compressed embeddings. As the computation of these embeddings is concentrated on the server, previously duplicated computing efforts across distributed machines can be reduced. However, pre-computing embeddings via batch processing can also present difficulties due to tiling and the indivisibility of embeddings, potentially limiting users to requesting data regions that align with the tiling grid. An alternative would be stream processing, which offers greater flexibility but at the expense of efficiency.

\textbf{Spatio-Temporal extent:} The extent of data captured by a single compressed embedding must also be considered in spectral and temporal dimensions. Greater compression may be achieved by jointly compressing more spectral bands and multiple temporally adjacent observations, but this reduces flexibility, as users may only require a subset of the data.

\textbf{Standards:} Additional challenges involve establishing standards and adapting geospatial data platforms. To fully realize the benefits of neurally compressed EO embeddings, EO cloud platforms need to support their processing. Currently, neither proprietary platforms like Google Earth Engine~\cite{GORELICK201718} nor the OpenEO standard~\cite{schramm2021openeo} support neurally compressed embeddings. Furthermore, an open standard is needed for storing and transferring these compressed embeddings between storage and processing environments. To ensure interoperability, such a format must contain all necessary information for decompression. However, to our knowledge, no such standard currently exists. The Cloud-Native Geospatial Foundation~\cite{cloudnativegeo2024survey} is surveying how the geospatial community is storing embeddings in GeoParquet to potentially develop guidelines or introduce a standard for storing and exchanging embeddings in the future. While this may lead to a solution for storing raw embeddings $\mathbf z$, it does not address the challenge of creating a standardized format for storing neurally compressed embeddings. In addition to the compressed data, metadata must be included to provide details such as the FM used to generate the embeddings $\mathbf z$ and the geographic area to which the embeddings apply. To ensure that neurally compressed EO data is findable according to the FAIR data principles~\cite{wilkinson2016fair}, it should be cataloged in a standardized manner, such as by registering it in a Spatio-Temporal Asset Catalog (STAC) with sufficient metadata.

We therefore argue that existing data catalogs and EO processing platforms must be extended to support the provision and processing of neurally compressed embeddings. Only then can we reap the benefits of using embedding representations over base imagery.

\subsection{Cost- and Energy-Efficiency \& Latency}

\textbf{Space Segment Efficiency.} Energy efficiency is crucial in RS. An example is the need to maximize the operational lifespan of nanosatellites, ensure effective data transfer within limited downlink windows, and reduce operational costs~\cite{furutanpey2024}. Efficient onboard data processing reduces the volume of data transfers, mitigating bandwidth constraints. Advances in \textit{Orbital Edge Computing} and neural feature compression may enable satellites to handle large data volumes without excessive energy use, enhancing overall system efficiency. This is essential for the sustainability and effectiveness of satellite constellations in capturing and transmitting valuable EO data. 

\textbf{Ground Segment Efficiency.} However, the approaches reviewed also have a significant impact for the ground operation of the data infrastructure. In this section, we aim to give back-of-the-envelope estimations that quantitatively assess the role that NC can have in geospatial analytics. A concise system analysis of the information and communication technology infrastructure as well as a detailed scenario analysis are left out of the scope of this review. However, we aim at providing a first orientation with a simple model calculation. The demonstrative case we study is the Copernicus program, for which the Sentinel satellites are the main data sources. We investigate the potential benefits of Copernicus data products leveraging neural (feature) compression on the server side, before transmission to clients. The focus of our analysis is energy efficiency, with aspects of cost-efficiency and latency also included. 

\textbf{Copernicus Data Volumes and Transfer Cost.}
According to the Copernicus Data Dashboard\footnote{\url{https://dashboard.dataspace.copernicus.eu}, accessed on 2024/06/27}, at the time of writing, the total volume of data products is growing by 759~TB per month, with 6.2~PB of data products downloaded in the same period. We use these figures and estimate approximate yearly values of 10 and 100~PB respectively, rounded up to account for expected growth. Price indications for data transport out of cloud storage (egress costs) are given on the website of AWS\footnote{\url{https://aws.amazon.com/s3/pricing/?trk=ap_card}}. While the detailed pricing depends on the site of the host and consumer, 20 USD/TB is a realistic lower bound for 2024 when expecting a high quality of service. Hence, a gross data transfer cost of 2.000.000 USD/year per data product download is a reasonable estimate. 

\textbf{Energy Spent on Transfer.} The energy footprint of data transfer is very difficult to assess. For example, a meta-study by \textcite{aslan2018electricity} summarizes 14 studies that deviate by a factor of ten or more, even after adjusting the system boundaries. The web page \emph{wholegraindigital} expands on the challenge of defining suitable system boundaries\footnote{\url{https://www.wholegraindigital.com/blog/website-energy-consumption/}}. The authors challenge the idea of a single metric measuring the energy of data transfer and point out that \textcite{aslan2018electricity} only evaluate the usage of a subsystem. With this in mind, as a starting point, we pick the number of 0.01 kWh/GB, carefully following the extrapolation in Fig.~3 of \textcite{aslan2018electricity} but correcting upwards. With this assumption, the annual energy cost of data transfer sums up to about 1 GWh/year.

\textbf{Compute Demand for Neural Compression.} NC could reduce the data transfer burden. However, the compression is associated with energy consumption as well. We propose a simple order-of-magnitude estimate to assess the consumption that transfers the insights from computer vision to RS using BigEarthNet as an intermediate, where multispectral data is available in a similar format to natural images~\cite{sumbul2019bigearthnet}. 
Typical convolutional encoder networks from computer vision may require up to several tens of billions of floating point operations, or GigaFLOPs, for processing an RGB image of the characteristic size 224x224 that has been abundantly used in computer vision datasets~\cite{liu2021swin, deng2009imagenet}.

For transformer architectures, the operation count can be as high as hundreds of GigaFLOPs per image. For simplicity, we assume here that compressing a multispectral image with resolution 120x120 is comparable to processing a 224x224 RGB image. For encoding the entire BigEarthNet-S1 archive consisting of 590,326 non-overlapping image patches with a total volume of 66 GB, assuming 100 GigaFLOPs per image, this adds up to $6 \dot 10^{16}$ FLOPs in total, or about $10^{15}$ FLOPs/GB. 

\textbf{Time and Energy Demand for Neural Compression.} The GPU that currently dominates AI compute centers, NVIDIA's A100 GPUs, can, according to~\textcite{kesselheim2021juwels},  
with moderate optimization, sustain 50\% of their nominal performance using fp16 accuracy throughout the ML workloads. Hence, they can achieve approximately 150 TeraFLOPs per second at a power consumption of 400 W\footnote{\url{https://www.nvidia.com/content/dam/en-zz/Solutions/Data-Center/a100/pdf/nvidia-a100-datasheet-us-nvidia-1758950-r4-web.pdf}}. 
Based on our previous assumptions, and adding an overhead of 50\% for server operations and cooling, we obtain a processing time of seven seconds and an energy consumption of about 1~Wh per GB. Scaling up to the yearly data generation of about 10~PB per year, this amounts a total energy requirement of approximately 0.011 GWh/year. The total processing time adds up to approximately two years, so with a single commercial eight-GPU server the continuous provision of compressed data products can be realized. 

\begin{table}[!ht]
\centering
\begin{tabular}{lrl}
\hline
\textbf{Copernicus Data} &  &  \\ \hline
Data volume to process & 10 & PB/year \\ 
Downloaded data volume & 100 & PB/year \\ 
Energy of data transfer & 1 & GWh/year \\ 
Egress cost & 2Mio & USD/year \\ \hline

\textbf{Cost of Compression}  & & \\ \hline
Compute time on A100 & 1.89	& years \\
Energy for compression & 10 & MWh/year \\
Compression to transfer energy ratio & 0.01 & \\ \hline

\end{tabular}
\caption{Key parameters and results of the analysis comparing energy needed for data compression relative to data transfer.}
\label{tab:compression_cost}
\end{table}

\textbf{Energy Demand for Neural Compression compared to Data Transfer.} Comparing the potential energy savings and the compression, despite all uncertainties, it seems apparent that a one-off compression of the data would be almost negligible with an estimated three orders of magnitude difference from the transfer consumption and thus, the utilization of compression is highly beneficial. The key parameters and results of this investigation are depicted in Table~\ref{tab:compression_cost}.

\textbf{Embedding vs. Image Reconstruction.} As a final factor, it is important to assess the consumer side. Here we distinguish two scenarios. In scenario (a) the consumer performs an ML operation directly on the transferred embeddings; in (b) the consumer reconstructs the data for other downstream tasks. In scenario (a) typically the user will save energy as the compact compressed representation is potentially even better suited for this class of tasks. In scenario (b) we must consider the energy consumption of the reconstruction. In many encoder-decoder architectures, the computational efforts of both are balanced. In this case, this would in turn produce a computation effort of approximately $10^{15}$ FLOPs/GB. It is important to consider with which computational devices (CPUs, older models of GPUs, etc.) and with which expertise the decompression is performed. Under non-ideal conditions, the reconstruction could offset the savings of the efficient transfer entirely. Encouragingly, projects such as \texttt{llama.cpp}\footnote{\url{https://github.com/ggerganov/llama.cpp}} show that techniques such as model quantization allow for even computationally demanding ML models to be executed on a wide variety of hardware systems efficiently.

\textbf{Neural Compression Benefit Scenarios.} Finally, we would like to share two estimates demonstrating the possible latency improvements when employing neural feature compression. Considering the results reviewed in \cref{sec:lossy_compression,sec:compression_RS}, we assume a neural feature compression system may be able to achieve a 10x improvement in compression ratio over JPEG2000. For our first example, we consider a researcher performing a spatiotemporal analysis over 10 years worth of multispectral imagery across all of Germany. For May 2024, all Sentinel-2 images available for the entire country require 470GB (L1C data product) when stored in JPEG2000 format. This extrapolates to an estimated volume of 56TB for one decade. With an internet download speed of 100MBit/s, the data is available with a delay of about 52 days, posing a significant obstacle to the progress of research. Neural feature compression may reduce the time to only about a week. For a second example, a researcher may want to create a mosaic of land cover information from Sentinel-2 imagery given a single timestamp. Assuming the global land mass covers an area of approximately \SI{148940000}{\km\squared}, and given a single Sentinel-2 tile covers about \SI{10000}{\km\squared}, that amounts to roughly $15,000$ tiles. With one tile consuming 0.8GB, that amounts to 12TB of data downloaded to generate a landcover product using JPEG200. With the same internet download speed from the previous example, we obtain a latency of 11 days, compared to about 1 day when neural feature compression is applied.

\textbf{Summary.} This analysis of the status quo of the data product generation and consumption of the Copernicus program shows tremendous potential for savings in cost and energy scale. Furthermore, we highlight how a reduction in latency can enable data-intensive applications for consumers. While the challenging number of variables involved in the analysis requires us to rely on coarse estimates and assumptions at times, we believe these figures are still illustrative of the potential impact of neural (feature) compression.

\textbf{Future Directions.} Future projections are extremely challenging. NVIDIA's Road Map presentation at Computex\footnote{\url{https://blogs.nvidia.com/blog/computex-2024-jensen-huang/}} indicates optimism about future performance and energy efficiency gains that make compute-heavy approaches more attractive. Comparably, the cost of data transfer decrease as well. However the complex interplay of efficiency and demand with focus on data transfer energy demand is explained in \textcite{koomey2021does}. The increasing energy efficiency can lead to reduced energy consumption, but a more accessible resource can generate increasing demand that overcompensates the energy savings.

\subsection{Democratization for Applications}

\textbf{Beyond Compression.} As discussed in the previous section, the optimized compression of the massive raster data generated by EO systems and climate simulators show potential for reducing the energy cost and transmission latency both for data-distributing platforms and their end users.
As a result, stakeholders with limited bandwidth may access scientific data previously out of reach for their resources.
Moreover, when applied to embeddings generated by large pretrained models, NC would also permit downstream users with modest compute and deep learning expertise to benefit from expressive feature representations without the need for training a backbone from scratch on their end.
Here, we propose to illustrate our point with four example applications that could directly benefit from such neurally compressed data.

\subsubsection{\textbf{Global Vegetation Structure Analysis}}

\begin{figure}[t]
    \centering
    \includegraphics[width=.98\linewidth]{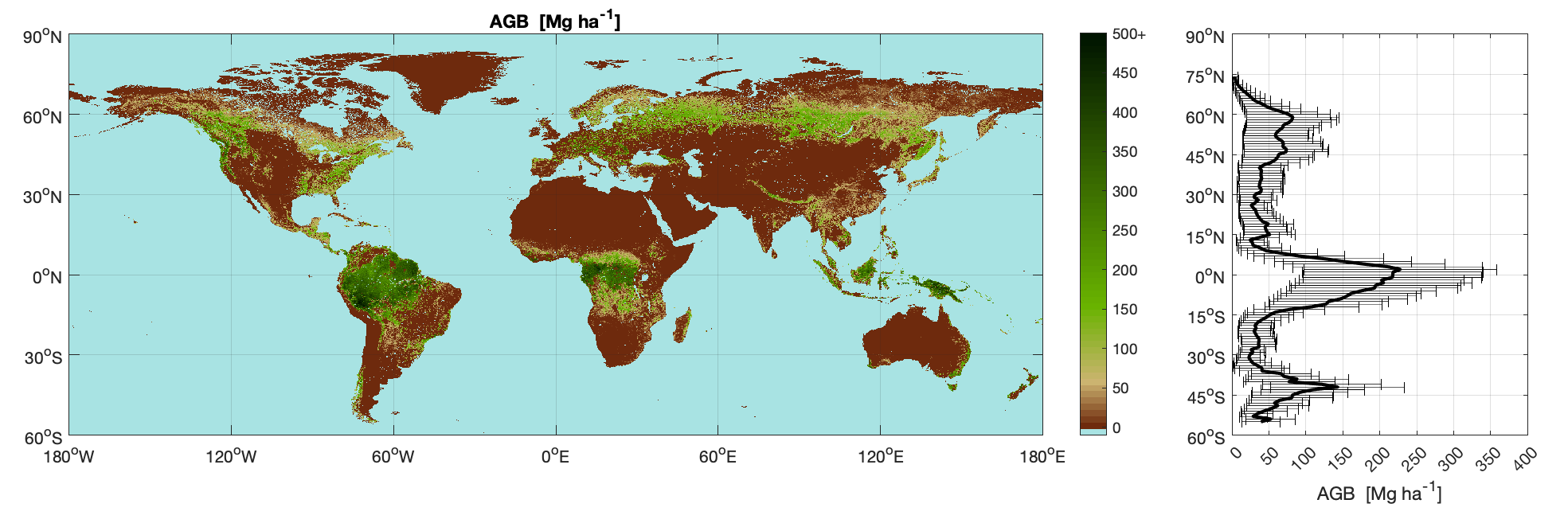}
    \caption{
        Aboveground biomass map from the Climate Change Initiative (CCI) Biomass project~\cite{santoro2021cci}. 
        Despite global coverage, this product has a 100~m ground sampling resolution and is only available for 2017, 2018, and 2020. 
        This is limiting for applications needing to monitor the evolution of vegetation structure at higher spatio-temporal resolution.
        Using neurally compressed satellite imagery or features would allow the computation and distribution of more frequent, higher-resolution, global vegetation structure products. 
    }
    \label{fig:biomass_cci}
\end{figure}

Worldwide mapping of vegetation properties is of prime importance for understanding the global carbon cycle~\cite{de2019global}, the impact of human activities on carbon emissions~\cite{hoang2021mapping}, and the study of ecosystem services~\cite{manning2018redefining}.
The accurate and frequent mapping of a small set of vegetation structure indicators such as canopy height (CH) and aboveground biomass (AGB) is key to the study of terrestrial ecosystem functions~\cite{jetz2019essential,migliavacca2021three}.

The traditional protocol for estimating such indicators requires in-situ--sometimes destructive--manual measurement surveys. 
Due to the poor spatio-temporal scalability of this approach, much research effort has been invested in characterising vegetation structure from RS data with terrestrial laser scanning (TLS) and aerial laser scanning (ALS)~\cite{brede2022non}.
While ALS provides accurate, dense, very high-resolution data, acquisition campaigns remain costly, limited to regional scales, with revisit rates of several years.
The ultimate need to scale vegetation mapping to global scale with revisit rates below one year and low-cost data hence calls for space-borne data.
Ideal satellite observations for global forest analysis need to capture vegetation properties at high spatial resolution with high revisit rate, and be freely available.
Several works have proposed to map forest structure from time-series of NASA/USGS Landsat or ESA Sentinel-1/2 acquisitions~\cite{potapov2021mapping}.
Recently, combining spaceborne LiDAR measurements from the NASA GEDI mission~\cite{dubayah2020global} with Sentinel imagery has shown great potential for regressing forest biophysical variables like AGB or CH at a global scale and 10~m resolution~\cite{lang2023high}. 

Still, \textcite{lang2023high} find that the prediction of a single, global map for the year of 2020 requires extensive computational power.
In order to cover the entire landmass of the Earth (excluding Antarctica), a total of $\sim 160$ terabytes of Sentinel-2 image data need to be downloaded. 
Running the model on these images takes $\sim 27,000$ GPU-hours ($\sim 3$ GPU years) of computation time, parallelized on a high-performance cluster to obtain the global map in \textit{ten days} real time.
Yet, the breakdown of the entire process reveals that more than half of the time is spent downloading and moving the data around. 

Besides, the rise of SSL leading to the current emergence of RS FMs~\cite{sun2022ringmo, smith2023earthpt, jakubik2023foundation, liu2024remoteclip} renders possible the distribution of expressive feature representations directly usable for downstream vegetation-related tasks~\cite{tolan2024very} without the need for the compute or AI expertise required to train the corresponding deep learning architecture.

Consequently, a pipeline capable of efficiently and accurately transmitting neurally-compressed sensor data or pretrained feature representations would allow producing and frequently updating vegetation structure maps.
By lowering the compute, bandwidth, and AI skills required for using deep learning models to regress vegetation structure variables from RS data, more stakeholders may take part in the production and analysis of such products.
This would in turn benefit crucial applications such as ecosystem protection and global carbon cycle monitoring.
What is more, new use cases may also emerge from the facilitated access to global vegetation structures.
For instance, numerous industrial actors are in need for tools for monitoring deforestation-free supply chains without investing in large-scale data storage, computer infrastructure, nor deep learning knowledge. 
A typical example are companies depending on commodities sourced in the tropics such as palm oil or cocoa~\cite{renier2023transparency, kalischek2023cocoa}. 


\subsubsection{\textbf{Ship Detection for Maritime Awareness}}

Ship detection is an important aspect of maritime awareness, as ships often carry valuable cargo and pose a potential threat to populations and infrastructure.
There are various methods for detecting ships, including SAR~\cite{margarit2009operational}, optical modalities~\cite{zhu2010novel} and Automatic identification system (AIS)~\cite{pelich2014ais}.
EO data allow ship traffic monitoring and the identification of potential security threats on large areas and the support of AIS data provides a technology used for maritime safety and security in near real-time to identify and track vessels.
Receiving timely, reliable, and meaningful information is therefore crucial.
In the last years, AI and ML have been used to detect, identify, and classify vessels in an automatic way~\cite{tang2014compressed}.
Vessel identification could benefit from neurally-compressed RS images or corresponding pretrained features in order to:
\begin{itemize}
    \item Compress the images to improve data transfer latency and facilitate access to relevant sources and collateral data (e.g. AIS).
    \item Support the creation tools for ship and port monitoring with minimal data labeling.
    \item Support the fusion of GeoData with AIS data for anomaly detection of ship movements.
\end{itemize}

\begin{figure}[t]
    \centering
    \includegraphics[width=.75\linewidth]{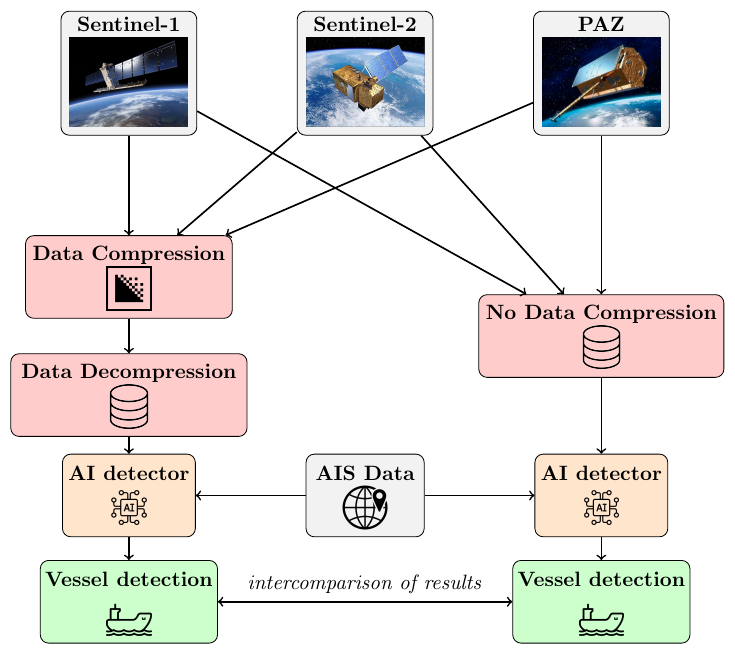}
    \caption{
         Workflow for ship detection using satellite imagery and artificial intelligence. 
    }
    \label{fig:maritime_awareness}
\end{figure}


\subsubsection{\textbf{Climate and Air Pollution Prediction}}

As described in Section~\ref{sec:compression_Climate}, high-resolution climate models are able to resolve key small-scale phenomena such as clouds and ocean eddies~\cite{schneider2017climate,stevens2019dyamond}.
An additional advantage of the increased resolution of modern climate models is that their generated data is now at the same resolution as the observations from RS devices such as geostationary satellites~\cite{stevens2019dyamond}.
However, the sheer volume of data generated poses challenges for full scientific exploitation, as the datasets are often unwieldy for efficient analysis and distribution.

The ability to compress the data into embeddings would significantly broaden the access to these datasets and enable new workflows.
For example, the potential use cases for the generated embeddings include:

\textbf{Training} cloud classification and air pollution prediction models directly in the embedding space, circumventing the need for complex and computationally expensive image processing methods. This includes identifying and tracking convective storms in the embedding space.

\textbf{Detecting} extreme events by modeling the distribution of embeddings and detecting out-of-distribution samples directly in the embedded space.

\textbf{Comparing} the outputs of climate model simulations with observational data. Meaningful embeddings make it possible to compute statistics about the occurrence of individual cloud types (e.g. deep convection and shallow convection) which is more difficult in the raw data space~\cite{mooers2023comparing}.

While the use of embeddings has shown some promising early applications~\cite{mooers2023comparing,kurihana2022aicca,denby2020Discovering,denby2023Chartinga}, their adoption is still in its infancy. 
It is important that any analysis run on the embeddings produced by a model lead to the same scientific conclusions as the analysis run on the full dataset.
Hence, any model for generating embeddings should be developed in collaboration with domain experts to ensure that the generated embeddings are fit for purpose.

\subsubsection{\textbf{Early Crop Stress and Yield Prediction}}
European agriculture is continuously affected by an increasing frequency of weather extremes~\cite{beillouin2020impact, hristov2020analysis}, which are expected to increase in magnitude and frequency in the near future.
How crops are affected by adverse weather conditions strongly depends on the crop's development stage.
Systems for timely monitoring of crop phenology are necessary to understand and assess the impact of climate change on crops~\cite{white2005global}.
The Sentinel missions~\cite{berger2012esa} contribute significantly to agricultural monitoring with its high temporal and spatial resolution.
Despite the development of crop maps and crop yield forecasting activities at the European scale~\cite{van2019performance}, integrating EO and climatological data is needed to capture the effect of increasing weather extremes on crops.

In particular, the early prediction of crop stress or crop yields at field, national, or continental level benefits from Sentinel-1/2 time series. Satellite time series also proved to be a valuable resource to improve crop type classification~\cite{Vuolo_2018} by capturing the dynamic changes in spectral and temporal signatures of crops during the growing season~\cite{Russwurm_2017_CVPR_Workshops}.
Comparatively, methods based on single-date imagery fail to accurately capture variations in phenology, biomass accumulation, and the effects of local conditions.

While crop-related tasks have proven to benefit from multi-modal, multi-date satellite imagery, mobilizing the necessary data and running models on it requires significant computational resources.
An efficient compression pipeline would allow the distribution of raw imagery or embeddings to stakeholders currently hindered by bandwidth and hardware requirements.
Such pipeline would support a range of actors in the agricultural community: farmers and agricultural organizations (e.g., improved monitoring/forecasting of field damage assessment), the public sector responsible for governing the transition of agriculture, the private sector, including agricultural technology and machinery industries, seed companies and agribusiness retailers, the agrochemical industry, and the insurance sector for risk management, and environmental agencies conducting crop forecasting activities.


\section{Perspectives \& Recommendations}
\label{sec:perspectives}
\begin{figure*}[t]
    \centering
    \resizebox{\textwidth}{!}{
    \begin{tikzpicture}
        \node[ellipse,
            inner color=yellow!60!black,
            fill=white,
            opacity=.3,
            minimum width=40ex,
            minimum height=15ex,
            rotate=-45,
            label={[orange!60!black, align=center]center:\it agriculture\\\it business},
        ] (grains) at (5.5,2.5) {};
        \node[ellipse,
            inner color=green!40!black,
            fill=white,
            opacity=.3,
            minimum width=40ex,
            minimum height=10ex,
            label={[green!40!black, align=center]center:\it biomass\\\it monitoring},
        ] (trees) at (3,4) {};
        \node[ellipse,
            inner color=blue!40!black,
            fill=white,
            opacity=.3,
            minimum width=15ex,
            minimum height=15ex,
            label={[blue!40!black, align=center]center:\it\qquad\relsize{-1}{maritime}\\\it\qquad \relsize{-1}{awareness}},
        ] (ships) at (1.2,1.2) {};
        \node[ellipse,
            inner color=red!60!black,
            fill=white,
            opacity=.3,
            minimum width=30ex,
            minimum height=15ex,
            label={[red!60!black, align=center]center:\it atmospheric\\\it dynamics},
        ] (clouds) at (8,1) {};
        \node (VQ-VAE/GAN) at (.5,.4) {\relsize{-3}{\cite{van2017neural,esser_taming_2021}}};
        \node (COIN++) at (6,1.2) {\relsize{-3}{\cite{dupont2022coin++}}};
        \node (RS_VIDEO) at (5,.9)  {\relsize{-3}{\cite{video_earth_observation,du2024earth+}}};
        \node (FINR) at (8,.5) {\relsize{-3}{\cite{huang2023compressing,mirowski2024neural}}};
        \node (HYPERSPEC) at (1.3,2.5) {\relsize{-3}{\cite{anuradha_efficient_2024,zhang_compressing_2024,Wang_2017,fuchs2023hyspecnet,mijaresiverduScalableReducedComplexityCompression2023}}};
        \draw [gray,thick,domain=90:-90, dashed] plot ({2.5*cos(\x)}, {3.5+2*sin(\x)});
        \node [text width=2cm] (FM) at (1.2,3.7) {\sc\color{gray}\relsize{-1}Geospatial Foundation Models};
        \draw [gray,thick,domain=0:180, dashed] plot ({6.2+1.5*cos(\x)}, {1*sin(\x)});
        \node [text width=3.5cm] (FM) at (1.4,2.8) {\begin{center}\color{black}\relsize{-3}{hyperspectral\\EO compr.} \end{center}};
        \node [text width=3.5cm] (EO_FM) at (.7,1.9) {\begin{center}\color{black}\relsize{-3}{EO compr.\\\cite{alves_de_oliveira_reduced-complexity_2021,kong_spectralspatial_2021,cao_spectralspatial_2022,chien_tensor-factorized_2018}}\end{center}};
        \node [text width=2cm] (VC) at (6.2,.5) {\begin{center}\sc\color{gray}\relsize{-1}Video\\Compression\end{center}};
        \draw [gray,thick,domain=90:0, dashed] plot ({1.3*cos(\x)}, {2*sin(\x)});
        \node [text width=2cm] (VC) at (.5,1) {\begin{center}\sc\color{gray}\relsize{-1}Image\\Compr.\end{center}};
        \draw[thick,->] (0,0) -- (10,0) node[anchor=north east] {\sf multi-temporal};
        \draw[thick,->] (0,0) -- (0,5) node[rotate=90, anchor=south east] {\sf multi-modal};
    \end{tikzpicture}}
    \caption{
        We represent downstream applications from the perspective of conceptual dimensions relevant for NC methodologies: multi-temporality and multi-modality.
        References to existing neural data compression literature are positioned with respect to these concepts.
        A gap in the literature can be observed for the compression of multimodal and multi-temporal data, showing potential for several downstream applications. 
    }
    \label{fig:AICompressorApplicationSpace}
\end{figure*}
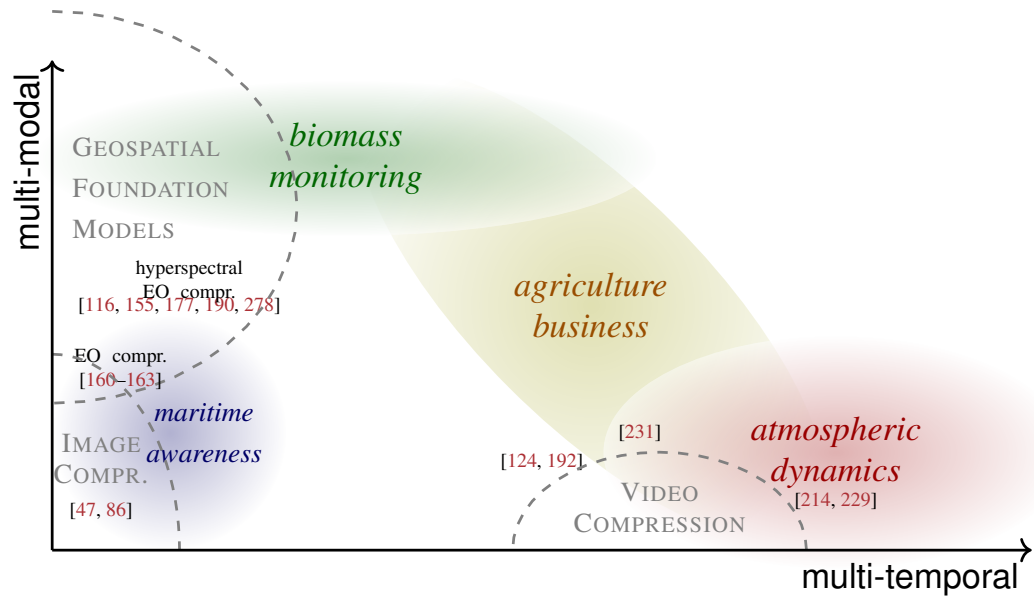

We conclude our literature survey on NC for geospatial analytics by a summary presented through \cref{fig:AICompressorApplicationSpace}. 
Novel methodologies to compress EO and ESM data need to cover a wide range of use cases---from single-image compression to embedding long time series, while incorporating information from a plethora of sensors and simulated physical quantities. 
However, existing (neural) compression algorithms only partially suit such needs. 
Image compression offers techniques to compress single-timestamp and single-modality data. 
Recent developments in FMs work towards joint representation learning of a variety of remote sensors. 
Video compression provides concepts to summarize time series of images as relevant to ESM applications. 
However, those algorithms currently lag support for multi-modal inputs.

\vspace{2ex}

\textbf{Earth Observation.} Given radar, LiDAR, and multi-spectral sensors operate on various bands of the electromagnetic spectrum, Hyperspectral EO data compression offers a direction toward multi-modal compression. However, a clear deficiency in the current research in the field is the lack of an established methodology to quantitatively compare methods.
Datasets meant as a benchmark for learned compression for the wide range of existing RS modalities are scarce or not used widely enough to enable systematized comparisons. The availability of standard training datasets, and perhaps more importantly, the alignment of the research community on standard evaluation datasets for different EO modalities is critical in enabling improvements in methodology in the field.

Further development of the transform $f$ for NC to better suit different modalities within RS data seems bound to continue to be a fruitful research direction. However, it should be noted that this direction poses the risk of incentivizing continuous small adaptations to methods proposed in the field of natural image compression with limited innovation regarding RS compression. On the other hand, other differentiating data characteristics in this domain seem relatively underexplored as of yet regarding their integration into NC methodologies. For instance, RS data is very rich in metadata. Specifically, geolocation and time of capture may be informative.

While neural video compression has been successfully employed in natural videos, a major difficulty is its usual reliance on \textit{optical flow} fields and their compression, which proves challenging in scenes with fast motion in uncorrelated directions. The adaptation of these techniques to EO, where such optical flow fields are mostly absent, has the potential to yield great compression ratios where the goal is to transmit a temporal sequence of samples.

\vspace{2ex}

\textbf{Earth System Modeling.} Compared to EO, NC for climate model data has received relatively little attention in the academic literature. Existing approaches for compressing the outputs of climate simulators mainly rely on hand-engineered transform coding schemes~\cite{lindstrom2014fixed,zender2016bit,liang2022sz3,ballester2019tthresh,klower2021compressing}. Data-driven NC offers an attractive alternative~\cite{huang2023compressing,mirowski2024neural,han2024cra5} but more work is needed to rigorously evaluate their impact on the scientific integrity of the datasets.

A major obstacle to designing and evaluating compression schemes is the lack of agreed-upon quantitative metrics that can be used to reliably assess whether lossy compression schemes preserve all the relevant aspects of the data for climate analysis~\cite{baker2016evaluating,cappello2020fulfilling,underwood2022understanding}. 
Hence, future work should not only focus on the development of new compression schemes, but also new metrics designed along with domain experts to meaningfully evaluate and compare different codecs.

\vspace{2ex}

\textbf{Foundational Models for Geospatial Analytics.} A desirable property of foundational models that is currently underexplored for NC would be to tightly integrate well-calibrated uncertainties by design. Model outputs with well-calibrated uncertainties could ease integration not only into downstream tasks based on deep learning but also, and more importantly, become a natural interface to Bayesian methods~\cite{wilson2020case,bishop2006pattern}, mechanistic modeling~\cite{kraslawski201323rd}, and the existing rich statistics toolbox including significance tests~\cite{lehmann1986testing}. Furthermore, well-calibrated uncertainties can work as a natural link to physics-based forward simulations in computational science~\cite{reichstein2019deep}, e.g. to tightly integrate radiative transfer models with learning-based approaches in RS~\cite{zerah2023}. Computing uncertainties along with model outputs would also act as a natural early alert if a given foundational model would be applied to new data far away from the original training distribution. 
In that way, FMs capable of uncertainty prediction could, for instance, identify strategical training samples within an active learning setting.

\textbf{Resulting Research Recommendations.} To sum up, we list our main research recommendations to advance the domain of NC in the EO and ESM domain.
\begin{enumerate}
    \item Extend NC for temporal and multi-modal EO and ESM data.
    \item Establish metrics to evaluate utility of NC results for specific applications.
    \item Compile methodologies and data sets to benchmark the progress of NC algorithms.
    \item Take advantage of FM development for NC.
    \item Establish guard rails to estimate compression quality in the context of distribution shift of input data.
\end{enumerate}

Advancing and solving these main issues will be key to establish NC as a widely accepted methodology, disrupting the EO and ESM domain with efficiency in data storage, transfer and computation. 

\section*{Acknowledgment}
\addcontentsline{toc}{section}{Acknowledgment}

This research is carried out as part of the Embed2Scale project and is co-funded by the EU Horizon Europe program under Grant Agreement No. 101131841. Additional funding for this project has been provided by the Swiss State Secretariat for Education, Research and Innovation (SERI) and UK Research and Innovation (UKRI).


\renewcommand*{\bibfont}{\footnotesize}
\printbibliography

@article{Zhang2024CVPRa,
    author = {Xinjie Zhang and Ren Yang and Dailan He and Xingtong Ge and Tongda Xu and Yan Wang and Hongwei Qin and Jun Zhang},
    title = {Boosting Neural Representations for Videos with a Conditional Decoder},
    journal = {Proceedings of the IEEE Conference on Computer Vision and Pattern Recognition (CVPR)},
    year = {2024},
    url = {https://arxiv.org/abs/2402.18152}
}

@article{Kim2023CVPR,
    author = {Hyunjik Kim and Matthias Bauer and Lucas Theis and Jonathan Richard Schwarz and Emilien Dupont},
    title = {C3: High-performance and low-complexity neural compression from a single image or video},
    journal = {Proceedings of the IEEE Conference on Computer Vision and Pattern Recognition (CVPR)},
    year = {2023},
    url = {https://arxiv.org/abs/2312.02753}
}

@article{Duan2024CVPR,
    author = {Zhihao Duan and Ming Lu and Justin Yang and Jiangpeng He and Zhan Ma and Fengqing Zhu},
    title = {Towards Backward-Compatible Continual Learning of Image Compression},
    journal = {Proceedings of the IEEE Conference on Computer Vision and Pattern Recognition (CVPR)},
    year = {2024},
    url = {https://arxiv.org/abs/2402.18862}
}

@article{Zhang2024CVPRb,
    author = {Zhe Zhang and Huairui Wang and Zhenzhong Chen and Shan Liu},
    title = {Learned Lossless Image Compression based on Bit Plane Slicing},
    journal = {Proceedings of the IEEE Conference on Computer Vision and Pattern Recognition (CVPR)},
    year = {2024},
    url = {https://openaccess.thecvf.com//content/CVPR2024/papers/Zhang_Learned_Lossless_Image_Compression_based_on_Bit_Plane_Slicing_CVPR_2024_paper.pdf}
}

@article{Khoshkhahtinat2024CVPR,
    author = {Atefeh Khoshkhahtinat and Ali Zafari and Piyush M. Mehta and Nasser M. Nasrabadi},
    title = {Laplacian-guided Entropy Model in Neural Codec with Blur-dissipated Synthesis},
    journal = {Proceedings of the IEEE Conference on Computer Vision and Pattern Recognition (CVPR)},
    year = {2024},
    url = {https://arxiv.org/abs/2403.16258}
}

@article{Li2024ECCV,
    author = {Jiahao Li and Bin Li and Yan Lu},
    title = {Neural Video Compression with Feature Modulation},
    journal = {Proceedings of the European Conference on Computer Vision (ECCV)},
    year = {2024},
    url = {https://arxiv.org/abs/2402.17414}
}

@article{czerkawski2024global,
  title={Global and Dense Embeddings of Earth: Major TOM Floating in the Latent Space},
  author={Czerkawski, Mikolaj and Kluczek, Marcin and Bojanowski, J{\k{e}}drzej S},
  journal={arXiv preprint arXiv:2412.05600},
  year={2024}
}

@article{mukkavilli2023ai,
  title={Ai foundation models for weather and climate: Applications, design, and implementation},
  author={Mukkavilli, S Karthik and Civitarese, Daniel Salles and Schmude, Johannes and Jakubik, Johannes and Jones, Anne and Nguyen, Nam and Phillips, Christopher and Roy, Sujit and Singh, Shraddha and Watson, Campbell and others},
  journal={arXiv preprint arXiv:2309.10808},
  year={2023}
}

@article{lu2024ai,
  title={AI foundation models in remote sensing: A survey},
  author={Lu, Siqi and Guo, Junlin and Zimmer-Dauphinee, James R and Nieusma, Jordan M and Wang, Xiao and VanValkenburgh, Parker and Wernke, Steven A and Huo, Yuankai},
  journal={arXiv preprint arXiv:2408.03464},
  year={2024}
}

@article{wang2022self,
  title={Self-supervised learning in remote sensing: A review},
  author={Wang, Yi and Albrecht, Conrad M and Braham, Nassim Ait Ali and Mou, Lichao and Zhu, Xiao Xiang},
  journal={IEEE Geoscience and Remote Sensing Magazine},
  volume={10},
  number={4},
  pages={213--247},
  year={2022},
  publisher={IEEE}
}

@article{braham2024spectralearth,
  title={SpectralEarth: Training Hyperspectral Foundation Models at Scale},
  author={Braham, Nassim Ait Ali and Albrecht, Conrad M and Mairal, Julien and Chanussot, Jocelyn and Wang, Yi and Zhu, Xiao Xiang},
  journal={arXiv preprint arXiv:2408.08447},
  year={2024}
}

@article{deng2020model,
  title={Model compression and hardware acceleration for neural networks: A comprehensive survey},
  author={Deng, Lei and Li, Guoqi and Han, Song and Shi, Luping and Xie, Yuan},
  journal={Proceedings of the IEEE},
  volume={108},
  number={4},
  pages={485--532},
  year={2020},
  publisher={IEEE}
}

@article{lu2024ai_EOFMreview,
    title={AI Foundation Models in Remote Sensing: A Survey},
    author={Lu, Siqi and Guo, Junlin and Zimmer-Dauphinee, James R and Nieusma, Jordan M and Wang, Xiao and VanValkenburgh, Parker and Wernke, Steven A and Huo, Yuankai},
    journal={arXiv preprint arXiv:2408.03464},
    year={2024},
}

@article{qian2022entroformer,
    title={Entroformer: A Transformer-based Entropy Model for Learned Image Compression}, 
    author={Qian, Yichen and Lin, Ming and Sun, Xiuyu and Tan, Zhiyu and Jin, Rong},
    journal={arXiv preprint arXiv:2202.05492},
    year={2022},
}

@article{Guerrisi2023,
    author={Guerrisi, Giorgia and Del Frate, F. and Schiavon, Giovanni},
    year={2023},
    title={Artificial Intelligence Based On-Board Image Compression for the Φ-Sat-2 Mission},
    journal={IEEE Journal of Selected Topics in Applied Earth Observations and Remote Sensing},
}

@article{Meoni2024,
    author={Meoni, Gabriele and Prete, Roberto and Serva, Federico and Beusscher, Alix and Colin, Olivier and Longépé, Nicolas},
    year={2024},
    title={Unlocking the Use of Raw Multispectral Earth Observation Imagery for Onboard Artificial Intelligence},
    journal={IEEE Journal of Selected Topics in Applied Earth Observations and Remote Sensing},
}

@article{ZHAO2021,
    title={Symmetrical lattice generative adversarial network for remote sensing images compression},
    journal={ISPRS Journal of Photogrammetry and Remote Sensing},
    year={2021},
    author={Shihui Zhao and Shuyuan Yang and Jing Gu and Zhi Liu and Zhixi Feng},
}

@article{cheng2020,
    title={Learned Image Compression with Discretized Gaussian Mixture Likelihoods and Attention Modules}, 
    author={Cheng, Zhengxue and Sun, Heming and Takeuchi, Masaru and Katto, Jiro},
    journal={CVPR},
    year={2020},
}

@article{strümpler2022implicitneural,
    title={Implicit neural representations for image compression},
    author={Str{\"u}mpler, Yannick and Postels, Janis and Yang, Ren and Gool, Luc Van and Tombari, Federico},
    journal={ECCV},
    year={2022},
}

@article{Martone22,
    author={Martone, Michele and Gollin, Nicola and Rizzoli, Paola and Krieger, Gerhard},
    journal={IEEE Transactions on Geoscience and Remote Sensing}, 
    title={Performance-Optimized Quantization for SAR and InSAR Applications}, 
    year={2022},
}

@article{Paras23,
    author={Maharjan, Paras and Li, Zhu},
    year={2023},
    title={Complex-valued SAR Image Compression: A Novel Approach for Amplitude and Phase Recovery},
    journal={International Conference on Visual Communications and Image Processing},
}

@article{balleEndtoendOptimizedImage2016,
    title={End-to-End {{Optimized Image Compression}}},
    author={Ballé, Johannes and Laparra, Valero and Simoncelli, Eero P.},
    journal={ICLR},
    year={2016},
}

@article{van2017neural,
    title={Neural discrete representation learning},
    author={Van Den Oord, Aaron and Vinyals, Oriol and others},
    journal={NeurIPS},
    year={2017},
}

@misc{copernicus2022,
    author={Copernicus},
    title={Observer: Copernicus Climate and Atmosphere Services provide 2021 climate insights at global and European level},
    year={2022},
}

@article{bommasani2021opportunities,
    title={{On the Opportunities and Risks of Foundation Models}},
    author={Bommasani, Rishi and Hudson, Drew A and Adeli, Ehsan and Altman, Russ and Arora, Simran and von Arx, Sydney and Bernstein, Michael S and Bohg, Jeannette and Bosselut, Antoine and Brunskill, Emma and others},
    journal={arXiv preprint arXiv:2108.07258},
    year={2021},
}

@article{autoregressive_hyperprior,
    author={Minnen, David and Ball\'{e}, Johannes and Toderici, George D},
    journal={NeurIPS},
    title={Joint Autoregressive and Hierarchical Priors for Learned Image Compression},
    year={2018},
}

@article{zeghidour2021soundstream,
    title={Soundstream: An end-to-end neural audio codec},
    author={Zeghidour, Neil and Luebs, Alejandro and Omran, Ahmed and Skoglund, Jan and Tagliasacchi, Marco},
    journal={IEEE/ACM Transactions on Audio, Speech, and Language Processing},
    year={2021},
}

@article{strumpler2022implicit,
    title={Implicit neural representations for image compression},
    author={Str{\"u}mpler, Yannick and Postels, Janis and Yang, Ren and Gool, Luc Van and Tombari, Federico},
    journal={ECCV},
    year={2022},
}

@article{dupont2021coin,
    title={Coin: Compression with implicit neural representations},
    author={Dupont, Emilien and Goli{\'n}ski, Adam and Alizadeh, Milad and Teh, Yee Whye and Doucet, Arnaud},
    journal={arXiv preprint arXiv:2103.03123},
    year={2021},
}

@article{mildenhall2021nerf,
    title={Nerf: Representing scenes as neural radiance fields for view synthesis},
    author={Mildenhall, Ben and Srinivasan, Pratul P and Tancik, Matthew and Barron, Jonathan T and Ramamoorthi, Ravi and Ng, Ren},
    journal={Communications of the ACM},
    year={2021},
}

@article{sitzmann2020implicit,
    title={Implicit neural representations with periodic activation functions},
    author={Sitzmann, Vincent and Martel, Julien and Bergman, Alexander and Lindell, David and Wetzstein, Gordon},
    journal={NeurIPS},
    year={2020},
}

@article{zhu2022transformerbased,
    title={Transformer-based Transform Coding},
    author={Yinhao Zhu and Yang Yang and Taco Cohen},
    journal={ICLR},
    year={2022},
}

@article{shannon_source_coding,
    author={Shannon, C. E.},
    journal={The Bell System Technical Journal}, 
    title={A mathematical theory of communication}, 
    year={1948},
}

@article{gomes2024compressed,
    title={Compressed Multi-task embeddings for Data-Efficient Downstream training and inference in Earth Observation},
    author={Gomes, Carlos and Brunschwiler, Thomas},
    journal={arXiv preprint arXiv:2403.17886},
    year={2024},
}

@article{bird20213d,
    title={3d scene compression through entropy penalized neural representation functions},
    author={Bird, Thomas and Ball{\'e}, Johannes and Singh, Saurabh and Chou, Philip A},
    journal={2021 Picture Coding Symposium (PCS)},
    year={2021},
}

@article{Gomes_2023_CVPR,
    author={Gomes, Carlos and Azevedo, Roberto and Schroers, Christopher},
    title={Video Compression With Entropy-Constrained Neural Representations},
    journal={CVPR},
    year={2023},
}

@article{lessig2023atmorep,
    title={{AtmoRep: A stochastic model of atmosphere dynamics using large scale representation learning}},
    author={Lessig, Christian and Luise, Ilaria and Gong, Bing and Langguth, Michael and Stadler, Scarlet and Schultz, Martin},
    journal={arXiv preprint arXiv:2308.13280},
    year={2023},
}

@article{deletang2023language,
    title={Language modeling is compression},
    author={Del{\'e}tang, Gr{\'e}goire and Ruoss, Anian and Duquenne, Paul-Ambroise and Catt, Elliot and Genewein, Tim and Mattern, Christopher and Grau-Moya, Jordi and Wenliang, Li Kevin and Aitchison, Matthew and Orseau, Laurent and others},
    journal={arXiv preprint arXiv:2309.10668},
    year={2023},
}

@article{mentzer2022vct,
    title={VCT: A Video Compression Transformer},
    author={Mentzer, Fabian and Toderici, George D and Minnen, David and Caelles, Sergi and Hwang, Sung Jin and Lucic, Mario and Agustsson, Eirikur},
    journal={NeurIPS},
    year={2022},
}

@article{wang2004image,
    title={Image quality assessment: from error visibility to structural similarity},
    author={Wang, Zhou and Bovik, Alan C and Sheikh, Hamid R and Simoncelli, Eero P},
    journal={IEEE transactions on image processing},
    year={2004},
}

@article{toderici2015variable,
    title={Variable rate image compression with recurrent neural networks},
    author={Toderici, George and O'Malley, Sean M and Hwang, Sung Jin and Vincent, Damien and Minnen, David and Baluja, Shumeet and Covell, Michele and Sukthankar, Rahul},
    journal={arXiv preprint arXiv:1511.06085},
    year={2015},
}

@article{todericiFullResolutionImage2017a,
    title={Full {{Resolution Image Compression}} with {{Recurrent Neural Networks}}},
    journal={{{IEEE Conference}} on {{Computer Vision}} and {{Pattern Recognition}}},
    author={Toderici, George and Vincent, Damien and Johnston, Nick and Hwang, Sung Jin and Minnen, David and Shor, Joel and Covell, Michele},
    year={2017},
}

@article{minnen2020channel,
    title={Channel-wise autoregressive entropy models for learned image compression},
    author={Minnen, David and Singh, Saurabh},
    journal={International Conference on Image Processing},
    year={2020},
}

@inbook{Bishop1998,
    author={Bishop, Christopher M.},
    title={Latent Variable Models},
    year={1998},
}

@article{balleVariationalImageCompression2018,
    title={Variational Image Compression with a Scale Hyperprior},
    author={Ballé, Johannes and Minnen, David and Singh, Saurabh and Hwang, Sung Jin and Johnston, Nick},
    year={2018},
    journal={ICLR},
}

@article{li2018learning,
    title={Learning convolutional networks for content-weighted image compression},
    author={Li, Mu and Zuo, Wangmeng and Gu, Shuhang and Zhao, Debin and Zhang, David},
    journal={CVPR},
    year={2018},
}

@article{agustsson2017soft,
    title={Soft-to-hard vector quantization for end-to-end learning compressible representations},
    author={Agustsson, Eirikur and Mentzer, Fabian and Tschannen, Michael and Cavigelli, Lukas and Timofte, Radu and Benini, Luca and Gool, Luc V},
    journal={NeurIPS},
    year={2017},
}

@article{mentzer2018conditional,
    title={Conditional probability models for deep image compression},
    author={Mentzer, Fabian and Agustsson, Eirikur and Tschannen, Michael and Timofte, Radu and Van Gool, Luc},
    journal={CVPR},
    year={2018},
}

@article{mijaresiverduScalableReducedComplexityCompression2023,
    title={A {{Scalable Reduced-Complexity Compression}} of {{Hyperspectral Remote Sensing Images Using Deep Learning}}},
    author={Mijares i Verdú, Sebastià and Ballé, Johannes and Laparra, Valero and Bartrina-Rapesta, Joan and Hernández-Cabronero, Miguel and Serra-Sagristà, Joan},
    year={2023},
    journal={Remote Sensing},
}

@article{bengio2013estimating,
    title={Estimating or propagating gradients through stochastic neurons for conditional computation},
    author={Bengio, Yoshua and L{\'e}onard, Nicholas and Courville, Aaron},
    journal={arXiv preprint arXiv:1308.3432},
    year={2013},
}

@article{park2019deepsdf,
    title={Deepsdf: Learning continuous signed distance functions for shape representation},
    author={Park, Jeong Joon and Florence, Peter and Straub, Julian and Newcombe, Richard and Lovegrove, Steven},
    journal={CVPR},
    year={2019},
}

@article{chen2019learning,
    title={Learning implicit fields for generative shape modeling},
    author={Chen, Zhiqin and Zhang, Hao},
    journal={CVPR},
    year={2019},
}

@article{mescheder2019occupancy,
    title={Occupancy networks: Learning 3d reconstruction in function space},
    author={Mescheder, Lars and Oechsle, Michael and Niemeyer, Michael and Nowozin, Sebastian and Geiger, Andreas},
    journal={CVPR},
    year={2019},
}

@article{blelloch2001introduction,
    title={Introduction to data compression},
    author={Blelloch, Guy E and others},
}

@book{elements_information_theory,
    author={Cover, Thomas M. and Thomas, Joy A.},
    title={Elements of Information Theory},
    year={2006},
}

@article{Hinton1994,
    author={Hinton, Geoffrey E. and Zemel, Richard S.},
    title={Autoencoders, Minimum Description Length and Helmholtz Free Energy},
    journal={NeurIPS},
    year={1994},
}

@article{Masci2011,
    author={Masci, Jonathan and Meier, Ueli and Cireşan, Dan and Schmidhuber, Jürgen},
    title={Stacked Convolutional Auto-Encoders for Hierarchical Feature Extraction},
    journal={International Conference on Artificial Neural Networks},
    year={2011},
}

@article{he2016deep,
    title={Deep residual learning for image recognition},
    author={He, Kaiming and Zhang, Xiangyu and Ren, Shaoqing and Sun, Jian},
    journal={CVPR},
    year={2016},
}

@article{9506497,
    author={Tsubota, Koki and Aizawa, Kiyoharu},
    journal={2021 IEEE International Conference on Image Processing (ICIP)}, 
    title={Comprehensive Comparisons Of Uniform Quantizers For Deep Image Compression}, 
    year={2021},
}

@article{bahdanau2014neural,
    title={Neural machine translation by jointly learning to align and translate},
    author={Bahdanau, Dzmitry and Cho, Kyunghyun and Bengio, Yoshua},
    journal={arXiv preprint arXiv:1409.0473},
    year={2014},
}

@article{balle2020nonlinear,
    title={Nonlinear transform coding},
    author={Ball{\'e}, Johannes and Chou, Philip A and Minnen, David and Singh, Saurabh and Johnston, Nick and Agustsson, Eirikur and Hwang, Sung Jin and Toderici, George},
    journal={IEEE Journal of Selected Topics in Signal Processing},
    year={2020},
}

@article{mp3,
    author={Brandenburg, Karlheinz and Stoll, Gerhard},
    title={ISO-MPEG-1 Audio: A Generic Standard for Coding of High-Quality Digital Audio},
    journal={Journal of the Audio Engineering Society},
    year={1994},
}

@article{hevc,    
    author={Sullivan, Gary J. and Ohm, Jens-Rainer and Han, Woo-Jin and Wiegand, Thomas},    
    journal={IEEE Transactions on Circuits and Systems for Video Technology},     
    title={Overview of the High Efficiency Video Coding (HEVC) Standard},     
    year={2012},    
}

@article{jpeg,
    author={Wallace, Gregory K.},
    title={The JPEG Still Picture Compression Standard},
    year={1991},
    journal={Commun. ACM},
}

@article{caron2021emerging,
    title={Emerging properties in self-supervised vision transformers},
    author={Caron, Mathilde and Touvron, Hugo and Misra, Ishan and J{\'e}gou, Herv{\'e} and Mairal, Julien and Bojanowski, Piotr and Joulin, Armand},
    journal={ICCV},
    year={2021},
}

@article{radford2021learning,
    title={Learning transferable visual models from natural language supervision},
    author={Radford, Alec and Kim, Jong Wook and Hallacy, Chris and Ramesh, Aditya and Goh, Gabriel and Agarwal, Sandhini and Sastry, Girish and Askell, Amanda and Mishkin, Pamela and Clark, Jack and others},
    journal={ICML},
    year={2021},
}

@article{wang2023ssl4eo,
    title={SSL4EO-S12: A large-scale multimodal, multitemporal dataset for self-supervised learning in Earth observation [Software and Data Sets]},
    author={Wang, Yi and Braham, Nassim Ait Ali and Xiong, Zhitong and Liu, Chenying and Albrecht, Conrad M and Zhu, Xiao Xiang},
    journal={IEEE Geoscience and Remote Sensing Magazine},
    year={2023},
}

@article{jacob2018quantization,
    title={Quantization and training of neural networks for efficient integer-arithmetic-only inference},
    author={Jacob, Benoit and Kligys, Skirmantas and Chen, Bo and Zhu, Menglong and Tang, Matthew and Howard, Andrew and Adam, Hartwig and Kalenichenko, Dmitry},
    journal={CVPR},
    year={2018},
}

@article{yu2023language,
    title={Language Model Beats Diffusion--Tokenizer is Key to Visual Generation},
    author={Yu, Lijun and Lezama, Jos{\'e} and Gundavarapu, Nitesh B and Versari, Luca and Sohn, Kihyuk and Minnen, David and Cheng, Yong and Gupta, Agrim and Gu, Xiuye and Hauptmann, Alexander G and others},
    journal={arXiv preprint arXiv:2310.05737},
    year={2023},
}

@article{downlink_latency,
    author={Vasisht, Deepak and Shenoy, Jayanth and Chandra, Ranveer},
    title={L2D2: low latency distributed downlink for LEO satellites},
    year={2021},
    journal={ACM SIGCOMM Conference},
}

@misc{clauson_etal_2024,
    author={Clauson, J. and Cantrell, S. and Vrabel, J. and Oeding, J. and Ranjitkar, B. and Rusten, T. and Ramaseri, S. and Casey, K.},
    title={Earth Observing Sensing Satellites Online Compendium: U.S. Geological Survey digital data},
    year={2024},
}

@article{du2021neural,
    title={Neural radiance flow for 4d view synthesis and video processing},
    author={Du, Yilun and Zhang, Yinan and Yu, Hong-Xing and Tenenbaum, Joshua B and Wu, Jiajun},
    journal={ICCV},
    year={2021},
}

@article{he2022masked,
    title={Masked autoencoders are scalable vision learners},
    author={He, Kaiming and Chen, Xinlei and Xie, Saining and Li, Yanghao and Doll{\'a}r, Piotr and Girshick, Ross},
    journal={CVPR},
    year={2022},
}

@article{furutanpey2024fool,
    title={FOOL: Addressing the Downlink Bottleneck in Satellite Computing with Neural Feature Compression},
    author={Furutanpey, Alireza and Zhang, Qiyang and Raith, Philipp and Pfandzelter, Tobias and Wang, Shangguang and Dustdar, Schahram},
    journal={arXiv preprint arXiv:2403.16677},
    year={2024},
}

@article{video_earth_observation,
    author={Wang, Xu and Hu, Ruimin and Wang, Zhongyuan and Xiao, Jing},
    journal={IEEE Signal Processing Letters}, 
    title={Virtual Background Reference Frame Based Satellite Video Coding}, 
    year={2018},
}

@article{du2024earth+,
    title={Earth+: on-board satellite imagery compression leveraging historical earth observations},
    author={Du, Kuntai and Cheng, Yihua and Olsen, Peder and Noghabi, Shadi and Chandra, Ranveer and Jiang, Junchen},
    journal={arXiv preprint arXiv:2403.11434},
    year={2024},
}

@article{furtuanpey2024frankensplit,
    title={FrankenSplit: Efficient Neural Feature Compression with Shallow Variational Bottleneck Injection for Mobile Edge Computing},
    author={Furtuanpey, Alireza and Raith, Philipp and Dustdar, Schahram},
    journal={IEEE Transactions on Mobile Computing},
    year={2024},
}

@book{bracewell1986fourier,
    title={The Fourier transform and its applications},
    author={Bracewell, Ronald Newbold and Bracewell, Ronald N},
    year={1986},
}

@article{matsubara2022supervised,
    title={Supervised compression for resource-constrained edge computing systems},
    author={Matsubara, Yoshitomo and Yang, Ruihan and Levorato, Marco and Mandt, Stephan},
    journal={Proceedings of the IEEE/CVF Winter Conference on Applications of Computer Vision},
    year={2022},
}

@book{wavelets,
    author={Daubechies, Ingrid},
    title={Ten lectures on wavelets},
    year={1992},
}

@article{neural_transform_coding,
    author={Goyal, V.K.},
    journal={IEEE Signal Processing Magazine}, 
    title={Theoretical foundations of transform coding}, 
    year={2001},
}

@article{ssim,
    author={Zhou Wang and Bovik, A.C. and Sheikh, H.R. and Simoncelli, E.P.},
    journal={IEEE Transactions on Image Processing}, 
    title={Image quality assessment: from error visibility to structural similarity}, 
    year={2004},
}

@article{early_nn_compression,
    author={Sonehara and Kawato and Miyake and Nakane},
    journal={International 1989 Joint Conference on Neural Networks}, 
    title={Image data compression using a neural network model}, 
    year={1989},
}

@article{yang2023introduction,
    title={An introduction to neural data compression},
    author={Yang, Yibo and Mandt, Stephan and Theis, Lucas and others},
    journal={Foundations and Trends{\textregistered} in Computer Graphics and Vision},
    year={2023},
}

@article{balle2016end,
    title={End-to-end optimization of nonlinear transform codes for perceptual quality},
    author={Ball{\'e}, Johannes and Laparra, Valero and Simoncelli, Eero P},
    journal={2016 Picture Coding Symposium (PCS)},
    year={2016},
}

@article{theis2022lossy,
    title={Lossy image compression with compressive autoencoders},
    author={Theis, Lucas and Shi, Wenzhe and Cunningham, Andrew and Husz{\'a}r, Ferenc},
    journal={ICLR},
    year={2022},
}

@article{liu_survey_2015,
	title={A survey of remote-sensing big data},
	journal={Frontiers in Environmental Science},
	author={Liu, Peng},
	year={2015},
}

@article{zhang2021universal,
    title={Universal rate-distortion-perception representations for lossy compression},
    author={Zhang, George and Qian, Jingjing and Chen, Jun and Khisti, Ashish},
    journal={NeurIPS},
    year={2021},
}

@article{agustsson2023multi,
    title={Multi-realism image compression with a conditional generator},
    author={Agustsson, Eirikur and Minnen, David and Toderici, George and Mentzer, Fabian},
    journal={CVPR},
    year={2023},
}

@article{singh2020end,
    title={End-to-end learning of compressible features},
    author={Singh, Saurabh and Abu-El-Haija, Sami and Johnston, Nick and Ball{\'e}, Johannes and Shrivastava, Abhinav and Toderici, George},
    journal={2020 IEEE International Conference on Image Processing (ICIP)},
    year={2020},
}

@article{chen2021nerv,
    title={Nerv: Neural representations for videos},
    author={Chen, Hao and He, Bo and Wang, Hanyu and Ren, Yixuan and Lim, Ser Nam and Shrivastava, Abhinav},
    journal={NeurIPS},
    year={2021},
}

@article{dubois2021lossy,
    title={Lossy compression for lossless prediction},
    author={Dubois, Yann and Bloem-Reddy, Benjamin and Ullrich, Karen and Maddison, Chris J},
    journal={NeurIPS},
    year={2021},
}

@article{Wang_2017, 
    title={Sparse Representation-Based Hyperspectral Data Processing: Lossy Compression}, 
    journal={IEEE Journal of Selected Topics in Applied Earth Observations and Remote Sensing}, 
    author={Wang, Hairong and Celik, Turgay}, 
    year={2017}, 
}

@article{Ertem_2020, 
    title={Superpixel based compression of hyperspectral image with modified dictionary and sparse representation}, 
    journal={International Journal of Remote Sensing}, 
    author={Ertem, Adem and Karaca, Ali Can and Urhan, Oğuzhan and Güllü, Mehmet Kemal}, 
    year={2020}, 
}

@article{Wu_2015, 
    title={Hyperspectral data compression using double sparsity model}, 
    journal={2015 7th Workshop on Hyperspectral Image and Signal Processing: Evolution in Remote Sensing (WHISPERS)}, 
    author={Wu, Qian and Zhang, Rong and Wang, Fan}, 
    year={2015},
}

@misc{lieberman_neural_2023,
      title={Neural Image Compression: Generalization, Robustness, and Spectral Biases}, 
      author={Kelsey Lieberman and James Diffenderfer and Charles Godfrey and Bhavya Kailkhura},
      year={2023},
      eprint={2307.08657},
      archivePrefix={arXiv},
      primaryClass={eess.IV},
      url={https://arxiv.org/abs/2307.08657}, 
}

@article{lu_onboard_2024,
	title = {Onboard {AI} for {Fire} {Smoke} {Detection} {Using} {Hyperspectral} {Imagery}: {An} {Emulation} for the {Upcoming} {Kanyini} {Hyperscout}-2 {Mission}},
	volume = {17},
	journal = {IEEE Journal of Selected Topics in Applied Earth Observations and Remote Sensing},
	author = {Lu, Sha and Jones, Eriita and Zhao, Liang and Sun, Yu and Qin, Kai and Liu, Jixue and Li, Jiuyong and Abeysekara, Prabath and Mueller, Norman and Oliver, Simon and O'Hehir, Jim and Peters, Stefan},
	year = {2024},
}

@book{H.264,
    author={Richardson, Iain E.},
    title={The H.264 Advanced Video Compression Standard},
    publisher={Wiley},
    year={2010},
    edition={2nd},
}

@article{zhang_compressing_2024,
    title={Compressing Hyperspectral Images Into Multilayer Perceptrons Using Fast-Time Hyperspectral Neural Radiance Fields},
    journal={{IEEE} Geoscience and Remote Sensing Letters},
    author={Zhang, L. and Pan, T. and Liu, J. and Han, L.},
    year={2024},
}

@article{xiang_discrete_2023,
    title={Discrete Wavelet Transform-Based Gaussian Mixture Model for Remote Sensing Image Compression},
    journal={{IEEE} Transactions on Geoscience and Remote Sensing},
    author={Xiang, S. and Liang, Q. and Fang, L.},
    year={2023},
}

@article{xiang_remote_2024,
    title={Remote Sensing Image Compression Based on High-Frequency and Low-Frequency Components},
    journal={{IEEE} Transactions on Geoscience and Remote Sensing},
    author={Xiang, S. and Liang, Q.},
    year={2024},
}

@article{kong_end--end_2021,
    title={End-to-end multispectral image compression framework based on adaptive multiscale feature extraction},
    journal={Journal of Electronic Imaging},
    author={Kong, F. and Zhao, S. and Li, Y. and Li, D.},
    year={2021},
}

@article{kong_spectralspatial_2021,
    title={Spectral–spatial feature partitioned extraction based on {CNN} for multispectral image compression},
    journal={Remote Sensing},
    author={Kong, F. and Hu, K. and Li, Y. and Li, D. and Zhao, S.},
    year={2021},
}

@article{zhu_research_2022,
    title={Research on {UAV} remote sensing multispectral image compression based on {CNN}},
    journal={International Conference on Geology, Mapping and Remote Sensing},
    author={Zhu, M. and Li, G. and Zhang, W.},
    year={2022},
}

@article{gao_mixed_2023,
    title={Mixed Entropy Model Enhanced Residual Attention Network for Remote Sensing Image Compression},
    journal={Neural Processing Letters},
    author={Gao, J. and Teng, Q. and He, X. and Chen, Z. and Ren, C.},
    year={2023},
}

@article{xiang_remote_2023,
    title={Remote sensing image compression with long-range convolution and improved non-local attention model},
    journal={Signal Processing},
    author={Xiang, S. and Liang, Q.},
    year={2023},
}

@article{cao_spectralspatial_2022,
    title={Spectral–Spatial Feature Completely Separated Extraction with Tensor {CNN} for Multispectral Image Compression},
    journal={Lecture Notes in Electrical Engineering},
    author={Cao, T. and Zhang, N. and Zhao, S. and Hu, K. and Wang, K.},
    year={2022},
}

@article{deng_synthetic_2024,
    title={Synthetic Aperture Radar Image Compression Based on Low-Frequency Rejection and Quality Map Guidance},
    journal={Remote Sensing},
    author={Deng, J. and Huang, L.},
    year={2024},
}

@article{fu_remote_2023,
    title={Remote Sensing Image Compression Based on the Multiple Prior Information},
    journal={Remote Sensing},
    author={Fu, C. and Du, B.},
    year={2023},
}

@article{potsdam,
	author={Rottensteiner, F. and Sohn, G. and Jung, J. and Gerke, M. and Baillard, C. and Benitez, S. and Breitkopf, U.},
	journal={ISPRS Annals of the Photogrammetry, Remote Sensing and Spatial Information Sciences},
	title={THE ISPRS BENCHMARK ON URBAN OBJECT CLASSIFICATION AND 3D BUILDING RECONSTRUCTION},
	year={2012},
}

@article{di_learned_2022,
    title={Learned Compression Framework with Pyramidal Features and Quality Enhancement for {SAR} Images},
    journal={{IEEE} Geoscience and Remote Sensing Letters},
    author={Di, Z. and Chen, X. and Wu, Q. and Shi, J. and Feng, Q. and Fan, Y.},
    year={2022},
}

@article{arithmetic_coding,
    author={Rissanen, J. and Langdon, G. G.},
    journal={IBM Journal of Research and Development}, 
    title={Arithmetic Coding}, 
    year={1979},
}

@article{huffman,
    author={Huffman, David A.},
    journal={Proceedings of the IRE}, 
    title={A Method for the Construction of Minimum-Redundancy Codes}, 
    year={1952},
}

@article{smith2023earthpt,
    title={EarthPT: a foundation model for Earth Observation},
    author={Smith, Michael J and Fleming, Luke and Geach, James},
    journal={NeurIPS Workshop on Tackling Climate Change with Machine Learning},
    year={2023},
}

@article{satmae2022,
    title={Sat{MAE}: Pre-training Transformers for Temporal and Multi-Spectral Satellite Imagery},
    author={Yezhen Cong and Samar Khanna and Chenlin Meng and Patrick Liu and Erik Rozi and Yutong He and Marshall Burke and David B. Lobell and Stefano Ermon},
    journal={NeurIPS},
    year={2022},
}

@article{jakubik2023foundation,
    title={Foundation models for generalist geospatial artificial intelligence},
    author={Jakubik, Johannes and Roy, Sujit and Phillips, CE and Fraccaro, Paolo and Godwin, Denys and Zadrozny, Bianca and Szwarcman, Daniela and Gomes, Carlos and Nyirjesy, Gabby and Edwards, Blair and others},
    journal={arXiv preprint arXiv:2310.18660},
    year={2023},
}

@article{sun2022ringmo,
    title={{RingMo: A remote sensing foundation model with masked image modeling}},
    author={Sun, Xian and Wang, Peijin and Lu, Wanxuan and Zhu, Zicong and Lu, Xiaonan and He, Qibin and Li, Junxi and Rong, Xuee and Yang, Zhujun and Chang, Hao and others},
    journal={IEEE Transactions on Geoscience and Remote Sensing},
    year={2022},
}

@article{liu2024remoteclip,
    title={{RemoteCLIP: A vision language foundation model for remote sensing}},
    author={Liu, Fan and Chen, Delong and Guan, Zhangqingyun and Zhou, Xiaocong and Zhu, Jiale and Ye, Qiaolin and Fu, Liyong and Zhou, Jun},
    journal={IEEE Transactions on Geoscience and Remote Sensing},
    year={2024},
}

@article{wang2022advancing,
    title={Advancing plain vision transformer toward remote sensing foundation model},
    author={Wang, Di and Zhang, Qiming and Xu, Yufei and Zhang, Jing and Du, Bo and Tao, Dacheng and Zhang, Liangpei},
    journal={IEEE Transactions on Geoscience and Remote Sensing},
    year={2022},
}

@article{guo2024skysense,
    title={{Skysense: A multi-modal remote sensing foundation model towards universal interpretation for earth observation imagery}},
    author={Guo, Xin and Lao, Jiangwei and Dang, Bo and Zhang, Yingying and Yu, Lei and Ru, Lixiang and Zhong, Liheng and Huang, Ziyuan and Wu, Kang and Hu, Dingxiang and others},
    journal={CVPR},
    year={2024},
}

@article{li2023assessment,
    title={{Assessment of IBM and NASA's geospatial foundation model in flood inundation mapping}},
    author={Li, Wenwen and Lee, Hyunho and Wang, Sizhe and Hsu, Chia-Yu and Arundel, Samantha T},
    journal={arXiv preprint arXiv:2309.14500},
    year={2023},
}

@article{li_intelligent_2023,
    title={An Intelligent Image Compression Method Based on Generative Adversarial Networks for Satellites},
    journal={International Symposium on Computer Engineering and Intelligent Communications},
    author={Li, J. and Ye, Y. and Liu, B.},
    year={2023},
}

@article{kuester_approach_2020,
    title={An approach to near-lossless hyperspectral data compression using deep autoencoder},
    journal={Proceedings of {SPIE} - The International Society for Optical Engineering},
    author={Kuester, J. and Gross, W. and Middelmann, W.},
    year={2020},
}

@article{alves_de_oliveira_reduced-complexity_2021,
    title={Reduced-Complexity End-to-End Variational Autoencoder for on Board Satellite Image Compression},
    journal={Remote Sensing},
    author={Alves de Oliveira, Vinicius and Chabert, Marie and Oberlin, Thomas and Poulliat, Charly and Bruno, Mickael and Latry, Christophe and Carlavan, Mikael and Henrot, Simon and Falzon, Frederic and Camarero, Roberto},
    year={2021},
}

@article{wilkinson_environmental_2024,
    title={Environmental impacts of earth observation data in the constellation and cloud computing era},
    journal={Science of The Total Environment},
    author={Wilkinson, R. and Mleczko, M. M. and Brewin, R. J. W. and Gaston, K. J. and Mueller, M. and Shutler, J. D. and Yan, X. and Anderson, K.},
    year={2024},
}

@article{miller_deep_2024,
	title={Deep {Learning} for {Satellite} {Image} {Time} {Series} {Analysis}: {A} {Review}},
	journal={IEEE Geoscience and Remote Sensing Magazine},
	author={Miller, Lynn and Pelletier, Charlotte and Webb, Geoffrey I.},
	year={2024},
}

@misc{WoCGraphic2024AICompressor,
    title={Citation Report graphic is derived from Clarivate Web of Science.},
    url={https://webofscience.com},
    year={2024},
}

@misc{WoCGraphic2000-2024AICompressorHistory,
    title={Certain data included herein are derived from Clarivate Web of Science.},
    url={https://webofscience.com},
    year={2024},
}

@article{guo_big_2017,
    title={Big Earth Data: a new challenge and opportunity for Digital Earth’s development},
    journal={International Journal of Digital Earth},
    author={Guo, Huadong and Liu, Zhen and Jiang, Hao and Wang, Changlin and Liu, Jie and Liang, Dong},
    year={2017},
}

@article{chien_tensor-factorized_2018,
    title={Tensor-factorized neural networks},
    journal={{IEEE} Transactions on Neural Networks and Learning Systems},
    author={Chien, Jen-Tzung and Bao, Yi-Ting},
    year={2018},
}

@article{xu_synthetic_2022,
    title={Synthetic Aperture Radar Image Compression Based on a Variational Autoencoder},
    journal={{IEEE} Geoscience and Remote Sensing Letters},
    author={Xu, Qihan and Xiang, Yunfan and Di, Zhixiong and Fan, Yibo and Feng, Quanyuan and Wu, Qiang and Shi, Jiangyi},
    year={2022},
}

@article{yeh_new_2005,
    title={The new {CCSDS} image compression recommendation},
    journal={{IEEE} Aerospace Conference},
    author={Yeh, Pen-Shu and Armbruster, P. and Kiely, A. and Masschelein, B. and Moury, G. and Schaefer, C. and Thiebaut, C.},
    year={2005},
}

@article{zhao_sparse_2021,
    title={Sparse Flow Adversarial Model For Robust Image Compression},
    journal={{IEEE} International Conference on Acoustics, Speech and Signal Processing},
    author={Zhao, Shihui and Yang, Shuyuan and Liu, Zhi and Feng, Zhixi and Liu, Xu},
    year={2021},
}

@article{mao_least_2017,
    title={Least Squares Generative Adversarial Networks},
    journal={ICCV},
    author={Mao, Xudong and Li, Qing and Xie, Haoran and Lau, Raymond Y.K. and Wang, Zhen and Smolley, Stephen Paul},
    year={2017},
}

@article{mirza_conditional_2014,
    title={Conditional Generative Adversarial Nets},
    journal={arXiv preprint arXiv:1411.1784},
    author={Mirza, Mehdi and Osindero, Simon},
    year={2014},
}

@article{yang_compression_2005,
    title={Compression of Remote Sensing Images Based on Ridgelet and Neural Network},
    journal={Advances in Neural Networks},
    author={Yang, Shuyuan and Wang, Min and Jiao, Licheng},
    year={2005},
}

@article{machairas_133_2020,
    title={A 13.3 Gbps 9/7M Discrete Wavelet Transform for {CCSDS} 122.0-B-1 Image Data Compression on a Space-Grade {SRAM} {FPGA}},
    journal={Electronics},
    author={Machairas, Elias and Kranitis, Nektarios},
    year={2020},
}

@article{zhang_compression_2015,
    title={Compression of hyperspectral remote sensing images by tensor approach},
    journal={Neurocomputing special issue Advances in Self-Organizing Maps},
    author={Zhang, Lefei and Zhang, Liangpei and Tao, Dacheng and Huang, Xin and Du, Bo},
    year={2015},
}

@article{li_tensor_2010,
    title={Tensor completion for on-board compression of hyperspectral images},
    journal={{IEEE} International Conference on Image Processing},
    author={Li, Nan and Li, Baoxin},
    year={2010},
}

@article{sidiropoulos_multi-way_2012,
    title={Multi-Way Compressed Sensing for Sparse Low-Rank Tensors},
    journal={{IEEE} Signal Processing Letters},
    author={Sidiropoulos, Nicholas D. and Kyrillidis, Anastasios},
    year={2012},
}

@article{karami_compression_2012,
    title={Compression of Hyperspectral Images Using Discerete Wavelet Transform and Tucker Decomposition},
    journal={{IEEE} Journal of Selected Topics in Applied Earth Observations and Remote Sensing},
    author={Karami, Azam and Yazdi, Mehran and Mercier, Grégoire},
    year={2012},
}

@online{akarami_hyperspectral_2010,
    title={Hyperspectral Image Compression Based on Tucker Decomposition and Discrete Cosine Transform},
    author={{A.Karami} and {M.Yazdi} and {A.Zolghadre} and {A.Zolghadre}},
    year={2010},
}

@article{li_multispectral_2019,
    title={Multispectral Transforms Using Convolution Neural Networks for Remote Sensing Multispectral Image Compression},
    journal={Remote Sensing},
    author={Li, Jin and Liu, Zilong},
    year={2019},
}

@article{li_compression_2014,
    title={Compression of Multispectral Images with Comparatively Few Bands Using Posttransform Tucker Decomposition},
    journal={Mathematical Problems in Engineering},
    author={Li, Jin and Xing, Fei and You, Zheng},
    year={2014},
}

@article{vali_deep_2020,
	title={Deep {Learning} for {Land} {Use} and {Land} {Cover} {Classification} {Based} on {Hyperspectral} and {Multispectral} {Earth} {Observation} {Data}: {A} {Review}},
	journal={Remote Sensing},
	author={Vali, Ava and Comai, Sara and Matteucci, Matteo},
	year={2020},
}

@article{qian_hyperspectral_2021,
	title={Hyperspectral {Satellites}, {Evolution}, and {Development} {History}},
	journal={IEEE Journal of Selected Topics in Applied Earth Observations and Remote Sensing},
	author={Qian, Shen-En},
	year={2021},
}

@article{bajwa_hyperspectral_2004,
	title={Hyperspectral image data mining for band selection in agricultural applications},
	journal={Transactions of the ASAE. American Society of Agricultural Engineers},
	author={Bajwa, Sreekala and Bajcsy, Peter and Groves, P. and Tian, L.F.},
	year={2004},
}

@article{zribi_analysis_2019,
	title={Analysis of {L}-{Band} {SAR} {Data} for {Soil} {Moisture} {Estimations} over {Agricultural} {Areas} in the {Tropics}},
	journal={Remote Sensing},
	author={Zribi, Mehrez and Muddu, Sekhar and Bousbih, Safa and Al Bitar, Ahmad and Tomer, Sat Kumar and Baghdadi, Nicolas and Bandyopadhyay, Soumya},
	year={2019},
}

@article{li_compression_2017,
    title={Compression of hyper-spectral images using an accelerated nonnegative tensor decomposition},
    journal={Open Physics},
    author={Li, Jin and Liu, Zilong},
    year={2017},
}

@article{huang2023compressing,
    title={Compressing multidimensional weather and climate data into neural networks},
    author={Langwen Huang and Torsten Hoefler},
    journal={ICLR},
    year={2023},
}

@article{klower2021compressing,
    title={Compressing atmospheric data into its real information content},
    author={Kl{\"o}wer, Milan and Razinger, Miha and Dominguez, Juan J and D{\"u}ben, Peter D and Palmer, Tim N},
    journal={Nature Computational Science},
    year={2021},
}

@article{stevens2019dyamond,
    title={DYAMOND: the DYnamics of the Atmospheric general circulation Modeled On Non-hydrostatic Domains},
    author={Stevens, Bjorn and Satoh, Masaki and Auger, Ludovic and Biercamp, Joachim and Bretherton, Christopher S and Chen, Xi and D{\"u}ben, Peter and Judt, Falko and Khairoutdinov, Marat and Klocke, Daniel and others},
    journal={Progress in Earth and Planetary Science},
    year={2019},
}

@article{liang2022sz3,
    title={Sz3: A modular framework for composing prediction-based error-bounded lossy compressors},
    author={Liang, Xin and Zhao, Kai and Di, Sheng and Li, Sihuan and Underwood, Robert and Gok, Ali M and Tian, Jiannan and Deng, Junjing and Calhoun, Jon C and Tao, Dingwen and others},
    journal={IEEE Transactions on Big Data},
    year={2022},
}

@article{ballester2019tthresh,
    title={TTHRESH: Tensor compression for multidimensional visual data},
    author={Ballester-Ripoll, Rafael and Lindstrom, Peter and Pajarola, Renato},
    journal={IEEE transactions on visualization and computer graphics},
    year={2019},
}

@article{lindstrom2014fixed,
    title={Fixed-rate compressed floating-point arrays},
    author={Lindstrom, Peter},
    journal={IEEE transactions on visualization and computer graphics},
    year={2014},
}

@article{dupont2022coin++,
    title={COIN++: Neural Compression Across Modalities},
    author={Dupont, Emilien and Loya, Hrushikesh and Alizadeh, Milad and Golinski, Adam and Teh, Yee Whye and Doucet, Arnaud},
    journal={Transactions on Machine Learning Research},
    year={2022},
}

@article{underwood2022understanding,
    title={Understanding the effects of modern compressors on the community earth science model},
    author={Underwood, Robert and Bessac, Julie and Di, Sheng and Cappello, Franck},
    journal={International Workshop on Data Analysis and Reduction for Big Scientific Data (DRBSD)},
    year={2022},
}

@article{zender2016bit,
    title={Bit Grooming: statistically accurate precision-preserving quantization with compression, evaluated in the netCDF Operators (NCO, v4. 4.8+)},
    author={Zender, Charles S},
    journal={Geoscientific Model Development},
    year={2016},
}

@article{gonzalez-conejero_jpeg2000_2010,
    title={{JPEG}2000 Encoding of Remote Sensing Multispectral Images With No-Data Regions},
    journal={{IEEE} Geoscience and Remote Sensing Letters},
    author={Gonzalez-Conejero, Jorge and Bartrina-Rapesta, Joan and Serra-Sagrista, Joan},
    year={2010},
}

@article{hou_improving_2000,
    title={Improving {JPEG} performance in conjunction with cloud editing for remote sensing applications},
    journal={{IEEE} Transactions on Geoscience and Remote Sensing},
    author={Hou, P. and Petrou, M. and Underwood, C.I. and Hojjatoleslami, A.},
    year={2000},
}

@article{luigi_dragotti_compression_2000,
    title={Compression of multispectral images by three-dimensional {SPIHT} algorithm},
    journal={{IEEE} Transactions on Geoscience and Remote Sensing},
    author={Luigi Dragotti, P. and Poggi, G. and Ragozini, A.R.P.},
    year={2000},
}

@article{du_low-complexity_2008,
    title={Low-Complexity Principal Component Analysis for Hyperspectral Image Compression},
    journal={The International Journal of High Performance Computing Applications},
    author={Du, Qian and Fowler, James E.},
    year={2008},
}

@article{du_hyperspectral_2007,
    title={Hyperspectral Image Compression Using {JPEG}2000 and Principal Component Analysis},
    journal={{IEEE} Geoscience and Remote Sensing Letters},
    author={Du, Qian and Fowler, James E.},
    year={2007},
}

@article{lim_compression_2001,
    title={Compression for hyperspectral images using three dimensional wavelet transform},
    journal={IGARSS},
    author={Lim, Sunghyun and Sohn, Kwanghoon and Lee, Chulhee},
    year={2001},
}

@article{markman_hyperspectral_2001,
    title={Hyperspectral image coding using 3D transforms},
    journal={International Conference on Image Processing},
    author={Markman, D. and Malah, D.},
    year={2001},
}

@article{fuchs2023hyspecnet,
    title={Hyspecnet-11k: A large-scale hyperspectral dataset for benchmarking learning-based hyperspectral image compression methods},
    author={Fuchs, Martin Hermann Paul and Demir, Beg{\"u}m},
    journal={IGARSS},
    year={2023},
}

@techreport{bodnar2024aurora,
    author={Bodnar, Cristian and Bruinsma, Wessel and Lucic, Ana and Stanley, Megan and Brandstetter, Johannes and Garvan , Patrick and Riechert, Maik and Weyn, Jonathan and Dong, Haiyu and Vaughan, Anna and Gupta, Jayesh and Thambiratnam, Kit and Archibald, Alex and Heider, Elizabeth and Welling, Max and Turner, Richard and Perdikaris, Paris},
    title={Aurora: A Foundation Model of the Atmosphere},
    year={2024},
}

@article{hong2023spectralgpt,
    title={SpectralGPT: Spectral foundation model},
    author={Hong, Danfeng and Zhang, Bing and Li, Xuyang and Li, Yuxuan and Li, Chenyu and Yao, Jing and Yokoya, Naoto and Li, Hao and Jia, Xiuping and Plaza, Antonio and others},
    journal={arXiv preprint arXiv:2311.07113},
    year={2023},
}

@article{kay2015community,
    title={The Community Earth System Model (CESM) large ensemble project: A community resource for studying climate change in the presence of internal climate variability},
    author={Kay, Jennifer E and Deser, Clara and Phillips, A and Mai, A and Hannay, Cecile and Strand, Gary and Arblaster, Julie Michelle and Bates, SC and Danabasoglu, Gokhan and Edwards, James and others},
    journal={Bulletin of the American Meteorological Society},
    year={2015},
}

@article{hersbach2020era5,
    title={The ERA5 global reanalysis},
    author={Hersbach, Hans and Bell, Bill and Berrisford, Paul and Hirahara, Shoji and Hor{\'a}nyi, Andr{\'a}s and Mu{\~n}oz-Sabater, Joaqu{\'\i}n and Nicolas, Julien and Peubey, Carole and Radu, Raluca and Schepers, Dinand and others},
    journal={Quarterly Journal of the Royal Meteorological Society},
    year={2020},
}

@article{mooers2023comparing,
    title={Comparing storm resolving models and climates via unsupervised machine learning},
    author={Mooers, Griffin and Pritchard, Mike and Beucler, Tom and Srivastava, Prakhar and Mangipudi, Harshini and Peng, Liran and Gentine, Pierre and Mandt, Stephan},
    journal={Scientific Reports},
    year={2023},
}

@article{schneider2017climate,
    title={Climate goals and computing the future of clouds},
    author={Schneider, Tapio and Teixeira, Jo{\~a}o and Bretherton, Christopher S and Brient, Florent and Pressel, Kyle G and Sch{\"a}r, Christoph and Siebesma, A Pier},
    journal={Nature Climate Change},
    year={2017},
}

@article{palmer2014climate,
    title={Climate forecasting: Build high-resolution global climate models},
    author={Palmer, Tim},
    journal={Nature},
    year={2014},
}

@article{stevens2013climate,
    title={What are climate models missing?},
    author={Stevens, Bjorn and Bony, Sandrine},
    journal={science},
    year={2013},
}

@article{gorski2005healpix,
    title={HEALPix: A framework for high-resolution discretization and fast analysis of data distributed on the sphere},
    author={Gorski, Krzysztof M and Hivon, Eric and Banday, Anthony J and Wandelt, Benjamin D and Hansen, Frode K and Reinecke, Mstvos and Bartelmann, Matthia},
    journal={The Astrophysical Journal},
    year={2005},
}

@article{rasp2024weatherbench,
    title={WeatherBench 2: A benchmark for the next generation of data-driven global weather models},
    author={Rasp, Stephan and Hoyer, Stephan and Merose, Alexander and Langmore, Ian and Battaglia, Peter and Russell, Tyler and Sanchez-Gonzalez, Alvaro and Yang, Vivian and Carver, Rob and Agrawal, Shreya and others},
    journal={Journal of Advances in Modeling Earth Systems},
    year={2024},
}

@article{baker2017toward,
    title={Toward a multi-method approach: Lossy data compression for climate simulation data},
    author={Baker, Allison H and Xu, Haiying and Hammerling, Dorit M and Li, Shaomeng and Clyne, John P},
    journal={High Performance Computing},
    year={2017},
}

@article{poppick2018statistical,
    title={A statistical analysis of compressed climate model data},
    author={Poppick, Andrew and Nardi, Joseph and Feldman, Noah and Baker, A and Hammerling, D},
    journal={Proc. DRBSD},
    year={2018},
}

@article{tancik2020fourfeat,
    title={Fourier Features Let Networks Learn High Frequency Functions in Low Dimensional Domains},
    author={Matthew Tancik and Pratul P. Srinivasan and Ben Mildenhall and Sara Fridovich-Keil and Nithin Raghavan and Utkarsh Singhal and Ravi Ramamoorthi and Jonathan T. Barron and Ren Ng},
    journal={NeurIPS},
    year={2020},
}

@article{garcia-vilchez_extending_2009,
    title={Extending the {CCSDS} Recommendation for Image Data Compression for Remote Sensing Scenarios},
    journal={{IEEE} Transactions on Geoscience and Remote Sensing},
    author={Garcia-Vilchez, Fernando and Serra-Sagrista, Joan},
    year={2009},
}

@article{lee_hyperspectral_2002,
    title={Hyperspectral image cube compression combining {JPEG}-2000 and spectral decorrelation},
    journal={IGARSS},
    author={Lee, H.S. and Younan, N.H. and King, R.L.},
    year={2002},
}

@article{yu_image_2009,
    title={Image compression systems on board satellites},
    journal={Acta Astronautica},
    author={Yu, Guoxia and Vladimirova, Tanya and Sweeting, Martin N.},
    year={2009},
}

@article{hacihaliloglu_dct_2004,
    title={{DCT} and {DWT} based image compression in remote sensing images},
    journal={{IEEE} Antennas and Propagation Society Symposium},
    author={Hacihaliloglu, I. and Karta, M.},
    year={2004},
}

@article{rahimi2007random,
    title={Random features for large-scale kernel machines},
    author={Rahimi, Ali and Recht, Benjamin},
    journal={NeurIPS},
    year={2007},
}

@article{allen2023fewshot,
    title={Fewshot learning on global multimodal embeddings for earth observation tasks},
    author={Allen, Matt and Dorr, Francisco and Gallego-Mejia, Joseph A and Mart{\'\i}nez-Ferrer, Laura and Jungbluth, Anna and Kalaitzis, Freddie and Ramos-Poll{\'a}n, Ra{\'u}l},
    journal={arXiv preprint arXiv:2310.00119},
    year={2023},
}

@article{ye_gfscompnet_2024,
    title={{GFSCompNet}: remote sensing image compression network based on global feature-assisted segmentation},
    journal={Multimedia Tools and Applications},
    author={Ye, W. and Lei, W. and Zhang, W. and Yu, T. and Feng, X.},
    year={2024},
}

@article{anuradha_efficient_2024,
    title={Efficient Compression for Remote Sensing: Multispectral Transform and Deep Recurrent Neural Networks for Lossless Hyper-Spectral Imagine},
    journal={International Journal of Advanced Computer Science and Applications},
    author={Anuradha, D. and Sekhar, G.C. and Mishra, A. and Thapar, P. and El-Ebiary, Y.A.B. and Syamala, M.},
    year={2024},
}

@article{li_remote_2023,
    title={Remote Sensing Image Compression Method Based on Implicit Neural Representation},
    journal={International Conference on Computing and Pattern Recognition},
    author={Li, Xin and Sun, Baile and Liao, Jixiu and Zhao, Xiaofei},
    year={2023},
}

@misc{cloudnativegeo2024survey,
    author={{Cloud-Native Geospatial Foundation}},
    title={Survey for current practices storing embeddings in GeoParquet},
    url={https://github.com/cloudnativegeo/geo-embeddings-survey},
    year={2024}
}

@article{GORELICK201718,
    title={Google Earth Engine: Planetary-scale geospatial analysis for everyone},
    journal={Remote Sensing of Environment},
    year={2017},
    author={Noel Gorelick and Matt Hancher and Mike Dixon and Simon Ilyushchenko and David Thau and Rebecca Moore},
}

@article{schramm2021openeo,
    author={Schramm, Matthias and Pebesma, Edzer and Milenković, Milutin and Foresta, Luca and Dries, Jeroen and Jacob, Alexander and Wagner, Wolfgang and Mohr, Matthias and Neteler, Markus and Kadunc, Miha and Miksa, Tomasz and Kempeneers, Pieter and Verbesselt, Jan and Gößwein, Bernhard and Navacchi, Claudio and Lippens, Stefaan and Reiche, Johannes},
    title={The openEO API–Harmonising the Use of Earth Observation Cloud Services Using Virtual Data Cube Functionalities},
    journal={Remote Sensing},
    year={2021},
}

@article{ahmed1974,
    author={Nasir Ahmed and T. Natarajan and K.R. Rao},
    title={Discrete Cosine Transform},
    journal={IEEE Transactions on Computers},
    year={1974},
}

@article{tucker1966,
    author={L.R. Tucker},
    title={Some Mathematical Notes on Three-Mode Factor Analysis},
    journal={Psychometrika},
    year={1966},
}

@article{candes1999,
    author={Emmanuel J. Candès and David L. Donoho},
    title={Ridgelets: A Key to Higher-Dimensional Intermittency?},
    journal={Philosophical Transactions of the Royal Society of London},
    year={1999},
}

@article{esser_taming_2021,
    title={Taming Transformers for High-Resolution Image Synthesis},
    journal={CVPR},
    author={Esser, Patrick and Rombach, Robin and Ommer, Björn},
    year={2021},
}

@article{wilkinson2016fair,
    title={The FAIR Guiding Principles for scientific data management and stewardship},
    author={Wilkinson, Mark D and Dumontier, Michel and Aalbersberg, IJsbrand Jan and Appleton, Gabrielle and Axton, Myles and Baak, Arie and Blomberg, Niklas and Boiten, Jan-Willem and da Silva Santos, Luiz Bonino and Bourne, Philip E and others},
    journal={Scientific data},
    year={2016},
}

@article{rodriguez_mapping_2021,
    title={Mapping oil palm density at country scale: An active learning approach},
    journal={Remote Sensing of Environment},
    author={Rodríguez, Andrés C. and D'Aronco, Stefano and Schindler, Konrad and Wegner, Jan D.},
    year={2021},
}

@article{li_forest_2020,
    title={Forest aboveground biomass estimation using Landsat 8 and Sentinel-1A data with machine learning algorithms},
    journal={Scientific Reports},
    author={Li, Yingchang and Li, Mingyang and Li, Chao and Liu, Zhenzhen},
    year={2020},
}

@misc{oms_2024,
    author={Copernicus},
    year={2024},
    title={Copernicus European Flood Awareness System},
}

@article{de2019global,
    title={Global buffering of temperatures under forest canopies},
    author={De Frenne, Pieter and Zellweger, Florian and Rodr{\'\i}guez-S{\'a}nchez, Francisco and Scheffers, Brett R and Hylander, Kristoffer and Luoto, Miska and Vellend, Mark and Verheyen, Kris and Lenoir, Jonathan},
    journal={Nature Ecology \& Evolution},
    year={2019},
}

@article{hoang2021mapping,
    title={Mapping the deforestation footprint of nations reveals growing threat to tropical forests},
    author={Hoang, Nguyen Tien and Kanemoto, Keiichiro},
    journal={Nature Ecology \& Evolution},
    year={2021},
}

@article{potapov2021mapping,
    title={Mapping global forest canopy height through integration of GEDI and Landsat data},
    author={Potapov, Peter and Li, Xinyuan and Hernandez-Serna, Andres and Tyukavina, Alexandra and Hansen, Matthew C and Kommareddy, Anil and Pickens, Amy and Turubanova, Svetlana and Tang, Hao and Silva, Carlos Edibaldo and others},
    journal={Remote Sensing of Environment},
    year={2021},
}

@article{dubayah2020global,
    title={The Global Ecosystem Dynamics Investigation: High-resolution laser ranging of the Earth’s forests and topography},
    author={Dubayah, Ralph and Blair, James Bryan and Goetz, Scott and Fatoyinbo, Lola and Hansen, Matthew and Healey, Sean and Hofton, Michelle and Hurtt, George and Kellner, James and Luthcke, Scott and others},
    journal={Science of remote sensing},
    year={2020},
}

@article{manning2018redefining,
    title={Redefining ecosystem multifunctionality},
    author={Manning, Peter and Van Der Plas, Fons and Soliveres, Santiago and Allan, Eric and Maestre, Fernando T and Mace, Georgina and Whittingham, Mark J and Fischer, Markus},
    journal={Nature ecology \& evolution},
    year={2018},
}

@article{migliavacca2021three,
    title={The three major axes of terrestrial ecosystem function},
    author={Migliavacca, Mirco and Musavi, Talie and Mahecha, Miguel D and Nelson, Jacob A and Knauer, J{\"u}rgen and Baldocchi, Dennis D and Perez-Priego, Oscar and Christiansen, Rune and Peters, Jonas and Anderson, Karen and others},
    journal={Nature},
    year={2021},
}

@article{jetz2019essential,
    title={Essential biodiversity variables for mapping and monitoring species populations},
    author={Jetz, Walter and McGeoch, Melodie A and Guralnick, Robert and Ferrier, Simon and Beck, Jan and Costello, Mark J and Fernandez, Miguel and Geller, Gary N and Keil, Petr and Merow, Cory and others},
    journal={Nature ecology \& evolution},
    year={2019},
}

@article{brede2022non,
    title={Non-destructive estimation of individual tree biomass: Allometric models, terrestrial and UAV laser scanning},
    author={Brede, Benjamin and Terryn, Louise and Barbier, Nicolas and Bartholomeus, Harm M and Bartolo, Ren{\'e}e and Calders, Kim and Derroire, G{\'e}raldine and Moorthy, Sruthi M Krishna and Lau, Alvaro and Levick, Shaun R and others},
    journal={Remote Sensing of Environment},
    year={2022},
}

@article{lang2023high,
    title={A high-resolution canopy height model of the Earth},
    author={Lang, Nico and Jetz, Walter and Schindler, Konrad and Wegner, Jan Dirk},
    journal={Nature Ecology \& Evolution},
    year={2023},
}

@article{tolan2024very,
    title={Very high resolution canopy height maps from RGB imagery using self-supervised vision transformer and convolutional decoder trained on aerial lidar},
    author={Tolan, Jamie and Yang, Hung-I and Nosarzewski, Benjamin and Couairon, Guillaume and Vo, Huy V and Brandt, John and Spore, Justine and Majumdar, Sayantan and Haziza, Daniel and Vamaraju, Janaki and others},
    journal={Remote Sensing of Environment},
    year={2024},
}

@article{renier2023transparency,
    title={Transparency, traceability and deforestation in the Ivorian cocoa supply chain},
    author={Renier, C{\'e}cile and Vandromme, Mathil and Meyfroidt, Patrick and Ribeiro, Vivian and Kalischek, Nikolai and Zu Ermgassen, Erasmus KHJ},
    journal={Environmental Research Letters},
    year={2023},
}

@article{kalischek2023cocoa,
    title={Cocoa plantations are associated with deforestation in C{\^o}te d’Ivoire and Ghana},
    author={Kalischek, Nikolai and Lang, Nico and Renier, C{\'e}cile and Daudt, Rodrigo Caye and Addoah, Thomas and Thompson, William and Blaser-Hart, Wilma J and Garrett, Rachael and Schindler, Konrad and Wegner, Jan D},
    journal={Nature Food},
    year={2023},
}

@article{karami_hyperspectral_2011,
    title={Hyperspectral image compression based on tucker decomposition and wavelet transform},
    journal={2011 3rd Workshop on Hyperspectral Image and Signal Processing: Evolution in Remote Sensing ({WHISPERS})},
    author={Karami, A. and Yazdi, M. and Mercier, G.},
    year={2011},
}

@article{chen_low-rank_2016,
    title={A Low-Rank Tensor Decomposition Based Hyperspectral Image Compression Algorithm},
    author={Zhang, Mengfei and Du, Bo and Zhang, Lefei and Li, Xuelong},
    year={2016},
}

@article{du_pltd_2017,
    title={{PLTD}: Patch-Based Low-Rank Tensor Decomposition for Hyperspectral Images},
    journal={{IEEE} Transactions on Multimedia},
    author={Du, Bo and Zhang, Mengfei and Zhang, Lefei and Hu, Ruimin and Tao, Dacheng},
    year={2017},
}

@article{santoro2021cci,
    title={ESA Biomass Climate Change Initiative (Biomass CCI): Global datasets of forest above-ground biomass for the years 2010, 2017 and 2018, v2},
    author={Santoro, Maurizio and Cartus, Oliver},
    journal={Centre for Environmental Data Analysis},
    year={2021},
}

@article{margarit2009operational,
    title={Operational ship monitoring system based on synthetic aperture radar processing},
    author={Margarit, Gerard and Barba Milan{\'e}s, Jos{\'e} A and Tabasco, Antonio},
    journal={Remote Sensing},
    year={2009},
}

@article{zhu2010novel,
    title={A novel hierarchical method of ship detection from spaceborne optical image based on shape and texture features},
    author={Zhu, Changren and Zhou, Hui and Wang, Runsheng and Guo, Jun},
    journal={IEEE Transactions on geoscience and remote sensing},
    year={2010},
}

@article{pelich2014ais,
    title={AIS-based evaluation of target detectors and SAR sensors characteristics for maritime surveillance},
    author={Pelich, Ramona and Long{\'e}p{\'e}, Nicolas and Mercier, Gr{\'e}goire and Hajduch, Guillaume and Garello, Ren{\'e}},
    journal={IEEE Journal of Selected Topics in Applied Earth Observations and Remote Sensing},
    year={2014},
}

@article{tang2014compressed,
    title={Compressed-domain ship detection on spaceborne optical image using deep neural network and extreme learning machine},
    author={Tang, Jiexiong and Deng, Chenwei and Huang, Guang-Bin and Zhao, Baojun},
    journal={IEEE transactions on geoscience and remote sensing},
    year={2014},
}

@article{bank2023autoencoders,
    title={Autoencoders},
    author={Bank, Dor and Koenigstein, Noam and Giryes, Raja},
    journal={Machine learning for data science handbook: data mining and knowledge discovery handbook},
    year={2023},
}

@book{mallat1999wavelet,
    title={A wavelet tour of signal processing},
    author={Mallat, St{\'e}phane},
    year={1999},
}

@article{furutanpey2024,
    title={FOOL: Addressing the Downlink Bottleneck in Satellite Computing with Neural Feature Compression}, 
    author={Alireza Furutanpey and Qiyang Zhang and Philipp Raith and Tobias Pfandzelter and Shangguang Wang and Schahram Dustdar},
    year={2024},
    journal={arXiv preprint arXiv:2403.16677},
}

@article{Vuolo_2018, 
    title={How much does multi-temporal Sentinel-2 data improve crop type classification?}, 
    journal={International Journal of Applied Earth Observation and Geoinformation}, 
    author={Vuolo, Francesco and Neuwirth, Martin and Immitzer, Markus and Atzberger, Clement and Ng, Wai-Tim}, 
    year={2018}, 
}

@article{beillouin2020impact,
    title={Impact of extreme weather conditions on European crop production in 2018},
    author={Beillouin, Damien and Schauberger, Bernhard and Bastos, Ana and Ciais, Phillipe and Makowski, David},
    journal={Philosophical Transactions of the Royal Society B},
    year={2020},
}

@article{ramanath_color_2005,
	title = {Color image processing pipeline},
	volume = {22},
	url = {http://ieeexplore.ieee.org/document/1407713/},
	journal = {IEEE Signal Processing Magazine},
	author = {Ramanath, R. and Snyder, W.E. and Yoo, Y. and Drew, M.S.},
	month = jan,
	year = {2005},
}

@article{hristov2020analysis,
    title={Analysis of climate change impacts on EU agriculture by 2050},
    author={Hristov, Jordan and Toreti, Andrea and P{\'e}rez Dom{\'\i}nguez, I and Dentener, Franciscus and Fellmann, Thomas and Elleby, Christian and Ceglar, Andrej and Fumagalli, Davide and Niemeyer, Stefan and Cerrani, Iacopo and others},
    journal={Publications Office of the European Union, Luxembourg, Luxembourg},
    year={2020},
}

@article{white2005global,
    title={A global framework for monitoring phenological responses to climate change},
    author={White, Michael A and Hoffman, Forrest and Hargrove, William W and Nemani, Ramakrishna R},
    journal={Geophysical Research Letters},
    year={2005},
}

@article{berger2012esa,
    title={ESA's sentinel missions in support of Earth system science},
    author={Berger, Michael and Moreno, Jose and Johannessen, Johnny A and Levelt, Pieternel F and Hanssen, Ramon F},
    journal={Remote sensing of environment},
    year={2012},
}

@article{van2019performance,
    title={Performance of the MARS-crop yield forecasting system for the European Union: Assessing accuracy, in-season, and year-to-year improvements from 1993 to 2015},
    author={Van der Velde, M and Nisini, L},
    journal={Agricultural Systems},
    year={2019},
}

@article{KANSAKAR201646,
    title={A review of applications of satellite earth observation data for global societal benefit and stewardship of planet earth},
    journal={Space Policy},
    year={2016},
    author={Pratistha Kansakar and Faisal Hossain},
}

@article{creswell2018generative,
    title={Generative adversarial networks: An overview},
    author={Creswell, Antonia and White, Tom and Dumoulin, Vincent and Arulkumaran, Kai and Sengupta, Biswa and Bharath, Anil A},
    journal={IEEE signal processing magazine},
    year={2018},
}

@article{aslan2018electricity,
    title={Electricity intensity of internet data transmission: Untangling the estimates},
    author={Aslan, Joshua and Mayers, Kieren and Koomey, Jonathan G and France, Chris},
    journal={Journal of industrial ecology},
    year={2018},
}

@article{sumbul2019bigearthnet,
    title={Bigearthnet: A large-scale benchmark archive for remote sensing image understanding},
    author={Sumbul, Gencer and Charfuelan, Marcela and Demir, Beg{\"u}m and Markl, Volker},
    journal={IGARSS},
    year={2019},
}

@article{liu2021swin,
    title={Swin transformer: Hierarchical vision transformer using shifted windows},
    author={Liu, Ze and Lin, Yutong and Cao, Yue and Hu, Han and Wei, Yixuan and Zhang, Zheng and Lin, Stephen and Guo, Baining},
    journal={ICCV},
    year={2021},
}

@article{deng2009imagenet,
    title={Imagenet: A large-scale hierarchical image database},
    author={Deng, Jia and Dong, Wei and Socher, Richard and Li, Li-Jia and Li, Kai and Fei-Fei, Li},
    journal={CVPR},
    year={2009},
}

@article{kesselheim2021juwels,
    title={Juwels booster--a supercomputer for large-scale ai research},
    author={Kesselheim, Stefan and Herten, Andreas and Krajsek, Kai and Ebert, Jan and Jitsev, Jenia and Cherti, Mehdi and Langguth, Michael and Gong, Bing and Stadtler, Scarlet and Mozaffari, Amirpasha and others},
    journal={High Performance Computing: ISC High Performance Digital 2021 International Workshops},
    year={2021},
}

@article{koomey2021does,
    title={Does not compute: Avoiding pitfalls assessing the Internet's energy and carbon impacts},
    author={Koomey, Jonathan and Masanet, Eric},
    journal={Joule},
    year={2021},
}

@book{bishop2006pattern,
    title={Pattern recognition and machine learning},
    author={Bishop, Christopher M and Nasrabadi, Nasser M},
    year={2006},
}

@article{wilson2020case,
    title={The case for Bayesian deep learning},
    author={Wilson, Andrew Gordon},
    journal={arXiv preprint arXiv:2001.10995},
    year={2020},
}

@book{kraslawski201323rd,
    title={23rd European Symposium on Computer Aided Process Engineering},
    author={Kraslawski, Andrzej and Turunen, Ilkka},
    year={2013},
}

@book{lehmann1986testing,
    title={Testing statistical hypotheses},
    author={Lehmann, Erich Leo and Romano, Joseph P and Casella, George},
    year={1986},
}

@article{reichstein2019deep,
    title={Deep learning and process understanding for data-driven Earth system science},
    author={Reichstein, Markus and Camps-Valls, Gustau and Stevens, Bjorn and Jung, Martin and Denzler, Joachim and Carvalhais, Nuno and Prabhat, F},
    journal={Nature},
    year={2019},
}

@article{zerah2023,
    title={Physics-driven probabilistic deep learning for the inversion of physical models with application to phenological parameter retrieval from satellite times series},
    author={Zérah, Yoël and Valero, Silvia and Inglada, Jordi},
    journal={IEEE Transactions on Geoscience and Remote Sensing},
    year={2023},
}

@article{Reza2023,
    author={Asiyabi, Reza Mohammadi and Anghel, Andrei and Rizzoli, Paola and Martone, Michele and Datcu, Mihai},
    journal={IGARSS}, 
    title={Complex-Valued Autoencoder for Multi-Polarization SLC SAR Data Compression with Side Information}, 
    year={2023},
}

@article{Pilikos2022,
    author={Georgios Pilikos and Mario Azcueta and Roberto Camarero and Nicolas Floury},
    journal={International Workshop on On-board Payload Data Compression}, 
    title={Raw Data Compression for Synthetic Aperture Radar using Deep Learning}, 
    year={2022},
}

@article{Fu2023_sar_compression,
    author={Fu, Chuan and Du, Bo and Zhang, Liangpei},
    journal={IEEE Geoscience and Remote Sensing Letters}, 
    title={SAR Image Compression Based on Multi-Resblock and Global Context}, 
    year={2023},
}

@article{Gollin2023,
    author={Gollin, Nicola and Martone, Michele and Krieger, Gerhard and Rizzoli, Paola},
    journal={IGARSS}, 
    title={AI-Based Performance-Optimized Quantization for Future SAR Systems}, 
    year={2023},
}

@article{Kwok1989_baq,
    author={Kwok, R. and Johnson, W.T.K.},
    journal={IEEE Transactions on Geoscience and Remote Sensing}, 
    title={Block adaptive quantization of Magellan SAR data}, 
    year={1989},
}

@phdthesis{Martone2019,
    school={Karlsruhe Institut f{\"u}r Technologie (KIT)},
    title={Onboard Quantization for Interferometric and Multichannel Synthetic Aperture Radar (SAR) Systems},
    year={2019},
    author={Martone, Michele},
}

@article{Attema2010,
    author={Attema, Evert and Cafforio, Ciro and Gottwald, Michael and Guccione, Pietro and Monti Guarnieri, Andrea and Rocca, Fabio and Snoeij, Paul},
    journal={IEEE Geoscience and Remote Sensing Letters}, 
    title={Flexible Dynamic Block Adaptive Quantization for Sentinel-1 SAR Missions}, 
    year={2010},
}

@article{blumenstiel2024multi,
    title={Multi-Spectral Remote Sensing Image Retrieval Using Geospatial Foundation Models},
    author={Blumenstiel, Benedikt and Moor, Viktoria and Kienzler, Romeo and Brunschwiler, Thomas},
    journal={arXiv preprint arXiv:2403.02059},
    year={2024},
}

@article{mirowski2024neural,
    title={Neural Compression of Atmospheric States},
    author={Mirowski, Piotr and Warde-Farley, David and Rosca, Mihaela and Grimes, Matthew Koichi and Hasson, Yana and Kim, Hyunjik and Rey, M{\'e}lanie and Osindero, Simon and Ravuri, Suman and Mohamed, Shakir},
    journal={arXiv preprint arXiv:2407.11666},
    year={2024},
}

@article{han2024cra5,
    title={Cra5: Extreme compression of era5 for portable global climate and weather research via an efficient variational transformer},
    author={Han, Tao and Guo, Song and Xu, Wanghan and Bai, Lei and others},
    journal={arXiv preprint arXiv:2405.03376},
    year={2024},
}

@article{kunkel2014exascale,
    title={Exascale storage systems: an analytical study of expenses},
    author={Kunkel, Julian Martin and Kuhn, Michael and Ludwig, Thomas},
    journal={Supercomputing frontiers and innovations},
    year={2014},
}

@article{cappello2020fulfilling,
    title={Fulfilling the promises of lossy compression for scientific applications},
    author={Cappello, Franck and Di, Sheng and Gok, Ali Murat},
    journal={Driving Scientific and Engineering Discoveries Through the Convergence of HPC, Big Data and AI: 17th Smoky Mountains Computational Sciences and Engineering Conference, SMC 2020, Oak Ridge, TN, USA, August 26-28, 2020, Revised Selected Papers 17},
    year={2020},
}

@article{baker2023structural,
    title={On a structural similarity index approach for floating-point data},
    author={Baker, Allison H and Pinard, Alexander and Hammerling, Dorit M},
    journal={IEEE Transactions on Visualization and Computer Graphics},
    year={2023},
}

@article{danabasoglu2020community,
    title={The community earth system model version 2 (CESM2)},
    author={Danabasoglu, Gokhan and Lamarque, J-F and Bacmeister, J and Bailey, DA and DuVivier, AK and Edwards, Jim and Emmons, LK and Fasullo, John and Garcia, R and Gettelman, Andrew and others},
    journal={Journal of Advances in Modeling Earth Systems},
    year={2020},
}

@article{jungclaus2022icon,
    title={The ICON earth system model version 1.0},
    author={Jungclaus, Johann H and Lorenz, Stephan J and Schmidt, Hauke and Brovkin, Victor and Br{\"u}ggemann, Nils and Chegini, Fatemeh and Cr{\"u}ger, Traute and De-Vrese, Philipp and Gayler, Veronika and Giorgetta, Marco A and others},
    journal={Journal of Advances in Modeling Earth Systems},
    year={2022},
}

@article{mauritsen2019developments,
    title={Developments in the MPI-M Earth System Model version 1.2 (MPI-ESM1. 2) and its response to increasing CO2},
    author={Mauritsen, Thorsten and Bader, J{\"u}rgen and Becker, Tobias and Behrens, J{\"o}rg and Bittner, Matthias and Brokopf, Renate and Brovkin, Victor and Claussen, Martin and Crueger, Traute and Esch, Monika and others},
    journal={Journal of Advances in Modeling Earth Systems},
    year={2019},
}

@article{baker2016evaluating,
    title={Evaluating lossy data compression on climate simulation data within a large ensemble},
    author={Baker, Allison H and Hammerling, Dorit M and Mickelson, Sheri A and Xu, Haiying and Stolpe, Martin B and Naveau, Phillipe and Sanderson, Ben and Ebert-Uphoff, Imme and Samarasinghe, Savini and De Simone, Francesco and others},
    journal={Geoscientific Model Development},
    year={2016},
}

@article{bauer2021digital,
    title={A digital twin of Earth for the green transition},
    author={Bauer, Peter and Stevens, Bjorn and Hazeleger, Wilco},
    journal={Nature Climate Change},
    year={2021},
}

@article{sather2022can,
    title={What can real information content tell us about compressing climate model data?},
    author={Sather, Hayden and Pinard, Alexander and Baker, Allison H and Hammerling, Dorit M},
    journal={International Workshop on Data Analysis and Reduction for Big Scientific Data (DRBSD)},
    year={2022},
}

@article{inness2019cams,
    title={The CAMS reanalysis of atmospheric composition},
    author={Inness, Antje and Ades, Melanie and Agust{\'\i}-Panareda, Anna and Barr{\'e}, J{\'e}r{\^o}me and Benedictow, Anna and Blechschmidt, Anne-Marlene and Dominguez, Juan Jose and Engelen, Richard and Eskes, Henk and Flemming, Johannes and others},
    journal={Atmospheric Chemistry and Physics},
    year={2019},
}

@article{kurihana2022aicca,
    title={AICCA: AI-driven cloud classification atlas},
    author={Kurihana, Takuya and Moyer, Elisabeth J and Foster, Ian T},
    journal={Remote Sensing},
    year={2022},
}

@article{denby2020Discovering,
    title={Discovering the {{Importance}} of {{Mesoscale Cloud Organization Through Unsupervised Classification}}},
    author={Denby, L.},
    year={2020},
    journal={Geophysical Research Letters},
}

@article{denby2023Chartinga,
    title={Charting the {{Realms}} of {{Mesoscale Cloud Organisation}} Using {{Unsupervised Learning}}},
    author={Denby, Leif},
    journal={arXiv preprint arXiv:2309.08567},
    year={2023},
}

@article{Russwurm_2017_CVPR_Workshops,
    author={Russwurm, Marc and Korner, Marco},
    title={Temporal Vegetation Modelling Using Long Short-Term Memory Networks for Crop Identification From Medium-Resolution Multi-Spectral Satellite Images},
    journal={CVPR Workshop},
    year={2017},
}

@article{eaton2003netcdf,
    title={NetCDF Climate and Forecast (CF) metadata conventions},
    author={Eaton, Brian and Gregory, Jonathan and Drach, Bob and Taylor, Karl and Hankin, Steve and Caron, John and Signell, Rich and Bentley, Phil and Rappa, Greg and H{\"o}ck, Heinke and others},
    journal={URL: http://cfconventions. org/Data/cf-conventions/cf-conventions-1.8/cf-conventions. pdf},
    year={2003},
}

@software{cherian2023cfxarray,
    author={Cherian, Deepak and Almansi, Mattia and Bourgault, Pascal and Thyng, Kristen and Thielen, Jonathan and Magin, Justus and Aoun, Abel and Buntemeyer, Lars and Caneill, Romain and Davis, Luke and Fernandes, Filipe and Hauser, Matthias and Heerdegen, Aidan and Kent, Julia and Mankoff, Ken and Müller, Sebastian and Schupfner, Martin and Vo, Tom and Haëck, Clément and Boutte, Jason},
    title={{cf\_xarray}},
    url={https://github.com/xarray-contrib/cf-xarray},
    year={2023},
}

@article{roberts2018climate,
    title={Climate model configurations of the ECMWF Integrated Forecasting System (ECMWF-IFS cycle 43r1) for HighResMIP},
    author={Roberts, Christopher D and Senan, Retish and Molteni, Franco and Boussetta, Souhail and Mayer, Michael and Keeley, Sarah PE},
    journal={Geoscientific model development},
    year={2018},
}

@article{schar2020kilometer,
    title={Kilometer-scale climate models: Prospects and challenges},
    author={Sch{\"a}r, Christoph and Fuhrer, Oliver and Arteaga, Andrea and Ban, Nikolina and Charpilloz, Christophe and Di Girolamo, Salvatore and Hentgen, Laureline and Hoefler, Torsten and Lapillonne, Xavier and Leutwyler, David and others},
    journal={Bulletin of the American Meteorological Society},
    year={2020},
}

@article{palmer2014real,
    title={The real butterfly effect},
    author={Palmer, TN and D{\"o}ring, Andreas and Seregin, G},
    journal={Nonlinearity},
    year={2014},
}

@article{palmer2019scientific,
    title={The scientific challenge of understanding and estimating climate change},
    author={Palmer, Tim and Stevens, Bjorn},
    journal={Proceedings of the National Academy of Sciences},
    year={2019},
}

@article{freischem2024multifractal,
    title={Multifractal analysis for evaluating the representation of clouds in global kilometer-scale models},
    author={Freischem, Lilli J and Weiss, Philipp and Christensen, Hannah M and Stier, Philip},
    journal={Geophysical Research Letters},
    year={2024},
}

@article{mandorli2024assessment,
    title={Assessment of object-based indices to identify convective organization},
    author={Mandorli, Giulio and Stubenrauch, Claudia J},
    journal={Geoscientific Model Development},
    year={2024},
}

@article{vavna2017single,
    title={Single precision in weather forecasting models: An evaluation with the IFS},
    author={V{\'a}{\v{n}}a, Filip and D{\"u}ben, Peter and Lang, Simon and Palmer, Tim and Leutbecher, Martin and Salmond, Deborah and Carver, Glenn},
    journal={Monthly Weather Review},
    year={2017},
}

@article{tinto2019use,
    title={How to use mixed precision in ocean models: Exploring a potential reduction of numerical precision in NEMO 4.0 and ROMS 3.6},
    author={Tint{\'o} Prims, Oriol and Acosta, Mario C and Moore, Andrew M and Castrillo, Miguel and Serradell, Kim and Cort{\'e}s, Ana and Doblas-Reyes, Francisco J},
    journal={Geoscientific Model Development},
    year={2019},
}

@article{guo2023compression,
    title={Compression with bayesian implicit neural representations},
    author={Guo, Zongyu and Flamich, Gergely and He, Jiajun and Chen, Zhibo and Hern{\'a}ndez-Lobato, Jos{\'e} Miguel},
    journal={NeurIPS},
    year={2023},
}

@article{he2024recombiner,
    title={Recombiner: Robust and enhanced compression with bayesian implicit neural representations},
    author={He, Jiajun and Flamich, Gergely and Guo, Zongyu and Hern{\'a}ndez-Lobato, Jos{\'e} Miguel},
    journal={ICLR},
    year={2024},
}

@article{pham2024neural,
    title={Neural NeRF Compression},
    author={Pham, Tuan and Mandt, Stephan},
    journal={ICML},
    year={2024},
}

@article{yang2024lossy,
    title={Lossy image compression with conditional diffusion models},
    author={Yang, Ruihan and Mandt, Stephan},
    journal={NeurIPS},
    year={2024},
}

@article{li2024frequency,
    title={Frequency-Aware Transformer for Learned Image Compression},
    author={Li, Han and Li, Shaohui and Dai, Wenrui and Li, Chenglin and Zou, Junni and Xiong, Hongkai},
    journal={ICLR},
    year={2024},
}

@article{defossez2024high,
    title={High fidelity neural audio compression},
    author={D{\'e}fossez, Alexandre and Copet, Jade and Synnaeve, Gabriel and Adi, Yossi},
    journal={ICLR},
    year={2024},
}

@article{jiang2023mlic,
    title={{MLIC}: Multi-reference entropy model for learned image compression},
    author={Jiang, Wei and Yang, Jiayu and Zhai, Yongqi and Ning, Peirong and Gao, Feng and Wang, Ronggang},
    journal={ACM International Conference on Multimedia},
    year={2023},
}
\end{document}